\documentclass[a4paper,12pt,headsepline,bibliography=totocnumbered,BCOR12mm]{scrbook}
\setkomafont{captionlabel}{\rmfamily\bfseries\normalsize}
\setcapindent{0pt}

\usepackage[utf8]{inputenc}
\usepackage[T1]{fontenc}
\usepackage{lmodern}
\usepackage{textcomp}
\usepackage{graphicx}
\usepackage[ngerman]{babel}

\usepackage{microtype}

\usepackage[babel,german=quotes]{csquotes}

\usepackage{bibgerm}

\usepackage{multirow}

\usepackage{amsmath}
\usepackage{amssymb}

\usepackage{lscape}

\usepackage{upgreek}

\usepackage{booktabs}

\usepackage{pdfpages}

\usepackage[breaklinks=true,colorlinks=true,linkcolor=blue,urlcolor=blue,citecolor=blue]{hyperref}

\hypersetup{
  pdftitle={Diplomarbeit},
  pdfauthor={Benjamin Löwe},
  pdfsubject={Entwicklung eines Gasmoderators für Positronen},
  pdfkeywords={Antiteilchen, Antimaterie, Positron, Moderation, Remoderation, Gas, Stickstoff}
}

\begin{document}

\includepdf[pages={1}]{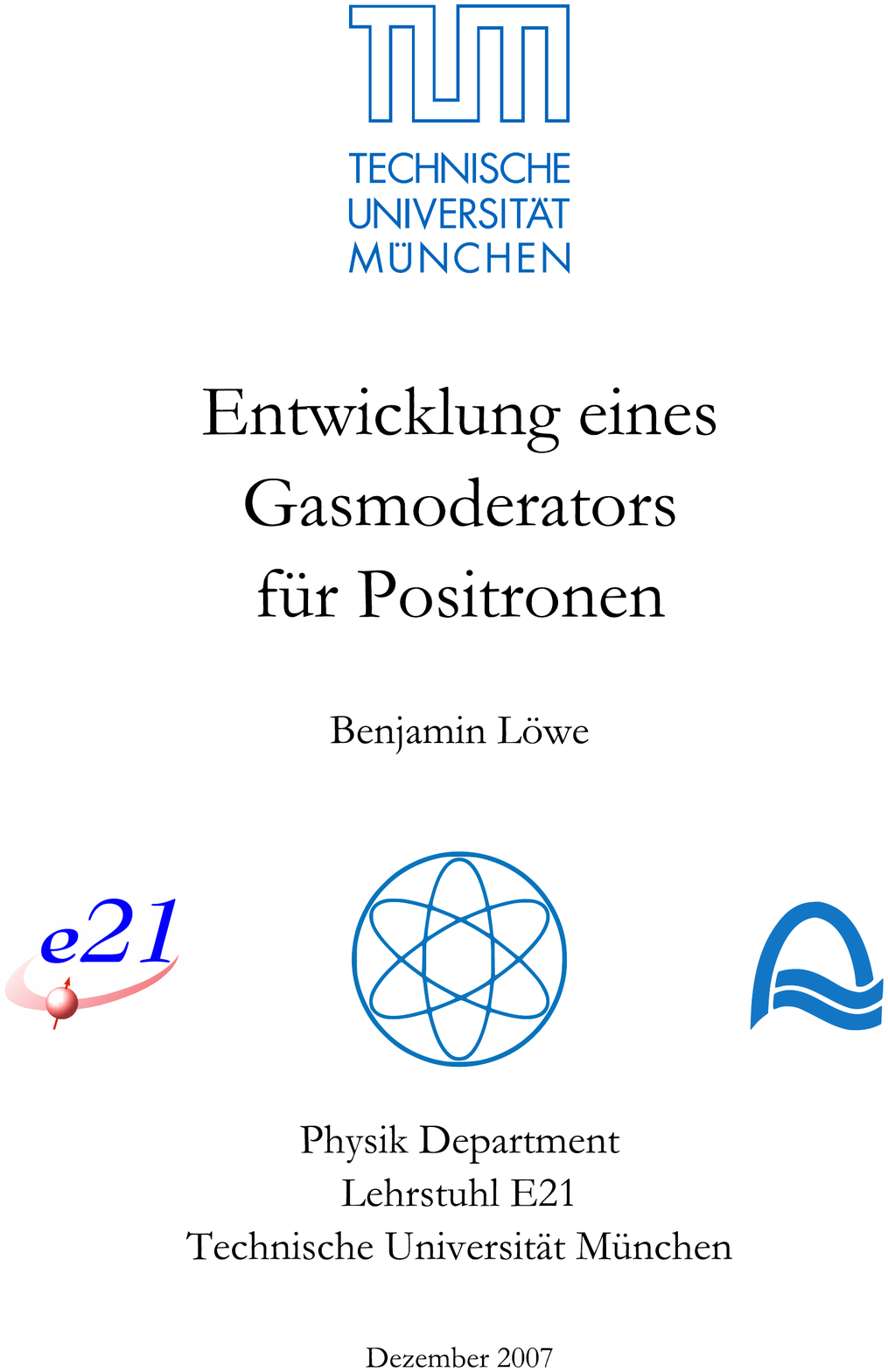}
\newpage
\thispagestyle{empty}

\hyphenation{Gas-atoms Po-si-tron Po-si-tro-nen Wech-sel-wir-kun-gen Po-si-tro-nen-im-plan-ta-tions-tie-fe Po-si-tro-nen-strahl-ex-pe-ri-ment An-ni-hi-la-tions-strah-lung}

\tableofcontents 
\clearpage{}\chapter{Einleitung}
Seit der Entdeckung des Positrons im Jahre 1932 durch Carl D. Anderson wurden im Laufe der Zeit verschiedenste erfolgreiche Methoden entwickelt, für die das Positron als hochsensibles Sondenteilchen für Materialuntersuchungen verwendet wird. Aber auch in der Medizin werden heutzutage die Eigenschaften des Positrons in der sogenannten Positronen-Emissions-Tomographie (PET), beispielsweise zur Krebsdiagnose oder zur Untersuchung von Stoffwechselerkrankungen, genutzt.

Die entscheidende Eigenschaft des Positrons ist, dass es als Antiteilchen des Elektrons mit diesem im Schwerpunktsystem unter Aussendung zweier Gammaquanten im Winkel von 180° zerstrahlt. Über die Analyse dieser Annihilationsstrahlung kann man bei der PET Rückschlüsse auf den Ort der Annihilation ziehen. In der Materialforschung misst man dagegen die Lebensdauer eines Positrons im Festkörper, die Winkelabweichung von 180° oder die Energie der Gammaquanten, um Aussagen über Materialstruktur und -defekte auf atomaren Skalen treffen zu können. Aufgrund der großen Sensitivität der Positronen für Leerstellen, lassen sich Konzentrationen von Einfachleerstellen von $10^{-7}$ nachweisen.

Für viele Experimente, die sich mit der Untersuchung von Materialien beschäftigen, ist es essentiell einen hochbrillanten Strahl an Positronen zur Verfügung zu haben. Anders ausgedrückt entspricht dies einem hochintensiven Positronenstrahl mit möglichst geringem Durchmesser, einer präzise definierten Energie und geringer Winkeldivergenz. Mit \textsc{Nepomuc} existiert am Forschungsreaktor München II der Technischen Universität München bereits eine hochintensive Quelle an langsamen Positronen. Der sogenannte Primärstrahl liefert ca. $5 \cdot 10^8$ Positronen pro Sekunde in einem Durchmesser von etwa 20\,mm. Die Energie dieses monoenergetischen Positronenstrahls kann an der Quelle im Bereich von 15\,eV bis einige keV eingestellt werden. Um die Brillanz dieses Strahls weiter zu verbessern, benutzt man in der Positronenphysik in der Regel keine $\vec{E} \times \vec{B}$-Filter, da dies die Intensität zu stark beeinträchtigen würde. Stattdessen wird das Prinzip der Moderation verwendet, bei der Positronen in einem geeigneten Festkörper wie Wolfram oder Platin auf thermische Energien abgebremst werden und aufgrund ihrer negativen Austrittsarbeit den Festkörper wieder verlassen können. Die Energieverschmierung in Strahlrichtung und senkrecht dazu beträgt nach der Moderation nur noch einige meV. Dieser Effekt wird bei \textsc{Nepomuc} bereits für den Primärstrahl angewendet.

Bereits 1989 zeigten C. M. Surko et al. \cite{Surko1989}, dass es auch in einem geeigneten Gas möglich ist Positronen abzubremsen und auf thermische Energien abzukühlen. Sie speicherten die thermalisierten Positronen in einer Penning-Falle und erzeugten daraus einen gepulsten Strahl.

Ziel der Diplomarbeit ist die Entwicklung eines Moderators, der einen Positronenstrahl in einem Gas abbremst und einen kontinuierlichen, remoderierten Positronenstrahl liefert, ohne dass eine Speicherung nötig ist. Dazu ist es nötig das Konzept der Gasmoderation experimentell zu untersuchen, einen Versuchsaufbau zu entwickeln, die Energieverteilung des moderierten Strahls zu analysieren und die Effizienz des Moderators zu bestimmen. Anhand von ortsaufgelösten Messungen der Annihilationsstrahlung sollen die Verluste im Aufbau lokalisiert werden. Für diese Messungen stand die \textsc{Nepomuc}-Quelle am FRM II zur Verfügung.
\clearpage{}
\clearpage{}\chapter{Das Positron: Sondenteilchen in der Festkörperphysik}

1929 postulierte Paul A. M. Dirac erstmals die Existenz eines positiv geladenen Elektrons \cite{Dirac1930}. Er begründete dies mit Zuständen negativer Energie als Lösung seiner nach ihm benannten Dirac-Gleichung. 1932 entdeckte Carl D. Anderson auf Nebelkammeraufnahmen die Spuren eines Teilchens mit einfach positiver Ladung und einer Masse, die in derselben Größenordnung wie der des Elektrons liegen musste. Er nannte das Teilchen Positron.
\begin{figure}
 \centering
 \includegraphics[width=0.5\textwidth]{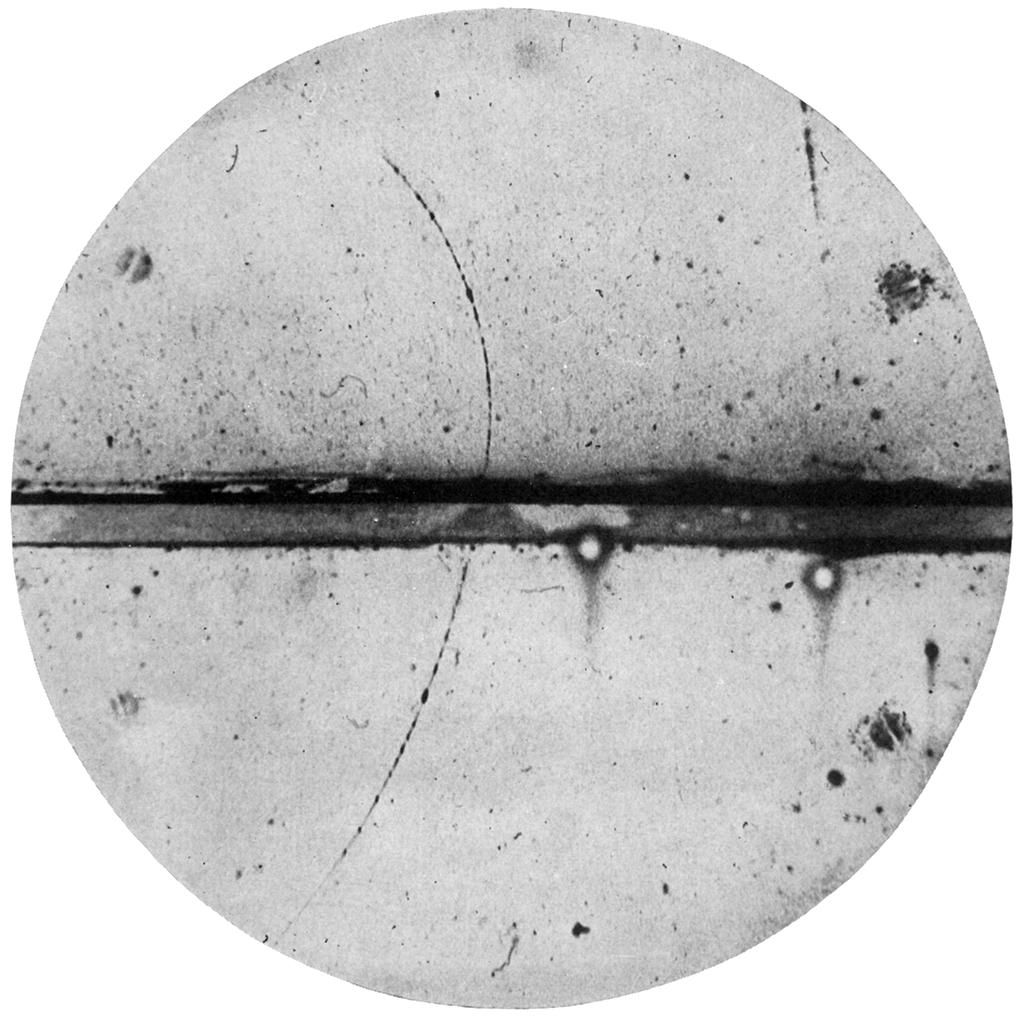}
 \caption{Erstmaliger Nachweis des Positrons in einer Nebelkammer durch Carl D. Anderson: Ein 63\,MeV-Positron dringt von unten durch eine 6\,mm dicke Bleiplatte (aus \cite{Anderson1933}).}
 \label{fig:Entdeckung}
\end{figure}

\section{Eigenschaften des Positrons}
Das Positron (abgekürzt e$^+$) ist das Antiteilchen des Elektrons und hat somit bis auf die entgegengesetzte Ladung dieselben Eigenschaften wie das Elektron (siehe Tabelle \ref{tbl:EigenschaftenPositron}). Das Positron ist im idealen Vakuum stabil. Trifft es jedoch auf ein Elektron, annihiliert es mit diesem - vorzugsweise unter Emission zweier $\upgamma$-Quanten mit einer Gesamtenergie von 1,022\,MeV. Da Positronen daher nicht \enquote{natürlich} vorkommen, müssen sie zunächst erzeugt werden, wenn man mit ihnen experimentieren will.
\begin{table}
 \centering
 \caption{Eigenschaften des Positrons}
 \label{tbl:EigenschaftenPositron}
 \begin{tabular}{l r} \toprule
 Ruhemasse & 511\,keV/c$^2$\\
 Ladung & $+1,6022 \cdot 10^{-19}$\,C\\
 magnetisches Moment & $+9,2848 \cdot 10^{-24}$\,A m$^2$\\
 Spin & 1/2\\
 Lebensdauer im Vakuum & $>10^{21}$\,a\\ \bottomrule
 \end{tabular}
\end{table}

\section{Positronenquellen}
Prinzipiell gibt es zwei Möglichkeiten Positronen zu erzeugen: Mit Hilfe von $\upbeta^+$-Emittern und durch Paarbildung.

\subsection{\texorpdfstring{$\upbeta^+$}{Positronen}-Emitter}
\begin{figure}
 \centering
 \includegraphics[width=0.7\textwidth]{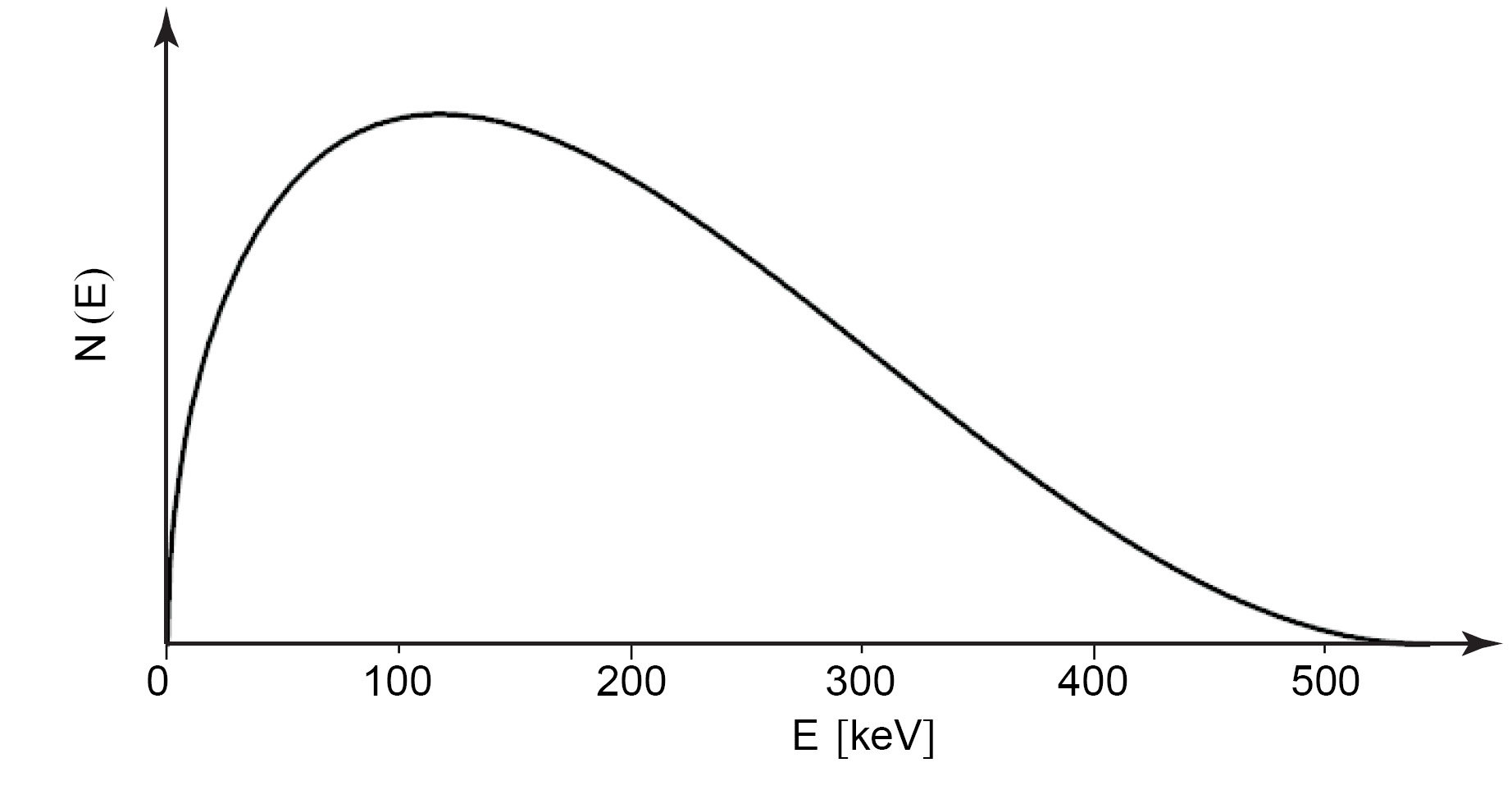}
 \caption{$\upbeta^+$-Spektrum von $^{22}$Na mit einer Endpunktenergie von 544\,keV \cite{DrBenno}.}
 \label{fig:Na22-Spektrum}
\end{figure}

Die Erzeugung von Positronen durch $\upbeta^+$-Emitter stellt die heute einfachste Möglichkeit dar ein Positronenstrahlexperiment aufzubauen. $\upbeta^+$-Emitter sind kommerziell erhältlich und können relativ einfach z.\,B. durch Protonenbeschuss eines Isotops hergestellt werden. Es entsteht dabei aufgrund des Protonenüberschusses ein instabiler Kern, der durch folgende Reaktion zerfallen kann:
\begin{equation}
 ^A_Z X \rightarrow ^A_{Z-1} Y + e^+ + \nu_e \label{beta+}
\end{equation}
Bei dieser Reaktion geht kinetische Energie nicht nur auf das Positron über, sondern auch an das Elektronneutrino. Somit erhält man ein kontinuierliches Spektrum an Positronen verschiedener Energie mit einer vom verwendeten $\upbeta^+$-Emitter abhängigen, maximalen Energie, der sogenannten Endpunktenergie. Abbildung \ref{fig:Na22-Spektrum} zeigt ein solches $\upbeta^+$-Spektrum am Beispiel von $^{22}$Na. Reaktion \ref{beta+} konkurriert jedoch mit dem sogenannten Elektroneneinfang, bei dem ein Proton mit einem Elektron aus der Atomhülle zu einem Neutron reagiert:
\begin{equation}
 p + e^- \rightarrow n + \nu_e 
\end{equation}
Die Positronenausbeute gibt an, mit welcher Wahrscheinlichkeit ein Kern unter $\upbeta^+$-Emission zerfällt.

\subsubsection{$^{22}$Na}
\begin{figure}
 \centering
 \includegraphics[width=0.6\textwidth]{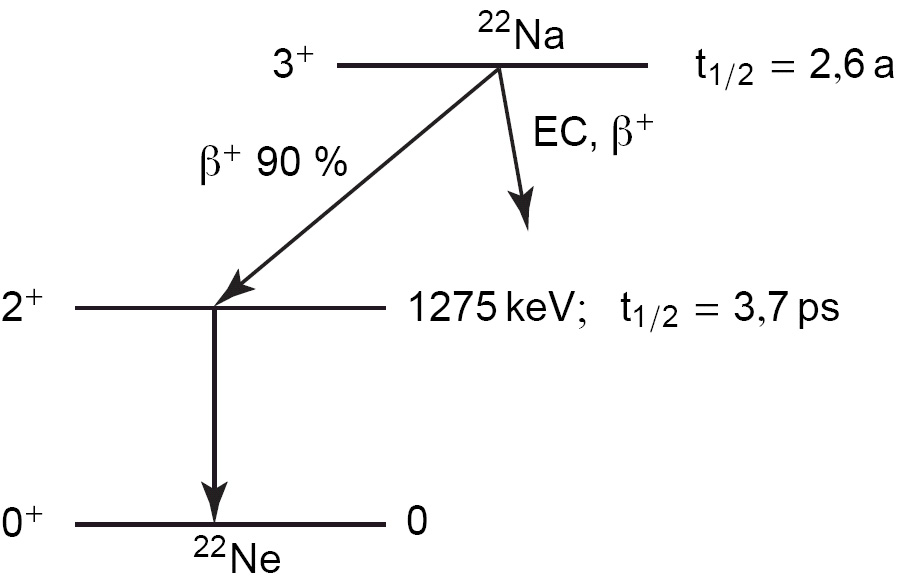}
 \caption{Zerfallsschema von $^{22}$Na \cite{SchatzWeidinger}}
 \label{fig:Zerfallsschema-Na22}
\end{figure}

Als Standardquelle für Laborpositronenstrahlen hat sich $^{22}$Na vor allem aufgrund seiner langen Halbwertszeit von 2,6 Jahren und seiner guten Positronenausbeute von 90\,\% durchgesetzt. Dies ermöglicht Messungen über einen längeren Zeitraum bei gleichbleibend hoher Positronenintensität. Abbildung \ref{fig:Zerfallsschema-Na22} zeigt das Zerfallsschema von $^{22}$Na. Darin wird eine weitere Eigenschaft des $^{22}$Na deutlich. Es zerfällt zunächst in einen kurzlebigen angeregten Zustand des $^{22}$Ne, um danach unter Aussendung eines $\upgamma$-Quants mit einer Energie von 1,275\,MeV in den Grundzustand zu zerfallen. Dieses $\upgamma$-Quant kann als Startsignal für Messungen der Positronenlebensdauer verwendet werden.

\subsubsection{$\upbeta^+$-Quellen in der Medizin}
In der Medizin werden Positronenemitter bei der Positronen-Emissions-To\-mo\-gra\-phie (PET) verwendet. Die PET ist ein bildgebendes Verfahren, bei dem einem Patienten eine mit einem $\upbeta^+$-Strahler markierte Substanz injiziert wird und mit Hilfe der Annihilationsstrahlung der Ort der Anreicherung dieses Radiopharmakon im Organismus sichtbar gemacht werden kann. Die PET wird vor allem zur Untersuchung stoffwechselbedingter Krankheiten eingesetzt. Es werden verschiedenste $\upbeta^+$-Emitter eingesetzt, deren Halbwertszeiten von nur 75 Sekunden ($^{82}$Rb) bis 20 Minuten ($^{11}$C) reichen.

\subsection{Positronen durch Paarbildung}
Die zweite Möglichkeit Positronen zu erzeugen ist die Paarbildung. Aufgrund der bekannten Masse-Energie-Beziehung $E = m c^2$ ist es möglich Teilchen aus elektromagnetischer Strahlung zu erzeugen. Wegen der Erhaltungssätze für Ladung und Leptonenzahl wird jedoch immer ein Positron zusammen mit seinem Antiteilchen, dem Elektron, erzeugt. Die benötigte Energie des $\upgamma$-Quants beträgt demnach mindestens der doppelten Ruheenergie des Positrons, nämlich 1,022\,MeV. Jedoch kann dieser Prozess aufgrund der Viererimpulserhaltung nur im Coulombfeld eines Atomkerns oder Elektrons geschehen, wobei der Wirkungsquerschnitt für Paarbildung durch $\upgamma$-Strahlung unterhalb von 20\,MeV in etwa proportional zum Quadrat der Kernladungszahl $Z$ ist.

\subsubsection{Beschleunigerquellen}
Bei Beschleunigerquellen wird die zur Paarbildung benötigte $\upgamma$-Strahlung aus der Bremsstrahlung von Elektronen gewonnen. Durch einen Linearbeschleuniger beschleunigte Elektronen werden dabei auf ein Target (z.\,B. Wolfram) geschossen, in dem die Positronen durch Paarbildung entstehen. Da der Strahl bei Beschleunigern gepulst ist, entsteht ein ebenfalls gepulster Positronenstrahl.

\subsubsection{Reaktorquellen}
\begin{figure}
 \centering
 \includegraphics[width=0.8\textwidth]{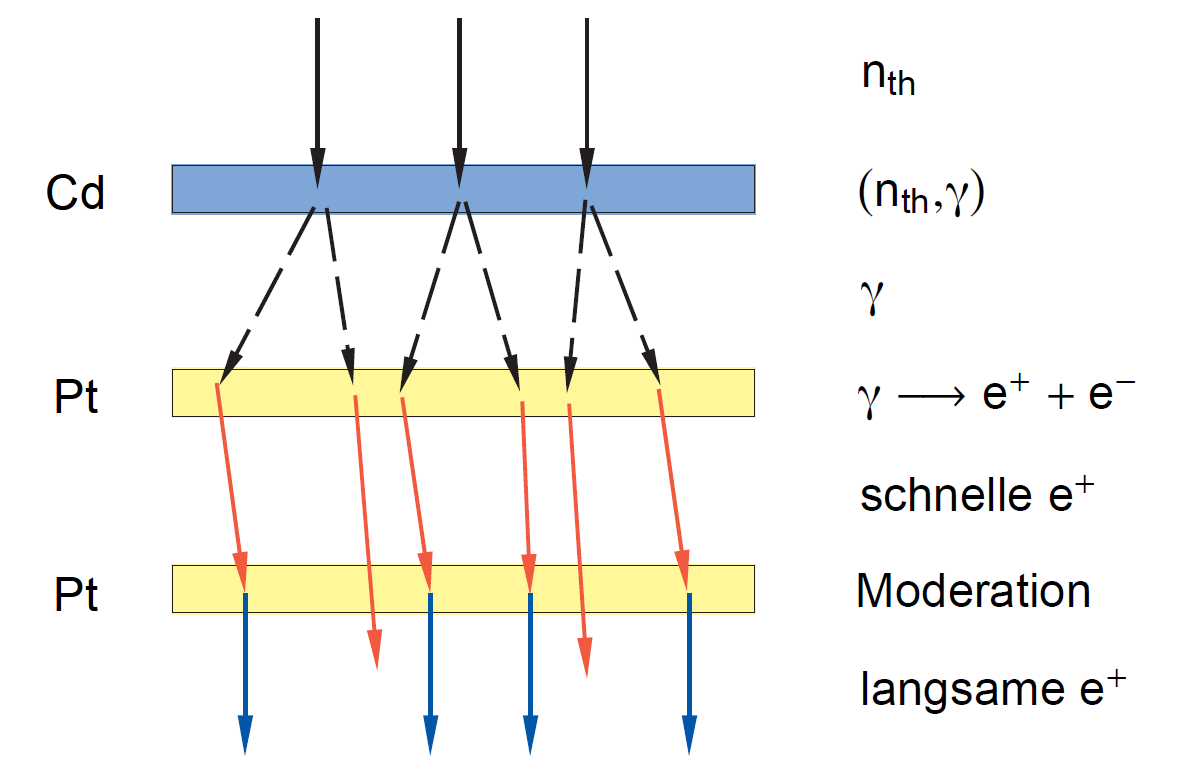}
 \caption{Erzeugung von Positronen durch thermische Neutronen \cite{DiplPio}}
 \label{fig:NEPOMUC-Schema}
\end{figure}

\begin{figure}
 \centering
 \includegraphics[width=0.98\textwidth]{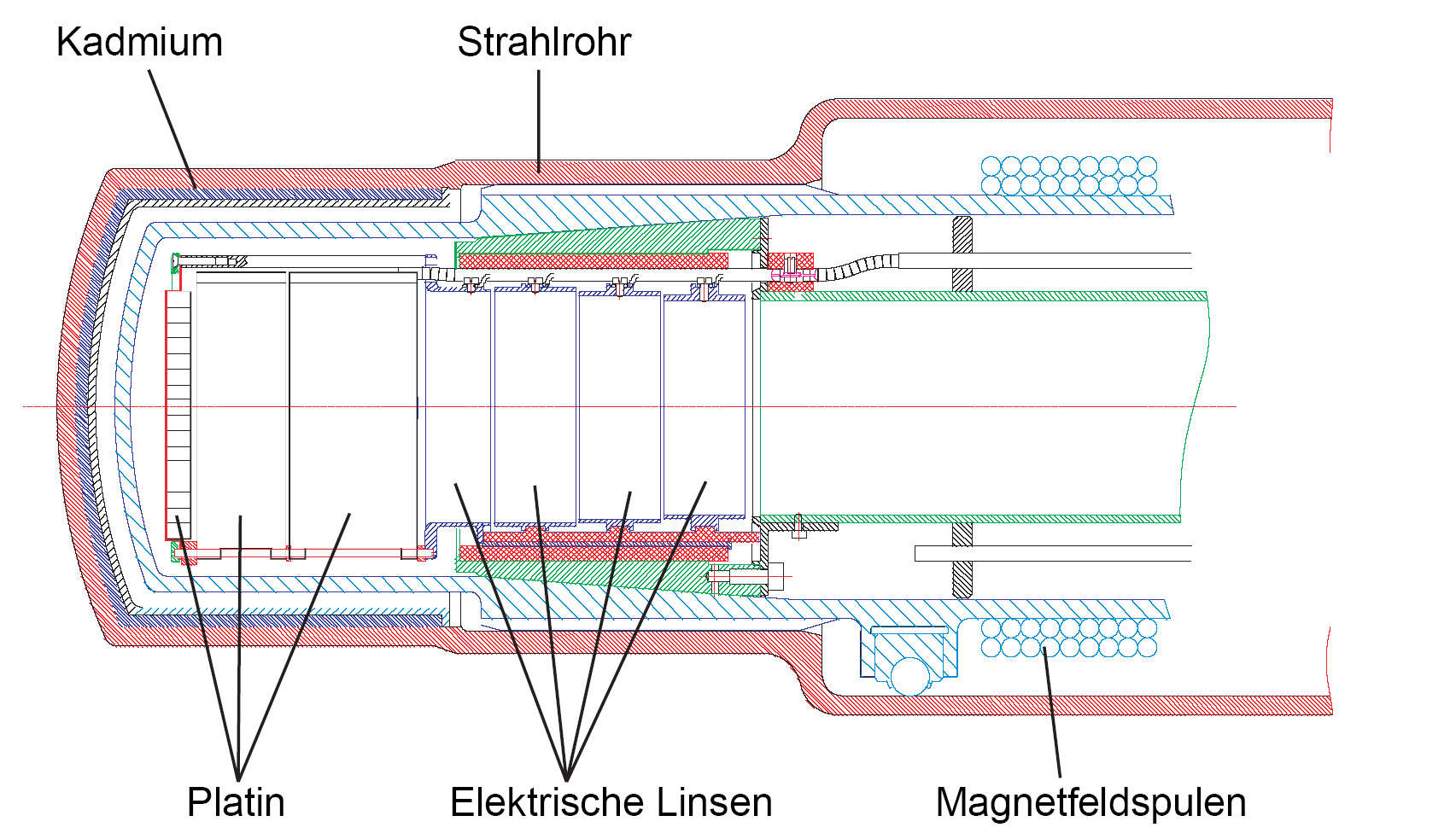}
 \caption{Quellkopf der Positronenquelle \textsc{Nepomuc} am Forschungsreaktor München~II (FRM~II) \cite{Hugenschmidt2002a}}
 \label{fig:NEPOMUC-Quellkopf}
\end{figure}

Reaktorquellen, wie die am Forschungsreaktor München~II (FRM~II) installierte Positronenquelle \textsc{Nepomuc}\footnote{\textbf{Ne}utron Induced \textbf{Po}sitron Source \textbf{Mu}ni\textbf{c}h}, nutzen $\upgamma$-Strahlung aus angeregten Atomkernen. Dabei gibt es wiederum zwei verschiedene Möglichkeiten zur Erzeugung der $\upgamma$-Strahlung: Entweder können die direkt im Reaktorkern entstehenden $\upgamma$-Quanten zur Paarbildung genutzt werden wie beispielsweise an der Uni Delft. Die andere Möglichkeit, die am FRM~II eingesetzt wird, besteht darin, über den Einfang thermischer Neutronen eine $\upgamma$-Kaskade auszulösen. Am Forschungsreaktor der TU München geschieht dies durch folgende Reaktion: $^{113}$Cd(n$_{\text{th}}$,~$\upgamma$)$^{114}$Cd. Cadmium-113 wurde aufgrund des großen Wirkungsquerschnitts für diese Einfangreaktion von 26000\,barn gewählt. Abbildung \ref{fig:NEPOMUC-Schema} zeigt die Erzeugung langsamer Positronen an \textsc{Nepomuc} schematisch. Durch den Einfang thermischer Neutronen im Cadmium wird die Neutronen-Bindungsenergie in Form von $\upgamma$-Strahlung freigesetzt. Die Erzeugung der Positronen-Elektronen-Paare geschieht in Platin, das gleichzeitig die schnellen Positronen auf niedrige Energien moderiert. Abbildung \ref{fig:NEPOMUC-Quellkopf} zeigt einen Schnitt durch den Quellkopf von \textsc{Nepomuc}, der sich im Moderatortank in der Nähe des Reaktorkerns befindet. Die aus dem Platin emittierten langsamen Positronen werden durch elektrische Linsen zu einem Strahl geformt und durch ein magnetisches Führungsfeld von etwa 8\,mT zu den einzelnen Experimenten geführt. Das Strahlrohr ist dabei bis auf einen Druck von ca. $10^{-8}$\,mbar evakuiert, um Annihilationsverluste im Restgas zu vermeiden. Dieser Primärstrahl lässt sich auf Energien von 15 bis 1000\,eV einstellen und liefert bis zu $5 \cdot 10^8$ moderierte Positronen/Sekunde.

\section{Positronenstrahlführung}
Um Positronen als Strahl führen zu können, müssen diese monoenergetisch sein, was bedeutet, dass alle Positronen die gleiche kinetische Energie besitzen müssen. Zur Erzeugung monoenergetischer Strahlen sei auf Kapitel \ref{Moderation}: \enquote{Moderation und Remoderation} verwiesen. Aufgrund ihrer elektrischen Ladung, gibt es zwei Möglichkeiten Positronenstrahlen zu führen: magnetisch und elektrisch.

\subsection{Magnetische Strahlführung}
Die magnetische Strahlführung ist die praktikabelste Art der Strahlführung und kann verwendet werden, um einen Positronenstrahl über eine längere Distanz zu führen. Dabei wird üblicherweise\footnote{Alternativ können auch Helmholtzspulen benutzt werden.} das Strahlrohr mit Spulen umwickelt, sogenannten Solenoidspulen, so dass durch den Spulenstrom $I$ ein Magnetfeld entsteht das parallel zum Strahl ist. Die magnetische Flussdichte $B$ im Inneren einer unendlich langen Spule erhält man gemäß:
\begin{equation}
 B = \mu_0 I n
\end{equation}
Dabei ist $\mu_0$ die magnetische Permeabilität und $n$ die Wicklungsdichte des Drahtes in der Einheit 1/m.
\begin{figure}
 \centering
 \includegraphics[width=0.98\textwidth]{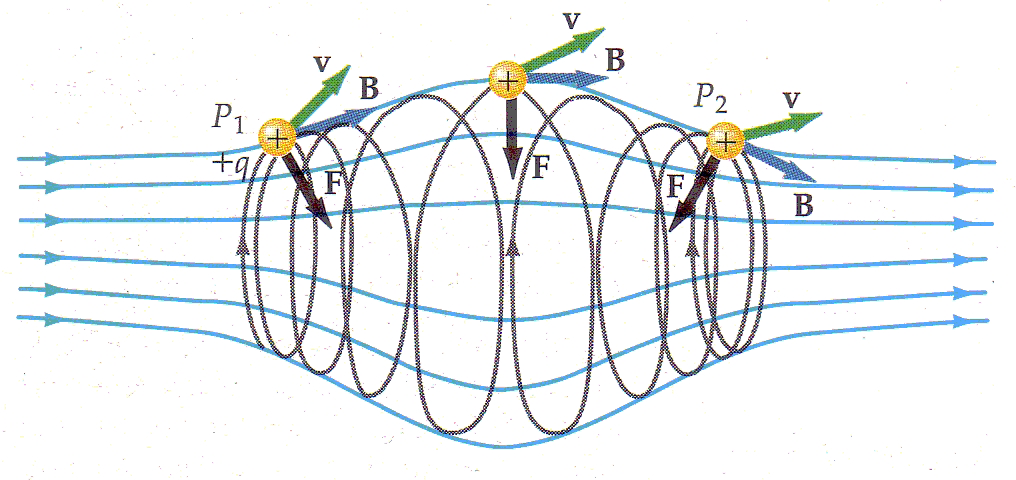}
 \caption{Bewegung eines geladenen Teilchens in einem inhomogenen Magnetfeld (aus \cite{Tippler})}
 \label{fig:B-Feld-inhomogen}
\end{figure}

Die Wirkungsweise der magnetischen Strahlführung basiert auf der resultierenden Kraft die ein geladenes Teilchen im Magnetfeld erfährt, der Lorentzkraft $\vec{F}$. Zusammen mit der Ladung des Teilchens $q$ und dessen Geschwindigkeit $\vec{v}$ gilt:
\begin{equation}
 \vec{F} = q \vec{v} \times \vec{B}.
\end{equation}
Auf ein Teilchen das sich in Magnetfeldrichtung bewegt wirkt daher, aufgrund des Kreuzprodukts, keine Kraft. Hat es jedoch eine Geschwindigkeitskomponente $v_{\perp}$ senkrecht dazu, wird es auf eine Kreisbahn um die Magnetfeldachse gezwungen, es gyriert. Die Gyration verhindert somit das Auseinanderlaufen eines Positronenstrahls. Durch Gleichsetzen der Lorentzkraft mit der Zentrifugalkraft $F_Z = \frac {m v_{\perp}^2} r$ lässt sich der Gyrationsradius $r$ bestimmen:
\begin{equation}
 r = \frac {m v_{\perp}} {e B} = \frac {\sqrt{2 E_{\perp} m}} {e B} = \frac {v_{\perp}} {\omega_c}
\end{equation}
Dabei ist $E_{\perp}$ die kinetische Energie senkrecht zur Magnetfeldrichtung und $\omega_c$ die Zyklotronfrequenz. Die Zeit für eine Gyration ist somit
\begin{equation}
 T = 2 \pi \frac 1 {\omega_c} = 2\pi \frac {m} {e B}
\end{equation}
Daraus lässt sich die Gyrationslänge $\lambda$, mit Hilfe der kinetischen Energie in Longitudinalrichtung $E_{\parallel}$ berechnen:
\begin{equation}
 \lambda = 2\pi \frac {\sqrt{2 E_{\parallel} m}} {e B}
\end{equation}
Sie gibt an, wie weit sich ein Positron während einer Gyration in Longitudinalrichtung fortbewegt. Man spricht von adiabatischer Strahlführung, wenn das Magnetfeld innerhalb einer Gyrationslänge nahezu konstant bleibt. Werden Positronen adiabatisch geführt, folgen sie dem Verlauf der Magnetfeldlinien auch in inhomogenen Magnetfeldern (siehe Abbildung \ref{fig:B-Feld-inhomogen}). Dadurch lässt sich ein Strahl adiabatisch komprimieren, und eine Führung der Positronen durch Bögen wird möglich. Die sogenannte Krümmungsdrift, die durch unterschiedliche Wicklungsdichten zwischen Innen- und Außenbereich des Bogens entsteht, muss jedoch durch ein Korrekturfeld, das senkrecht zur Krümmungsebene des Führungsfeld steht, ausgeglichen werden. Das Erdmagnetfeld lässt sich durch zusätzliche transversale Felder oder durch Ummantelung des Strahlrohrs mit $\upmu$-Metall abschirmen.

\subsection{Elektrische Strahlführung}
\begin{figure}
 \centering
 \includegraphics[width=0.98\textwidth]{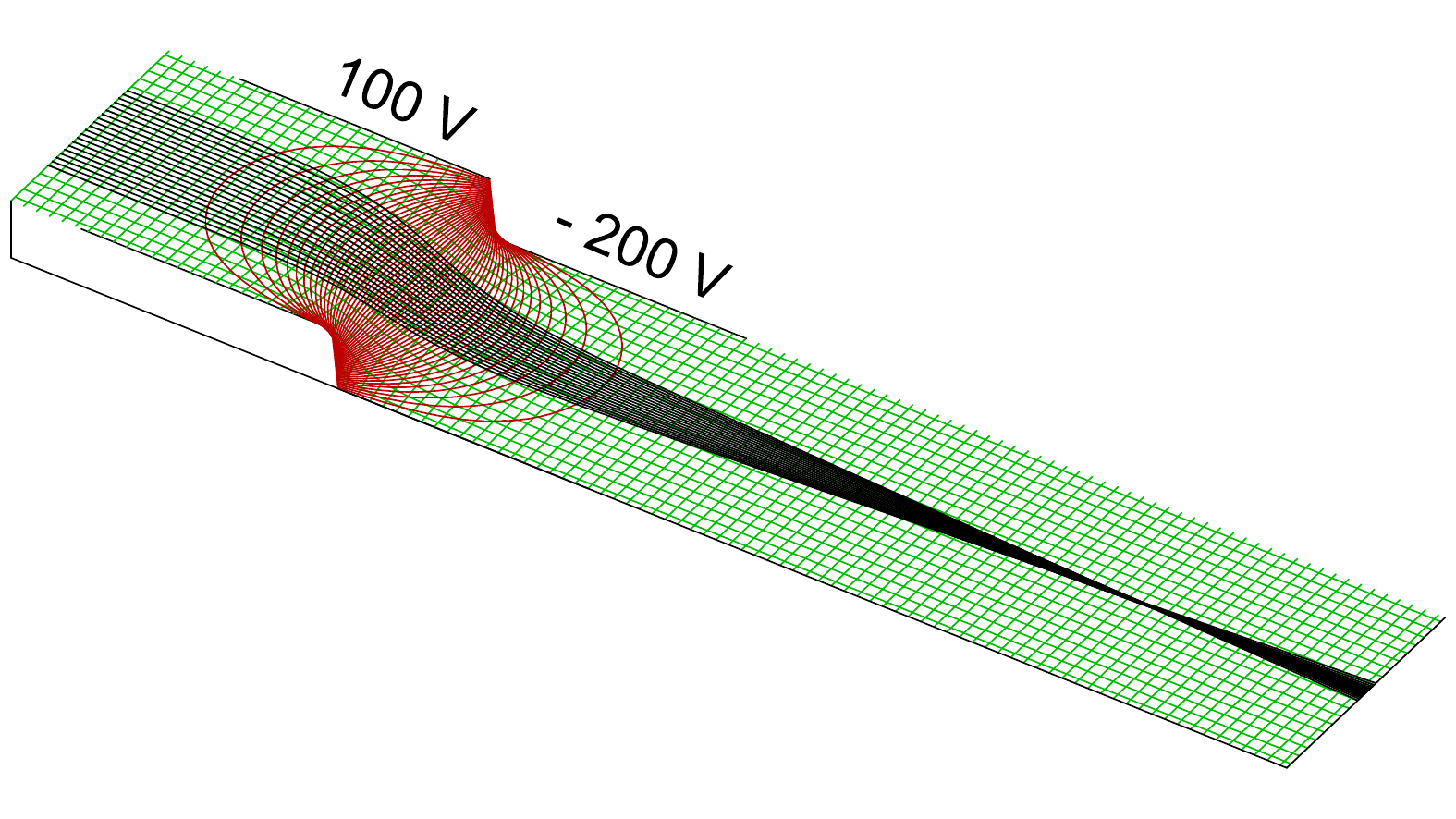}
 \caption{Fokussierung eines 100\,eV Positronenstrahls durch eine elektrische Linse, simuliert mit \textsc{Simion 3D}}
 \label{fig:Linse}
\end{figure}
Die elektrische Strahlführung basiert auf der Kraft $\vec{F} = e \vec{E}$, die ein Positron in einem elektrischen Feld erfährt und wird vor allem eingesetzt, um Positronen, z.\,B. auf eine Probe, zu fokussieren. Positronenstrahlen können durch elektrische Linsen fokussiert werden. Eine elektrische Linse besteht aus mindestens zwei hintereinander angeordneten, aber voneinander isolierten Metallringen, durch die der Strahl hindurchläuft.

Abbildung \ref{fig:Linse} zeigt eine aus zwei Metallringen bestehende elektrische Linse durch die ein 100\,eV Positronenstrahl hindurchläuft. Die perspektivisch dargestellte Höhe entspricht dabei dem Potential an diesem Ort. Äquipotentiallinien sind rot eingezeichnet. Die von links einlaufenden Positronen werden im Bereich des 100\,V-Ringes durch das elektrische Feld fokussiert. Im Bereich der zweiten Linse wirkt durch das elektrische Feld dieselbe Kraft, jedoch defokussierend. Aufgrund der Potentialstufe zwischen beiden Ringen werden die Positronen beschleunigt, wodurch die Aufenthaltsdauer im Bereich der zweiten Linse geringer ist und somit die Defokussierung schwächer ausfällt als die Fokussierung. Als Resultat wird der Strahl letztendlich fokussiert.

Die elektrostatische Fokussierung auf einen punktförmigen Fokus funktioniert jedoch nur mit einem monoenergetischen Strahl, da für unterschiedliche Energien aufgrund der chromatischen Aberation unterschiedliche Brennweiten zustande kommen. Ein weiterer Nachteil gegenüber der magnetischen Kompression besteht im verhältnismäßig komplizierten Aufbau der Linsen mit elektrischen Anschlüssen und Durchführungen im Vakuum.

Eine andere Möglichkeit Positronen zu führen, nämlich der über elektrische Driftfelder, wird in Abschnitt \ref{simion-potentiale} vorgestellt.

\section{Das Positron im Festkörper} \label{DasPositronimFestkoerper}
\begin{figure}
 \centering
 \includegraphics[viewport= 245 100 600 455,clip,width=0.9\textwidth]{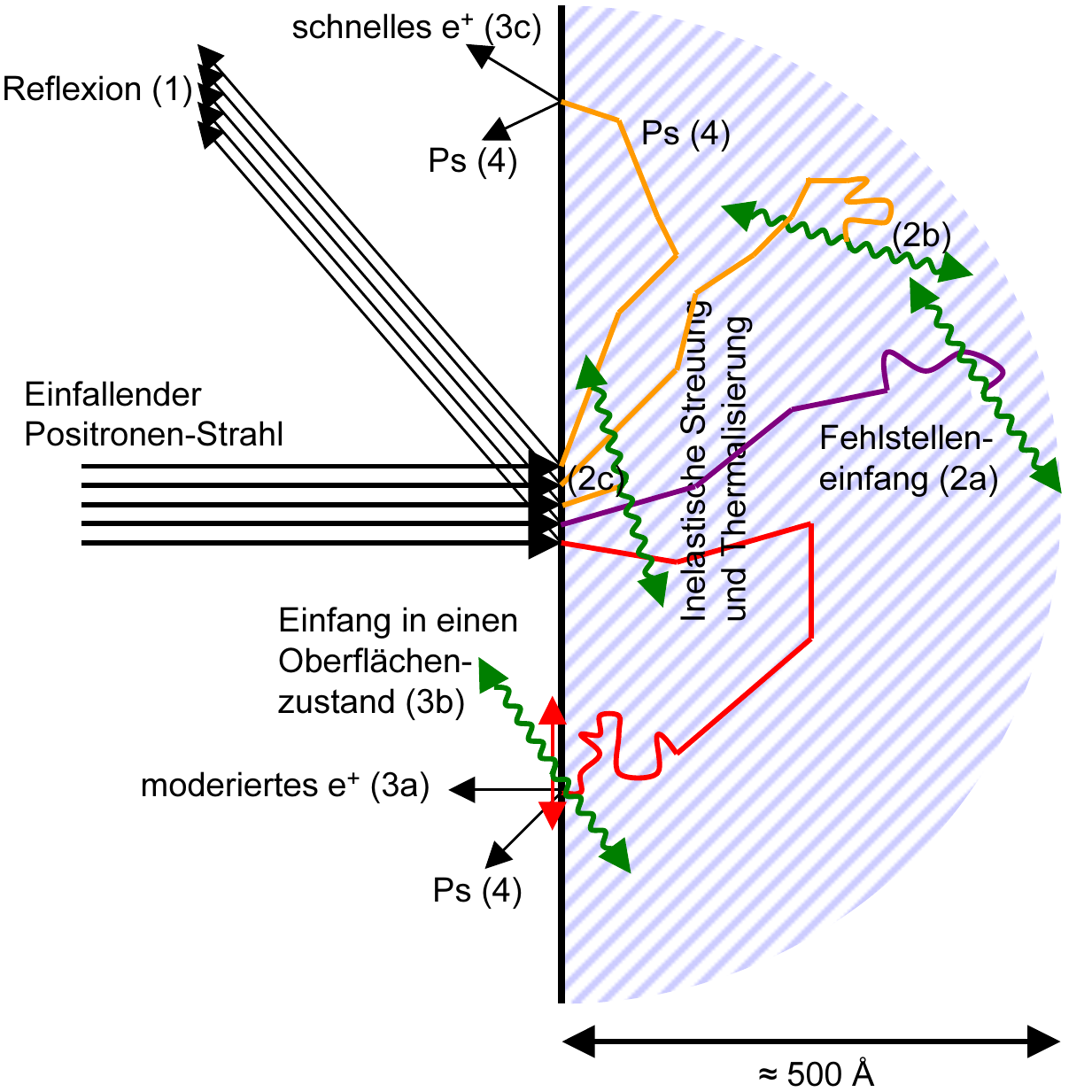}
 \caption{Übersicht über die möglichen Prozesse im Festkörper \cite{DiplJakob}, \cite{DiplPio}}
 \label{fig:ProzesseImFK}
\end{figure}

Treffen Positronen auf einen Festkörper, so sind verschiedene Prozesse in diesem möglich. Anhand von Abbildung \ref{fig:ProzesseImFK} werden diese im Folgenden erläutert. In Tabelle \ref{tbl:Zeitskalen} sind einige Zeitskalen der einzelnen Prozesse aufgeführt.
\begin{table}
 \centering
 \caption{Typische Zeitskalen \cite{Schultz1988}}
 \label{tbl:Zeitskalen}
 \begin{tabular}{lcl} \toprule
 Prozess/Lebensdauer &  \multicolumn{2}{c}{Zeit}\\ \midrule
 Lebensdauer im Vakuum & > & $2 \cdot 10^{+21}$\,a\\
 Streuung oder Beugung & $\approx$ & $1 \cdot 10^{-15}$\,s\\
 Abbremsung auf ca. Fermi-Energie & $\approx$ & $1 \cdot 10^{-13}$\,s\\
 Abbremsung auf thermische Energien & $\approx$ & $1 \cdot 10^{-12}$\,s\\
 Freie Diffusion & $\approx$ & $1 \cdot 10^{-10}$\,s\\
 Lebensdauer eingefangen in einzelner Leerstelle & $\approx$ & $2 \cdot 10^{-10}$\,s\\
 Lebensdauer eingefangen in Mehrfachleerstelle & $\approx$ & $4 \cdot 10^{-10}$\,s\\
 Lebensdauer in Oberflächenzustand & $\approx$ & $4-6 \cdot 10^{-10}$\,s\\
 Lebensdauer des Positronium Singlett-Zustands (Vakuum) & $\approx$ & $1,25 \cdot 10^{-10}$\,s\\
 Lebensdauer des Positronium Triplett-Zustands (Vakuum) & $\approx$ & $1,42 \cdot 10^{-7}$\,s\\ \bottomrule
 \end{tabular}
\end{table}
\subsubsection{Streuung, Thermalisierung}
Die erste Möglichkeit besteht darin, dass Positronen überhaupt nicht in den Festkörper eindringen und an ihm nur elastisch oder inelastisch gestreut werden (siehe (1) in Abbildung \ref{fig:ProzesseImFK}). Gelangen die Positronen in den Festkörper, können sie durch inelastische Stöße solange Energie verlieren, bis sie thermalisiert sind, also auf Energien von etwa $E = k_B T$ abgebremst worden sind. Dies geschieht innerhalb weniger Picosekunden ($10^{-12}$\,s). Schon nach etwa $10^{-15}$\,s haben sie jedoch bereits Energien von etwa 10\,eV. Dies geschieht durch Ionisation bzw. elektrische Anregung von Atomen oder Erzeugung von Exzitonen. Unterhalb von 10\,eV werden die Positronen durch Anregung von Plasmonen und durch Streuung mit Phononen abgebremst, was jedoch mit $10^{-12}$\,s deutlich länger dauert. Aufgrund der niedrigen Intensitäten heutiger Positronenquellen befindet sich immer nur ein Positron im Festkörper und kann dadurch den energetisch niedrigsten Zustand einnehmen. Es ist jedoch auch möglich, dass Positronen vor dem Erreichen thermischer Energien mit einem Elektron annihilieren (2c) oder den Festkörper als epithermische Positronen wieder verlassen (3c).

\subsubsection{Diffusion und Einfang}
\begin{figure}
 \centering
 \includegraphics[width=0.6\textwidth]{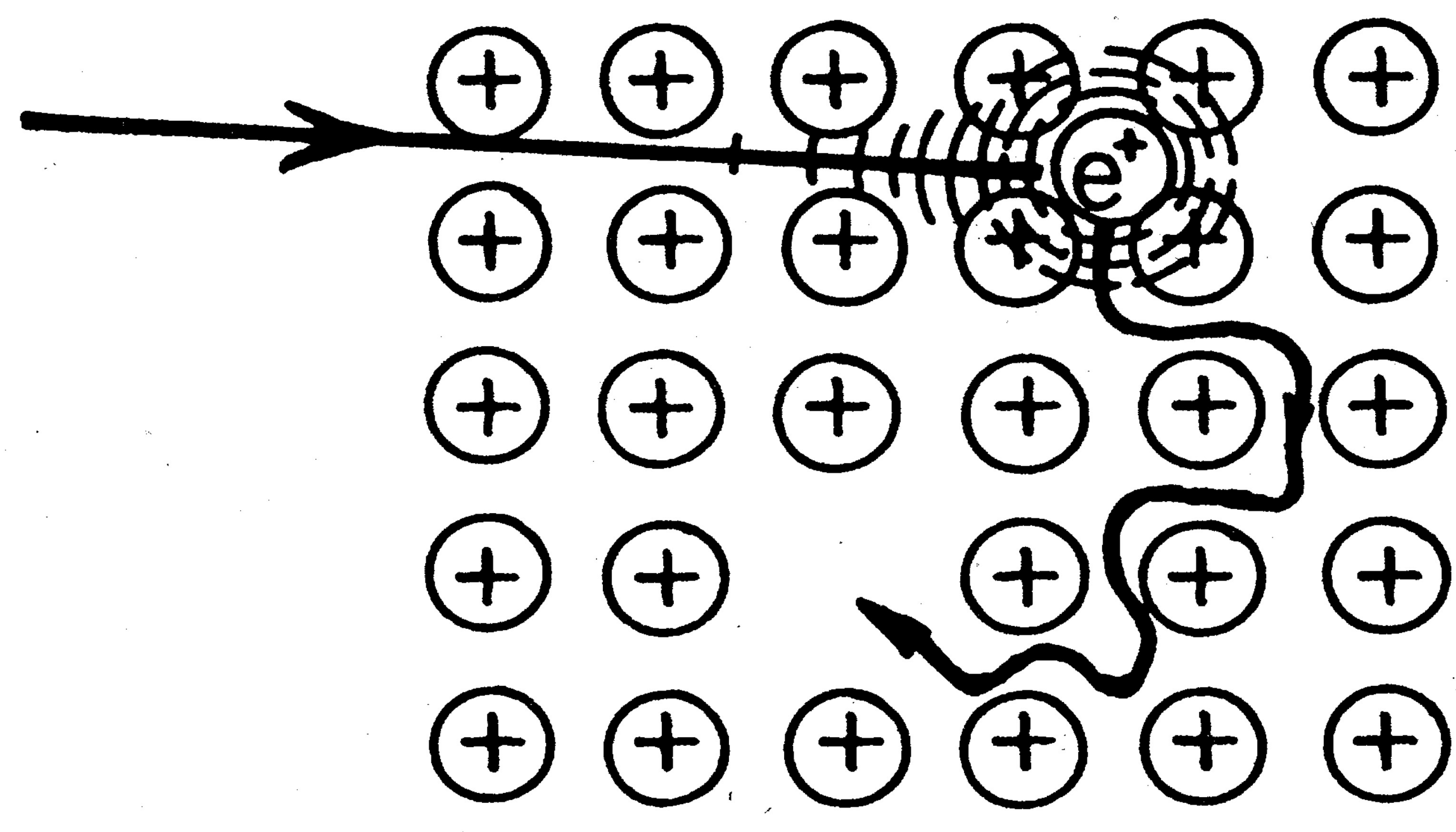}
 \caption{Thermalisierung, Diffusion und Einfang eines Positrons in einer Leerstelle \cite{Schultz1988}.}
 \label{fig:Trapping2}
\end{figure}

\begin{figure}
 \centering
 \includegraphics[width=0.6\textwidth]{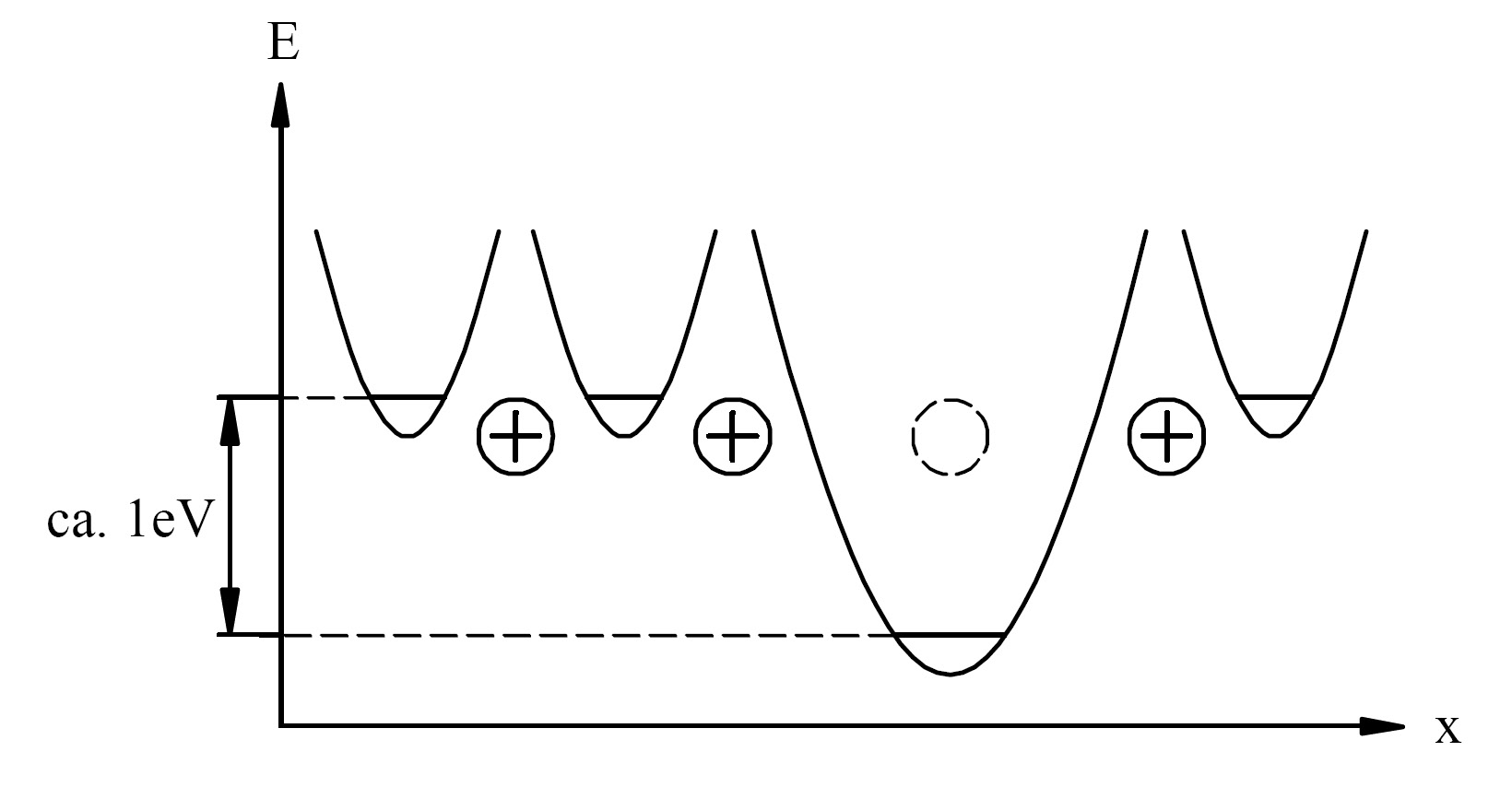}
 \caption{Aufgrund des für Positronen attraktiven Potentials einer Leerstelle können thermalisierte e$^+$ dort eingefangen (\enquote{getrappt}) werden \cite{DrBenno}.}
 \label{fig:Trapping}
\end{figure}

Thermalisierte Positronen können im Festkörper etwa $10^{-10}$\,s lang diffundieren, bevor sie mit einem Elektron annihilieren (2b). Dies entspricht einer Diffusionslänge von einigen 100\,nm. Während der Diffusion werden Positronen bevorzugt von sogenannten Haftstellen eingefangen (siehe (2a) und Abbildung \ref{fig:Trapping2}). Haftstellen sind Gitterfehler wie z.\,B. (Mehrfach-) Leerstellen, Korngrenzen oder Versetzungen. In einer Leerstelle existiert für Positronen wegen des fehlenden Atomrumpfes ein attraktives Potential von ca. 1\,eV Tiefe (siehe Abbildung \ref{fig:Trapping}). Aufgrund der geringen Energie eines thermalisierten Positrons von nur etwa 40\,meV wird es in diesem Zustand gebunden bis es annihiliert. Die Lebensdauer in einer Einfachleerstelle ist dabei um einige 10\,ps größer als für ein nicht eingefangenes Positron, was an der dort geringeren Elektronendichte liegt.

\subsubsection{Annihilation}
\begin{figure}
 \centering
 \includegraphics[width=0.98\textwidth]{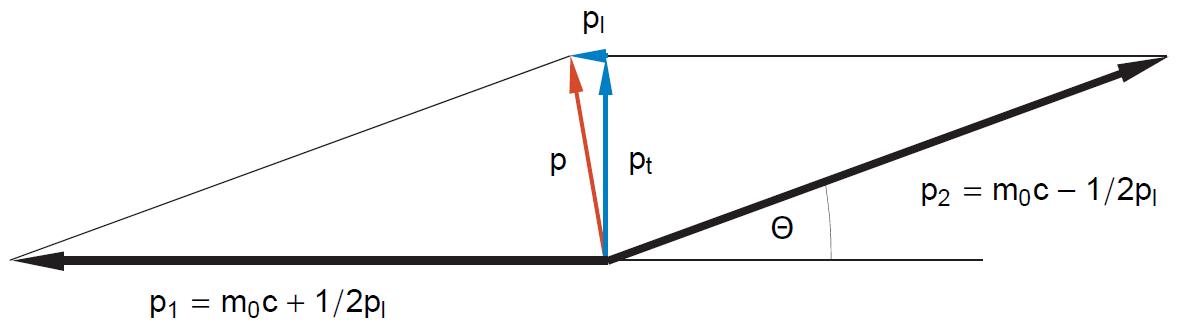}
 \caption{Annihilation eines Positrons mit einem Elektron \cite{DiplPio}: Die Transversalkomponente $p_t$ des Elektronenimpulses $p$ führt zu einer Winkelabweichung $\Theta$ der Gammaquanten von 180°. Der Longitudinalanteil $p_l$ führt zu einer Dopplerverschiebung der Gammaenergie \cite{Hering}.}
 \label{fig:Annihilation}
\end{figure}

Die Annihilation von Positron und Elektron in zwei Gammaquanten mit einer Energie von jeweils 511\,keV erfolgt im Schwerpunktsystem unter einem Winkel von 180°. Aufgrund der Transversalkomponente des Impulses des beteiligten Elektrons\footnote{Der Impuls des thermalisierten Positrons kann vernachlässigt werden.}  (siehe Abbildung \ref{fig:Annihilation}) misst man jedoch im Laborsystem einen Winkel, der um einige mrad von 180° abweicht. Die Longitudinalkomponente des Elektronenimpuls führt zu einer Dopplerverschiebung der Energien beider Gammaquanten, die bis zu einige keV betragen kann.

\subsubsection{Positroniumsbildung}
In Isolatoren, an Oberflächen und in Gasen ist die Bildung von Positronium möglich (4). Das Positronium ist ein wasserstoffähnlicher, gebundener Zustand aus einem Elektron und einem Positron. In Metallen ist Positroniumsbildung aufgrund der Abschirmwirkung der Leitungselektronen nur an der Oberfläche möglich. Eine genauere Behandlung des Positroniums erfolgt in Abschnitt \ref{Positroniumsbildung}.

\subsubsection{Reemission}
Besitzt der Festkörper eine negative Austrittsarbeit für Positronen, können thermalisierte Positronen, die zurück zur Oberfläche diffundieren, den Festkörper wieder verlassen. Dieser Effekt wird bei der Moderation von Positronen genutzt (siehe Kapitel \ref{Moderation}).

\section{Experimentelle Methoden}
Im Laufe der Zeit sind verschiedene experimentelle Methoden mit Positronen entwickelt worden, die auch miteinander kombiniert werden können.

\subsubsection{Angular correlation of annihilation radiation (ACAR)}
Bei der ACAR wird die Winkelabweichung der Annihilationsstrahlung von 180° gemessen, um Aussagen über den Elektronenimpuls im Festkörper machen zu können. Damit ist es möglich Fermiflächen zu vermessen.

\subsubsection{Doppler Broadening Spectroscopy (DBS)}
Die DBS (Dopplerspektroskopie) betrachtet dagegen die Dopplerverschiebung der Gammaenergie der Annihilationsquanten. Eine Weiterentwicklung ist die koinzidente Dopplerspektroskopie (CDBS), mit der die Energieverschiebung beider Gammas gleichzeitig gemessen wird. Dies resultiert in einer drastischen Verbesserung des Peak-to-Background-Verhältnisses.

\subsubsection{Lebensdauermessungen}
Mit Hilfe eines gepulsten Strahls oder einer $^{22}$Na-Quelle ist es möglich die Lebensdauer eines Positrons im Festkörper zu bestimmen. Damit lassen sich z.\,B. Aussagen über Defekttyp und -konzentration, beispielsweise nach mechanischer Beanspruchung eines Materials machen. Wird der Strahl über die Probe gescannt, ist dies auch ortsaufgelöst möglich.

\subsubsection{Positron Annihilation induced Auger Electron Spectroscopy (PAES)}
Bei der durch Positronenannihilation induzierten Auger-Elektronenspektroskopie (PAES) wird der Auger-Prozess nicht durch Stoßionisation ausgelöst, sondern durch die Annihilation eines kernnahen Elektrons mit einem Positron. Ein Vorteil gegenüber der konventionellen AES ist eine höhere Oberflächensensitivität aufgrund der Rückdiffusion der langsamen Positronen zur Oberfläche.
\clearpage{}
\clearpage{}\chapter{Moderation und Remoderation} \label{Moderation}

\section{Zweck der Moderation}
\begin{figure}
 \centering
 \includegraphics[width=0.55\textwidth]{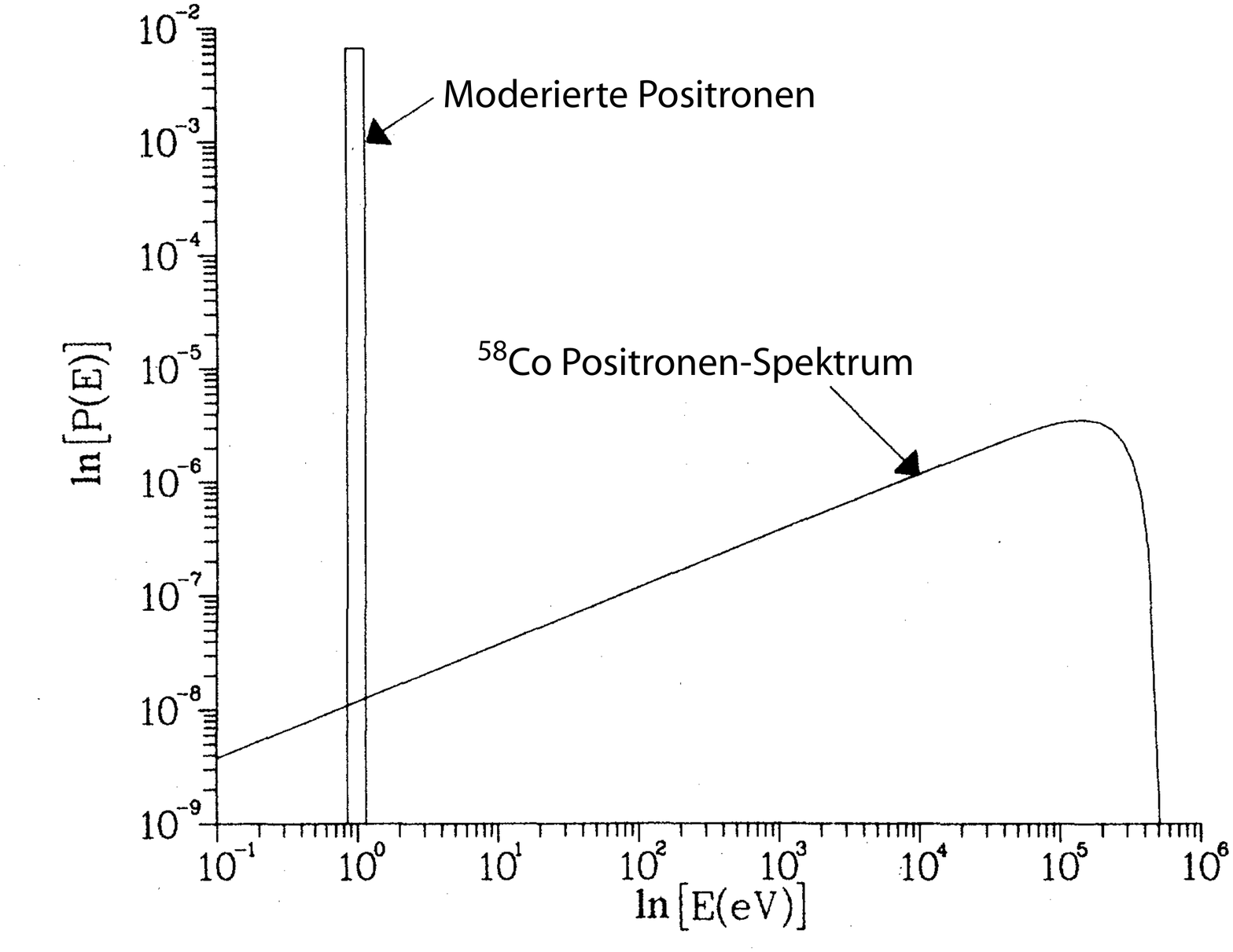}
 \caption{Moderiertes und nicht-moderiertes Positronen-Spektrum einer $^{58}$Co-Quelle (aus \cite{Schultz1988})}
 \label{fig:Co-Spek-Moderiert}
\end{figure}

Unter Moderation\footnote{lat.: moderare = mäßigen}, im physikalischen Sinn, versteht man im Allgemeinen das Abbremsen von Teilchen auf thermische Energien ($E_{kin} \approx k_B T \approx 25$\,meV). So werden z.\,B. in einem Kernreaktor Neutronen durch Wasser moderiert, um die Kernspaltung mit thermischen Neutronen am Laufen zu halten. Auch in der Positronenphysik ist es für eine Vielzahl an Experimenten essentiell, langsame Positronen - im Energiebereich von einigen eV - zur Verfügung zu haben. $\upbeta^+$-Emitter sowie Quellen die auf der Positronenerzeugung durch Paarbildung basieren (wie z.\,B. \textsc{Nepomuc}) liefern jedoch Positronen mit einem einige 100\,keV breiten Energiespektrum. Abbildung \ref{fig:Co-Spek-Moderiert} zeigt das $\upbeta^+$-Spektrum einer Cobalt-58-Quelle, mit einer Endpunktenergie von 472\,keV. Ebenfalls eingezeichnet ist das Spektrum moderierter Positronen aus dieser Quelle. Wie man sieht konnte die Intensität an Positronen im eV-Bereich durch einen Wolfram-Moderator (W(110)) um beinahe sechs Größenordnungen gesteigert werden. Hier wird die Notwendigkeit der Moderation deutlich, da ein einfaches Geschwindigkeitsfilter gerade im Bereich niedriger Energie zu erheblichen Intensitätseinbußen führen würde.

In der Positronenphysik werden Moderatoren aber nicht nur eingesetzt, um den Betrag der Geschwindigkeit zu reduzieren, sondern auch um die Winkeldivergenz der Positronen - und damit den Impuls in Transversalrichtung - zu minimieren. Mit anderen Worten: Erst durch Moderation lässt sich ein intensiver, monoenergetischer Strahl aus Positronen bilden.

\section{Die Brillanz}
\begin{figure}
 \centering
 \includegraphics[height=5cm]{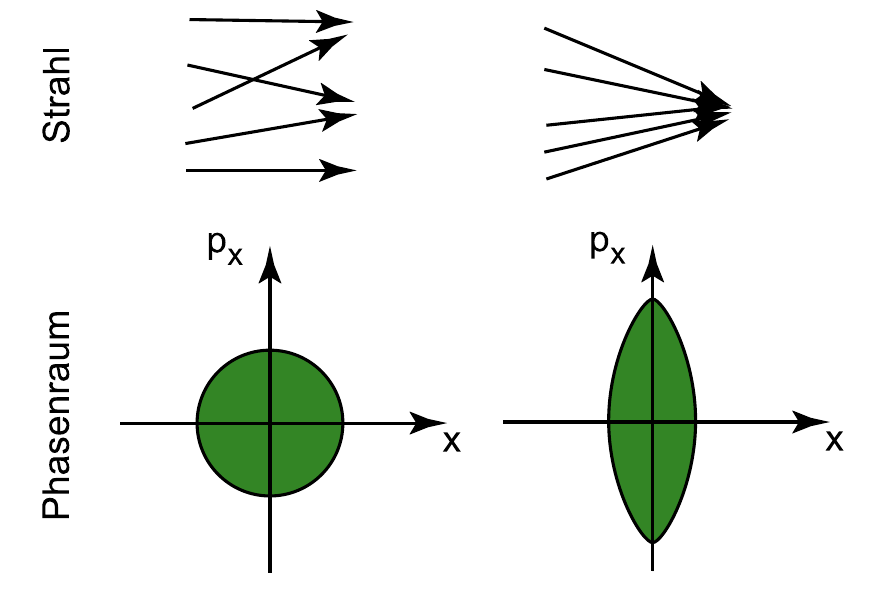}
 \caption{Bei der Fokussierung eines Strahls durch elektrostatische Linsen wird das Phasenraumvolumen in der Ortskoordinate (hier $x$) komprimiert. Aufgrund des Satzes von Liouville muss das Phasenraumvolumen eines Teilchenstrahls jedoch erhalten bleiben, weshalb die Impulsunschärfe zunimmt und somit die Dichte im Impulsraum ($p_x$) abnimmt \cite{DiplPio}.}
 \label{fig:Phasenraum-Fokussierung}
\end{figure}
Die Brillanz ist definiert als das Verhältnis der Anzahl an Positronen pro Sekunde zur Größe des besetzten Phasenraums. Um die Qualität eines Strahls zu verbessern, muss daher bei konstanter Intensität das Phasenraumvolumen verkleinert werden.

\begin{figure}
 \centering
 \includegraphics[height=7cm]{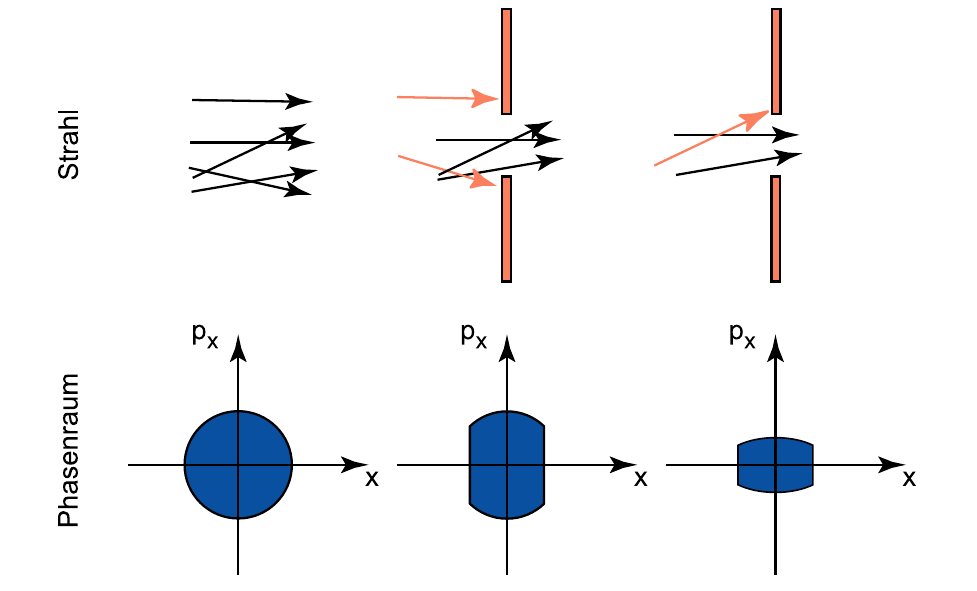}
 \caption{Mit der Beschneidung des Phasenraums durch mehrere Kollimatoren geht gleichzeitig eine erheblich Anzahl an Positronen verloren \cite{DiplPio}.}
 \label{fig:Phasenraum-Kollimierung}
\end{figure}

Der Phasenraum ist ein sechs-dimensionaler Raum der aus den drei Ortskomponenten ($x$, $y$, $z$) und den drei Impulskomponenten ($p_x$, $p_y$, $p_z$) besteht. Für jedes Positron im Strahl lässt sich - in den Grenzen, die durch die heisenbergsche Unschärferelation vorgegeben sind - die Lage in diesem sechs-dimensionalen Raum bestimmen. Bei einem idealen Strahl würden deshalb alle Positronen zu einem beliebigen Zeitpunkt denselben Punkt im Phasenraum besetzen oder anders ausgedrückt, sie würden von einem Punkt im Ortsraum mit gleicher Geschwindigkeit in dieselbe Richtung fliegen. In der Realität bewegen sich die Positronen jedoch auf unterschiedlichen Trajektorien im Phasenraum. Nach dem Satz von Liouville ist es nur durch nicht-konservative Kräfte möglich die Größe des Phasenraums zu verkleinern. Elektrostatische Linsen können somit nicht benutzt werden, da sie einen Strahl zwar im Ortsraum komprimieren können, die Größe im Impulsraum (und damit die Winkeldivergenz) dadurch aber zwangsläufig ansteigt (siehe Abbildung \ref{fig:Phasenraum-Fokussierung}). Natürlich lässt sich die Winkeldivergenz eines Strahls durch die Verwendung mehrerer hintereinander geschalteter Blenden verringern. Bei der im Vergleich zu z.\,B. Ionen- oder Laserstrahlen geringen Intensität an Positronen ist diese Vorgehensweise jedoch ineffizient, da hier der Phasenraum nur abgeschnitten und nicht komprimiert wird (siehe Abbildung \ref{fig:Phasenraum-Kollimierung}). Ziel der Moderation ist es den Phasenraum eines Strahls zu komprimieren.

Da es sich bei der Positronenquelle \textsc{Nepomuc} um eine kontinuierliche (nicht gepulste) Quelle handelt, ist die Ausdehnung des Strahls im Ortsraum in eine Richtung unendlich. Ist die Strahlachse parallel zur $z$-Achse, kann man den Phasenraum um die $z$-Richtung reduzieren. Die kinetische Energie eines Positrons in Longitudinalrichtung $E_{\parallel}$ lässt sich somit durch $p_z$ wie folgt ausdrücken:
\begin{equation}
E_{\parallel}=\frac{p_z^2}{2m_e}
\end{equation}
Für die Energie in Transversalrichtung $E_{\perp}$ gilt:
\begin{equation}
E_{\perp}=\frac {p_x^2 + p_y^2} {2m_e}
\end{equation}

\section{Der Festkörpermoderator}
Die Funktionsweise des Festköpermoderators basiert auf der Thermalisierung von Positronen in einem Festkörper. Wie in Abschnitt \ref{DasPositronimFestkoerper} dargestellt, thermalisieren Positronen aus einer $\beta^+$-Quelle im Festkörper innerhalb von etwa 1\,ps. Die thermalisierten Positronen diffundieren durch den Festkörper bis sie mit einem Elektron zerstrahlen. Gelangen sie davor jedoch an die Oberfläche, können sie den Festkörper wieder verlassen. Voraussetzung dafür ist jedoch eine negative Austrittsarbeit $\Phi^+$ des Materials für Positronen. In Frage kommen z.\,B. Wolfram, Nickel, Platin und feste Edelgase\footnote{Bei festen Edelgasen, wie z.\,B. Neon, beruht die Funktion der Moderation jedoch nicht auf der negativen Austrittsarbeit sondern auf der großen Bandlücke von über 5\,eV. Die Positronen können somit nur durch elektronische Anregung abgebremst werden, solange ihre Energie größer als die der Bandlücke ist. Darunter ist nur noch Phononenstreuung möglich, die bei weitem nicht so effektiv ist. Die noch \enquote{heißen} Positronen können somit trotz positiver Austrittsarbeit den Festkörper verlassen \cite{Bluhme2000}.}, wobei sich (einkristallines) Wolfram ($\Phi^+ = -2,9$\,eV) als Standardmoderationsmaterial durchgesetzt hat. Die kinetische Energie der Positronen in Longitudinalrichtung $E_{\parallel}$ entspricht nach dem Verlassen des Festkörpers (mit $U=0$\,Volt an der Oberfläche) der Austrittsarbeit, jedoch mit einer thermischen Verschmierung im Bereich von $k_B T$. Da die Positronen senkrecht zur Oberfläche den Festkörper verlassen, erhält man einen nahezu parallelen Strahl. Die thermische Energie des Festkörpers führt jedoch zu einer Energiekomponente in Transversalrichtung und damit zu einer Winkeldivergenz $\theta$ (= Öffnungswinkel des Strahls) \cite{Schultz1988}:
\begin{equation}
\theta = 2 \sqrt{ \frac {k_B T} {|\Phi^+|}}
\end{equation}
In der Praxis können jedoch Unebenheiten der Oberfläche des Festkörpers dazu führen, dass diese Winkeldivergenz deutlich größer ausfällt.

\subsection{Moderatorgeometrien}
\begin{figure}
 \centering
 \includegraphics[height=6cm]{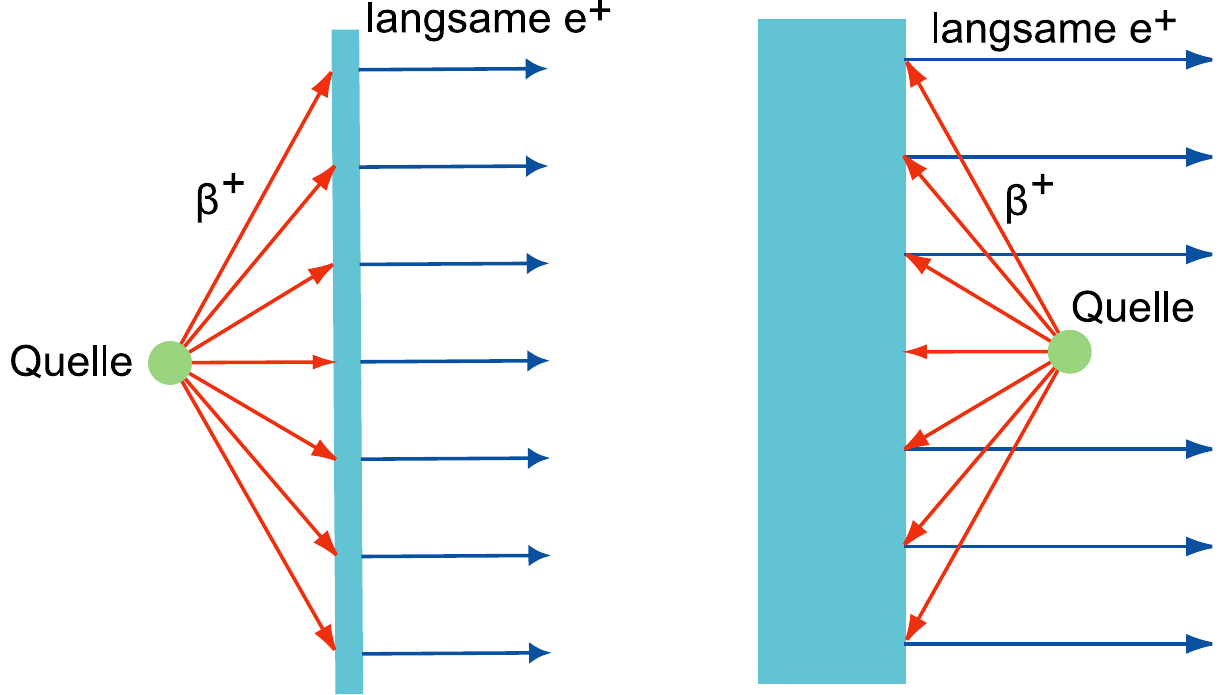}
 \caption{Verschiedene Moderatorgeometrien: links der Transmissionsmoderator, rechts der Reflexionsmoderator \cite{DiplPio}}
 \label{fig:Moderatortypen}
\end{figure}
\begin{figure}
 \centering
 \includegraphics[height=6cm]{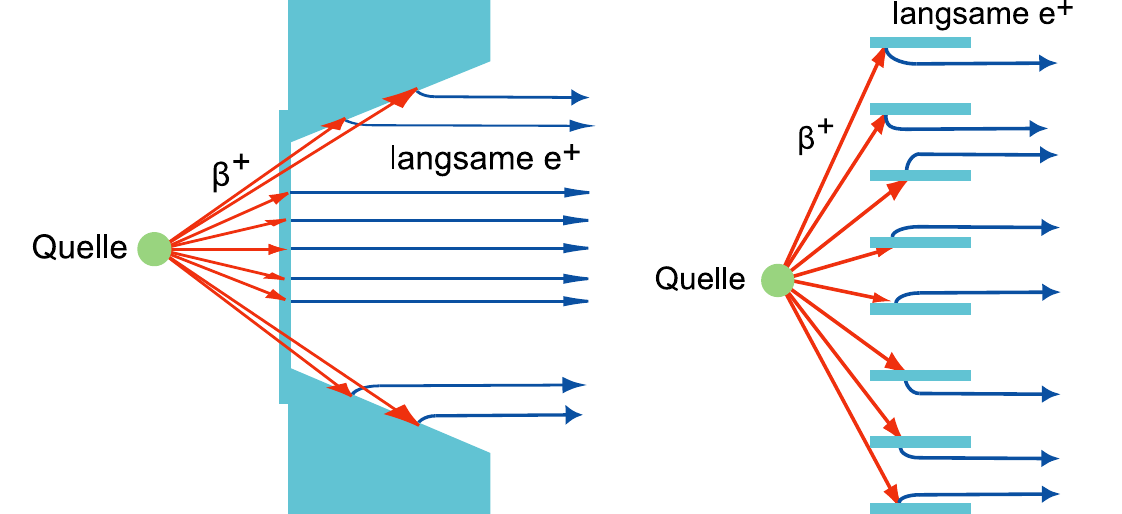}
 \caption{Zwei mögliche Kombinationen aus Transmissions- und Reflexionsmoderator \cite{DiplPio}. Links: Positronen die in der Moderatorfolie nicht (vollständig) moderiert werden, können vom Reflexionskonus  moderiert werden. Dadurch ist es möglich die Dicke der Folie zu reduzieren, um Transmissionsverluste zu minimieren \cite{Maenning2000}. Rechts: Sowohl reflektierte als auch transmittierte Positronen können durch ein elektrisches Feld zu einem Strahl geformt werden.}
 \label{fig:KombiReflTrans}
\end{figure}

Man unterscheidet grundsätzlich zwei verschiedene Moderatortypen. Beim Transmissionsmoderator wird eine etwa 1\,$\upmu$m dünne Folie vor die Quelle gespannt. Die moderierten Positronen verlassen die Folie an der der Eintrittsseite gegenüberliegenden Seite, während beim Reflexionsmoderator die Positronen auf der gleichen Seite des Moderatormaterials emittiert werden. Für den Transmissionsmoderator spricht der einfache Aufbau, während beim Reflexionsmoderator die Gefahr besteht, dass langsame Positronen durch die Quelle abgeschattet werden. Vorteil des Reflexionsmoderators ist jedoch das einfachere Handling des Moderatorkristalls im Vergleich zur dünnen Folie des Transmissionsmoderators. In den Abbildungen \ref{fig:Moderatortypen} und \ref{fig:KombiReflTrans} sind verschiedene Geometrien für den Aufbau von Moderatoren an einer $\upbeta^+$-Quelle abgebildet.

\section{Remoderation}
Je nach  Einsatz unterscheidet man zwischen Moderation und Remoderation. Ein Remoderator wird benutzt um einen bereits an der Quelle moderierten Strahl ein zweites Mal zu moderieren und die Brillanz weiter zu erhöhen. Auch hier gibt es wieder prinzipiell zwei verschiedene Geometrien: den Transmissionsremoderator und den Reflexionsremoderator.

Während die Grundgeometrien denen der Moderation entsprechen, kommt hier hinzu, dass der Strahl möglichst gut auf den Moderator fokussiert werden muss, um die Brillanz (durch Verkleinerung des Ortsraums) zu erhöhen. Beim Transmissionsremoderator muss zusätzlich aufgrund der geringen Energie des einfallenden Positronenstrahls die Dicke der Folie reduziert werden. Wolframfolien von der hier benötigten Dicke von 0,1\,$\upmu$m sind jedoch schwierig zu handhaben und neigen dazu sich zu \enquote{werfen}, was zu einer erheblichen Erhöhung der Winkeldivergenz führt \cite{DiplPio}. Mit aus Nickel hergestellten Folien lässt sich diesem Problem jedoch begegnen, da durch die niedrigere Kernladungszahl Z und der damit verbundenen größeren Positronenimplantationstiefe dickere Folien benutzt werden können. Mit Remoderatorfolien aus Nickel wurde eine kurzzeitige Moderationseffizienz von über 20\,\% erreicht (stabil: 6,5\,\%) \cite{FujinamiSLOPOS11}.

\begin{figure}
 \centering
 \includegraphics[width=0.8\textwidth]{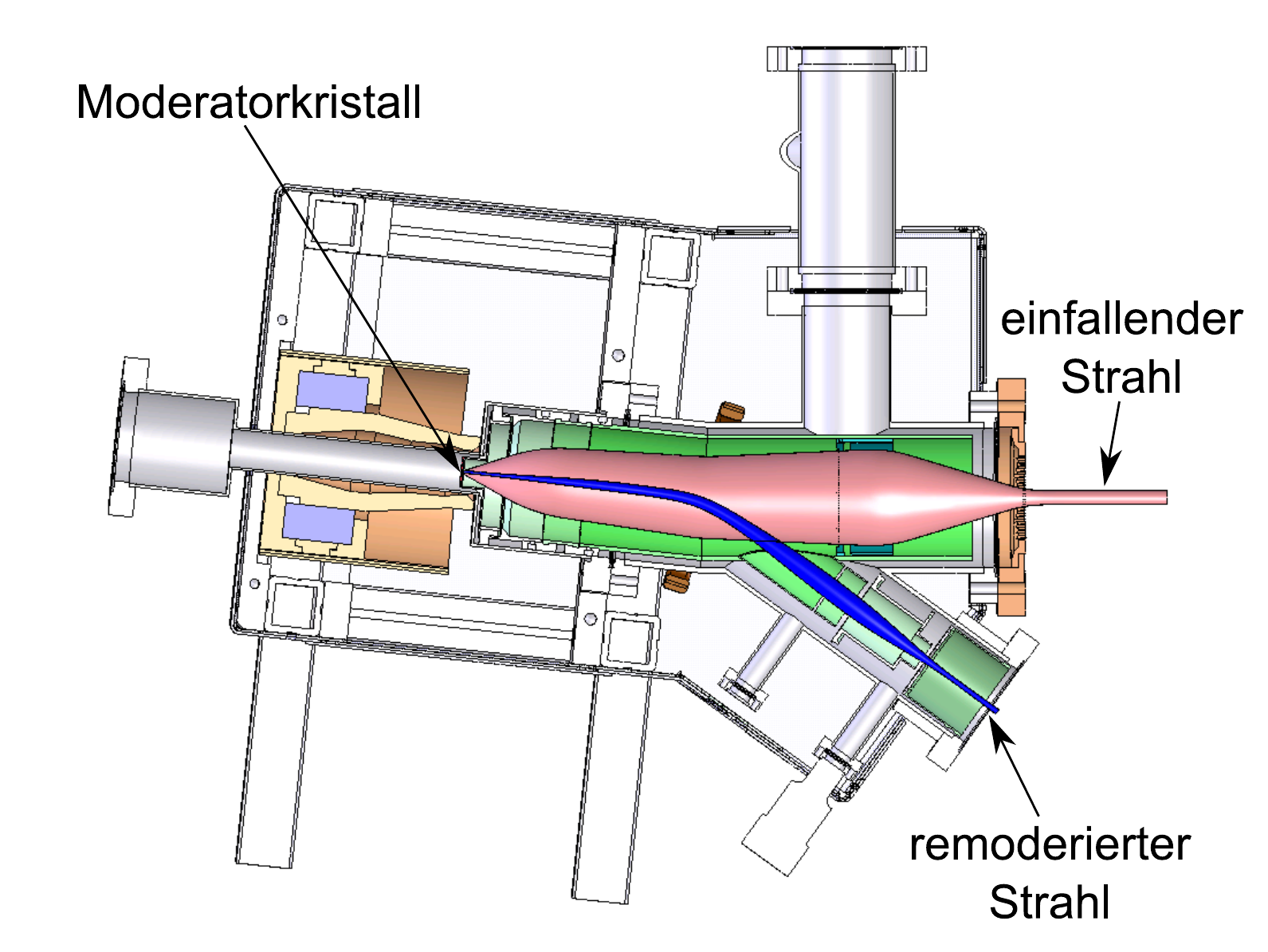}
 \caption{Reflexionsremoderator der Positronenquelle \textsc{Nepomuc} \cite{Piochacz2007}}
 \label{fig:PiosRemoderator}
\end{figure}
Demgegenüber hat man das Problem der Unebenheiten auf der Oberfläche beim Reflexionsremoderator nicht, da sich die hier eingesetzten Moderatorkristalle sehr präzise fertigen lassen. Hier liegt die Herausforderung jedoch bei der unterschiedlichen Strahlführung von eingehendem und ausfallendem Strahl. Abbildung \ref{fig:PiosRemoderator} zeigt die Schnittdarstellung des an \textsc{Nepomuc} eingesetzten Reflexionsremoderators. Der Primärstrahl ($E_{kin} \approx 1000$\,eV) tritt auf der rechten Seite in den Remoderator ein und wird von elektrostatischen Linsen auf den Wolfram(100)-Kristall fokussiert. Die remoderierten Positronen werden magnetisch nach unten geführt und verlassen den Remoderator. Ermöglicht wird diese Strahlführung durch die unterschiedlichen Energien von eingehendem und ausfallendem Strahl: Der schnelle einfallende Strahl wird durch das magnetische Feld am Ausgang nur wenig abgelenkt. Mit diesem Moderatortyp werden üblicherweise Effizienzen von $\approx 5$\,\% (siehe dazu Abschnitt \ref{Moderationseffizienz}) erreicht \cite{Piochacz2007}.

\section{Gasmoderation und Positronenakkumulatoren}
Ein anderer Ansatz ist die Benutzung von Gas als Moderator für Positronen. Bereits 1989 zeigten C. M. Surko et al., dass es möglich ist, Positronen effektiv in Stickstoff abzubremsen und in einer Falle zu speichern \cite{Surko1989}. Sein primäres Ziel war es jedoch, Positronen in der Penning-Falle zu sammeln, um einerseits Untersuchungen an einem Elektronen-Positronen-Plasma durchzuführen und andererseits alle gesammelten Positronen auf einmal aus der Falle heraus zu lassen. Somit handelt es sich hierbei auch um eine Brillanzerhöhung, da etwa $6 \cdot 10^7$ Positronen in eine Zeitspanne von gerade einmal 1\,ns \enquote{komprimiert} wurden. Damit war es zum ersten Mal möglich mit einem \enquote{Schuss} ein ganzes Positronium-Lebensdauerspektrum aufzunehmen - bei einer Zeitauflösung von 1,55\,ns \cite{Cassidy2006}. Die dieser Positronenfalle zugrunde liegende Physik des Abbremsens von Positronen in einem Gas wird im nächsten Kapitel vorgestellt.
\clearpage{}
\clearpage{}\chapter{Positronen im Gas}

\section{Wechselwirkungen von Positronen mit Gasmolekülen}
Im Folgenden werden verschiedene Möglichkeiten aufgeführt, wie Positronen mit Atomen und Molekülen eines Gases wechselwirken können. Exemplarisch wird dies am Beispiel des Stickstoffs gezeigt, da dieser für die späteren Messungen verwendet wurde.

\subsection{Positroniumsbildung}
\label{Positroniumsbildung}
Durch Stoßionisation eines Moleküls oder Atoms durch ein Positron kann das frei gewordene Elektron zusammen mit dem Positron einen neuen Bindungszustand eingehen. Dieser wasserstoffähnliche Zustand wird Positronium (abgekürzt Ps) genannt. Durch die identischen Massen von Elektron und Positron verringert sich die reduzierte Masse des Systems im Vergleich zum Wasserstoff um den Faktor Zwei. Die Bindungsenergie des Positroniums beträgt damit 6,8\,eV - genau halb soviel wie die Ionisierungsenergie beim Wasserstoffatom. Je nach Gesamtspin S unterscheidet man zwischen Ortho- und Parapositronium. Während das Parapositronium (S=0) mit einer Lebensdauer von 125\,ps in zwei Gammaquanten zerstrahlt, ist dies beim Orthopositronium (S=1) aus Spinerhaltungsgründen nicht möglich. Somit muss das Orthopositronium in mindestens drei Gammas mit einer Gesamtenergie von 1,022\,MeV zerfallen. Dadurch ist dessen Lebensdauer mit 142\,ns bedeutend länger.

\subsection{Direkte Annihilation}
Im Gegensatz zur Positroniumsbildung zerstrahlt ein Positron bei der direkten Annihilation mit einem Elektron eines Gasatoms ohne zuvor einen gebunden Zustand mit diesem einzugehen. Es lässt sich zeigen, dass der Wirkungsquerschnitt für die direkte Annihilation eines Positrons im Gas durch die folgende Formel gegeben ist (mit $r_0 = 2,82 \cdot 10^{-15}$\,m als klassischem Elektronenradius und der Relativgeschwindigkeit zwischen den Teilchen $v$):
\begin{equation}
 \sigma_{dir. Ann.}= \pi r^2_0 \frac c v \approx 1,26 \cdot 10^{-22} \text{cm}^2 / \sqrt{E/\text{eV}}
\end{equation}
Dieser Wirkungsquerschnitt liegt damit, wie man später sieht, um einige Größenordnungen unter denen der anderen Wechselwirkungen und kann daher vernachlässigt werden \cite{Bluhme2000}.

\subsection{Elastische Streuung} \label{elastischeStreuung}
Beim elastischen Stoß wird ein Teil der kinetischen Energie eines Stoßpartners in kinetische Energie des anderen Stoßpartners umgewandelt. Die gesamte kinetische Energie bleibt also erhalten. Die Übertragung von kinetischer Energie ist umso effektiver, je weniger sich die Massen der beiden Stoßpartner unterscheiden. Da die Masse des Stickstoffmoleküls jedoch die des Positrons um mehr als das 50000-fache übersteigt, ist die übertrage Energie pro Stoß nur etwa $3,9 \cdot 10^{-5}$ mal der Eingangsenergie \cite{Coleman1981}.

\subsection{Rotationsanregung}
Unter Rotationsanregung versteht man das Rotieren der Atome eines Moleküls um den gemeinsamen Schwerpunkt. Rotationsanregungen treten daher nur bei molekularen Gasen (H$_2$, O$_2$, N$_2$, etc.) und nicht bei atomaren Gasen wie z.\,B. Helium auf. Die Energieniveaus von Rotationsanregungen lassen sich nach \cite{Tippler} folgendermaßen berechnen:
\begin{equation}
 E_{rot} = \frac{J(J+1)\hbar^2}{2I} = J (J+1) B\text{ mit } J=0,1,2,...
\end{equation}
Dabei ist $I$ das Trägheitsmoment des Moleküls, $J$ die Rotationsquantenzahl und $B$ die charakteristische Rotationsenergie des Moleküls:
\begin{equation}
 B = \frac {\hbar^2} {2I}
\end{equation}
Für ein zweiatomiges Molekül gleicher Atome gilt für das Trägheitsmoment
\begin{equation}
 I=\frac 1 2 m r^2 \text{.}
\end{equation}
Wobei $r$ der Abstand der beiden Moleküle ist. Die sich daraus ergebenden charakteristischen Rotationsenergien liegen im Bereich von $10^{-4}$\,eV (für Stickstoff: $B=2,49 \cdot 10^{-4}$\,eV \cite{Coleman1981}) und damit unterhalb der thermischen Energie. Aufgrund der Auswahlregel $\Delta J = \pm 1$ darf sich die Rotationsquantenzahl $J$ bei Übergängen zwischen den Zuständen nur um 1 ändern.

\subsection{Vibrationsanregung}
Vibrationsanregungen sind Schwingungen der Atome eines Moleküls untereinander. Sie können daher wie Rotationsanregungen nicht bei atomaren Gasen auftreten. Die Energieniveaus lassen sich durch einen quantenmechanischen harmonischen Oszillator beschreiben:
\begin{equation}
 E_{vib}= \left(\nu + \frac1 2 \right) h f \text{ mit } \nu=0,1,2,3,...
\end{equation}
Dabei ist $\nu$ die Schwingungsquantenzahl, $f$ die Schwingungsfrequenz und $h$ das plancksche Wirkungsquantum. Es gilt analog zu den Rotationsanregungen die Auswahlregel $\Delta \nu = \pm 1$, wobei die Energie eines Übergangs etwa im Bereich von $10^{-1}$\,eV liegt.

\subsection{Stoßionisation} \label{Stossionisation}
Bei der Stoßionisation wird ein Elektron aus der Elektronenhülle eines Atoms herausgeschlagen. Geschieht dies durch ein anderes Elektron, so benötigt dieses mindestens eine kinetische Energie von der Größe der Ionisierungsenergie. Bei Positronen liegt diese Energieschwelle um 6,8\,eV niedriger, was der Bindungsenergie des Positroniums entspricht. Die für die Ionisation benötigte Energie wird nämlich zu einem Teil aus der durch Positroniumsbildung gewonnenen Bindungsenergie bereitgestellt und reduziert daher die Energieschwelle, ab der Ionisation erfolgt, um diesen Betrag. Demnach können bei der Ionisation eines Atoms A (oder, analog dazu, des Moleküls AB) zwei mögliche Reaktionen erfolgen:
\begin{align}
 \text{e}^+ + \text{A} &\rightarrow \text{A}^+ + \text{e}^- + \text{e}^+ \label{eq:Ionisation} \\
 \text{e}^+ + \text{A} &\rightarrow \text{A}^+ + \text{Ps} \label{eq:IonisationPs}
\end{align}
Die erste Reaktion wird als direkte Ionisation - oder auch, aus dem Englischen, als reine Ionisation (\enquote{pure ionisation}, \cite{Bluhme2000}) - bezeichnet, während die zweite Reaktion \enquote{Einfang} (\enquote{capture}) genannt wird.
Man unterscheidet daher zwischen den Wirkungsquerschnitten für die direkte Ionisation $\sigma_{dir}$ und dem für die totale $\sigma_{tot}$, wobei gilt
\begin{equation}
 \sigma_{tot} = \sigma_{dir} + \sigma_{Ps}
\end{equation}
Dabei ist $\sigma_{Ps}$ der Wirkungsquerschnitt für die Positroniumsbildung.
Ist die Energie des einfallenden Positrons hoch genug, kann das Atom oder das Molekül auch mehrfach ionisiert werden. Im Stickstoff liegt die Schwelle für die direkte einfache Ionisierung (also $N_2 \rightarrow N_2^+$) bei $E_{dir} = 15,58$\,eV und somit die für Positroniumsbildung um 6,8\,eV niedriger bei $E_{Ps} = 8,78$\,eV.

\subsection{Elektronische Anregung}
\begin{figure}
 \centering
 \includegraphics[width=0.65\textwidth]{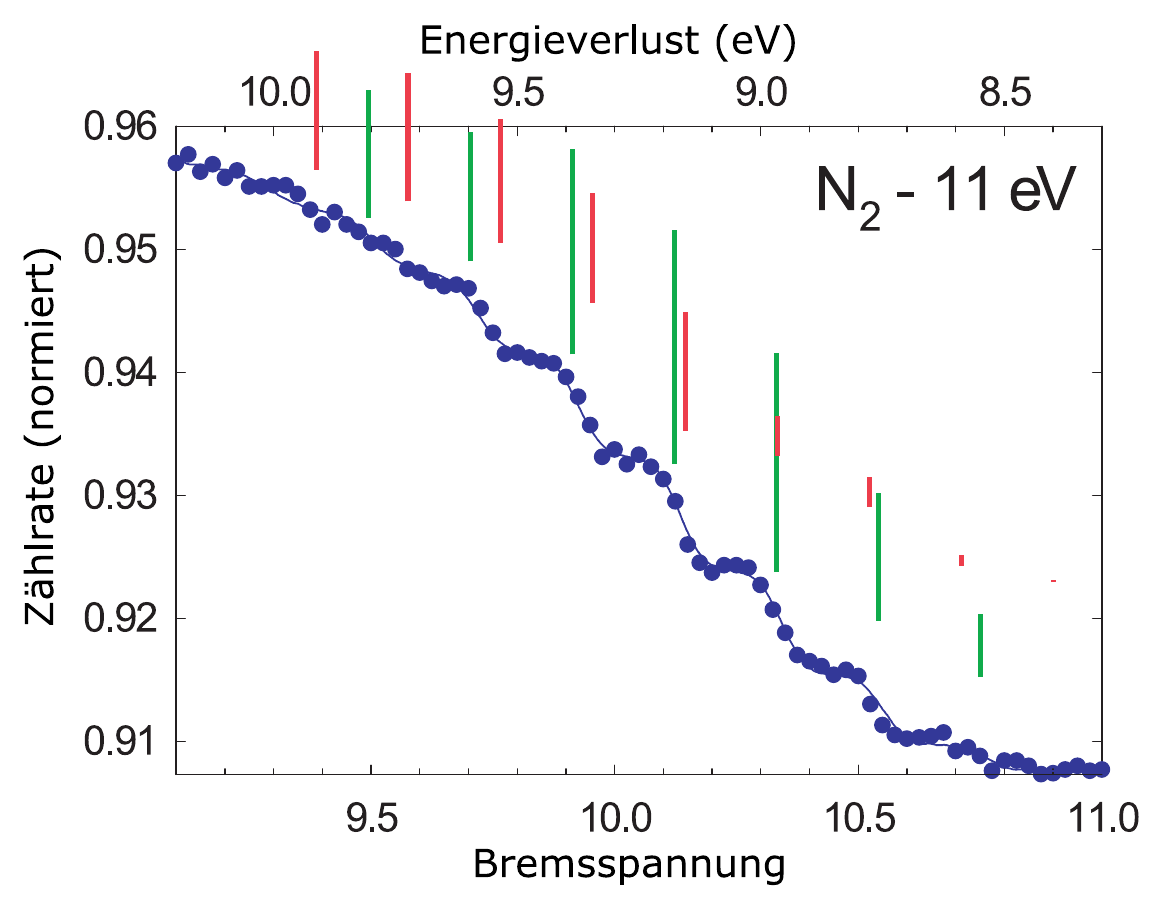}
 \caption{Elektronische Anregungen in Stickstoff (aus \cite{Marler2005PHD}): Ein 11\,eV Positronenstrahl wird durch eine mit Stickstoff gefüllte Zelle gelenkt und am Ausgang nach Durchlaufen einer variablen Gegenspannung durch einen Detektor analysiert. Aufgetragen ist die (normierte) Zählrate als Funktion der Bremsspannung. An den Positionen der Stufen im Spektrum lässt sich an der oben aufgetragenen x-Achse der Energieverlust der Positronen ablesen.}
 \label{fig:elektronischeAnregungenN2}
\end{figure}

Ähnlich wie bei der Stoßionisation gibt das Positron einen Teil seiner kinetischen Energie an ein Hüllenelektron des Gasmoleküls ab, das dadurch jedoch nicht die Elektronenhülle verlässt, sondern in einen angeregten Zustand gelangt. Unter Aussendung eines Photons fällt das Elektron wieder zurück in den Grundzustand. Dadurch, dass sich bei einem Übergang jedoch auch die Vibrationsquantenzahl $\nu$ sowie die Rotationsquantenzahl des Moleküls $J$ ändern können, wird aus einem elektronischen Übergang eine ganze Absorptionsbande. Abbildung \ref{fig:elektronischeAnregungenN2} zeigt ein (integrales) Absorptionsspektrum für 11\,eV-Positronen in Stickstoff.  Jede Stufe stellt dabei einen elektronischen Übergang da. Man sieht, dass Positronen durch elektronische Anregungen Energien von etwa 8,5 bis 10\,eV verlieren können. Die Aufweitung des elektronischen Spektrums wird hier durch zwei verschiedene Vibrationsanregungen sichtbar (grüne und rote Linien).

\section{Wirkungsquerschnitte und freie Weglängen}
Da die vorgestellten Wechselwirkungen in weiten Energiebereichen miteinander konkurrieren, benötigt man ihre energieabhängigen Wirkungsquerschnitte, um Aussagen über das Moderationsverhalten im Gas machen zu können. Der Wirkungsquerschnitt $\sigma$  ist definiert als das Verhältnis aus der Anzahl der pro Zeiteinheit und Gasmolekül stattfindenden Reaktionen $R$ und der Intensität der einfallenden Teilchen $I$ (in diesem Fall also der Positronen):
\begin{equation}
 \sigma = \frac R I
\end{equation}
Der Wirkungsquerschnitt lässt sich auch durch eine mittlere freie Weglänge $\lambda$ der Positronen ausdrücken. Sie gibt die Länge der Flugstrecke an, nach der die Wahrscheinlichkeit, dass keine Wechselwirkung des Positrons mit dem Gas stattgefunden hat nur noch $e^{-1}$\ ($\approx 37$\,\%) beträgt. Oder mit anderen Worten: 63\,\% aller Positronen wechselwirken mit dem Gas, bevor sie die Strecke $\lambda$ durchfliegen.
\begin{equation}
 \lambda = \frac 1 {\sigma \rho_T}
\end{equation}
Dabei ist $\rho_T$ die vom Druck $P$ und der Temperatur $T$ abhängige Teilchendichte des Gases:
\begin{equation}
 \rho_T = \frac P {k_B T}
\end{equation}

\section{Wirkungsquerschnitte im Stickstoff}
\begin{figure}
 \centering
 \includegraphics[width=0.7\textwidth]{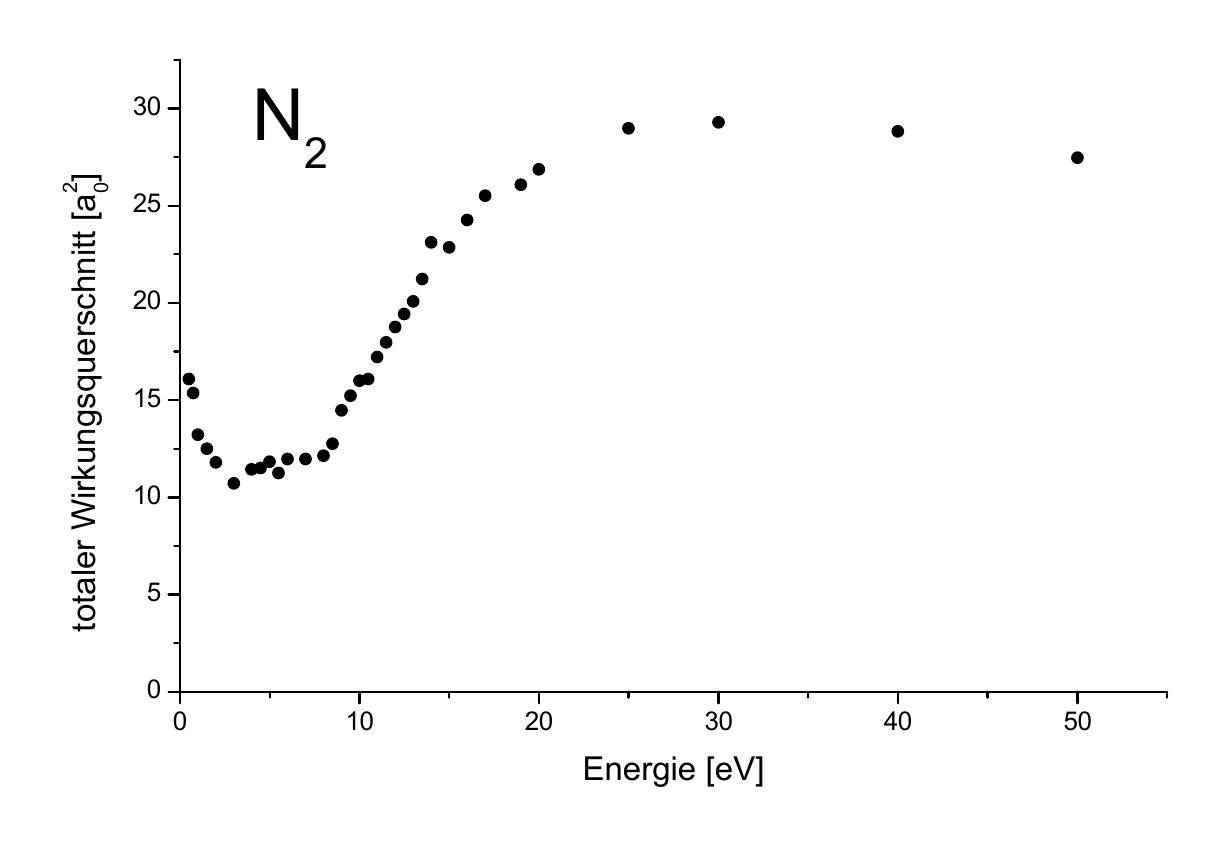}
 \caption{Totaler Wirkungsquerschnitt von Positronen im Stickstoff}
 \label{fig:TotalerWqs-N2}
\end{figure}
\begin{figure}
 \centering
 \includegraphics[width=0.7\textwidth]{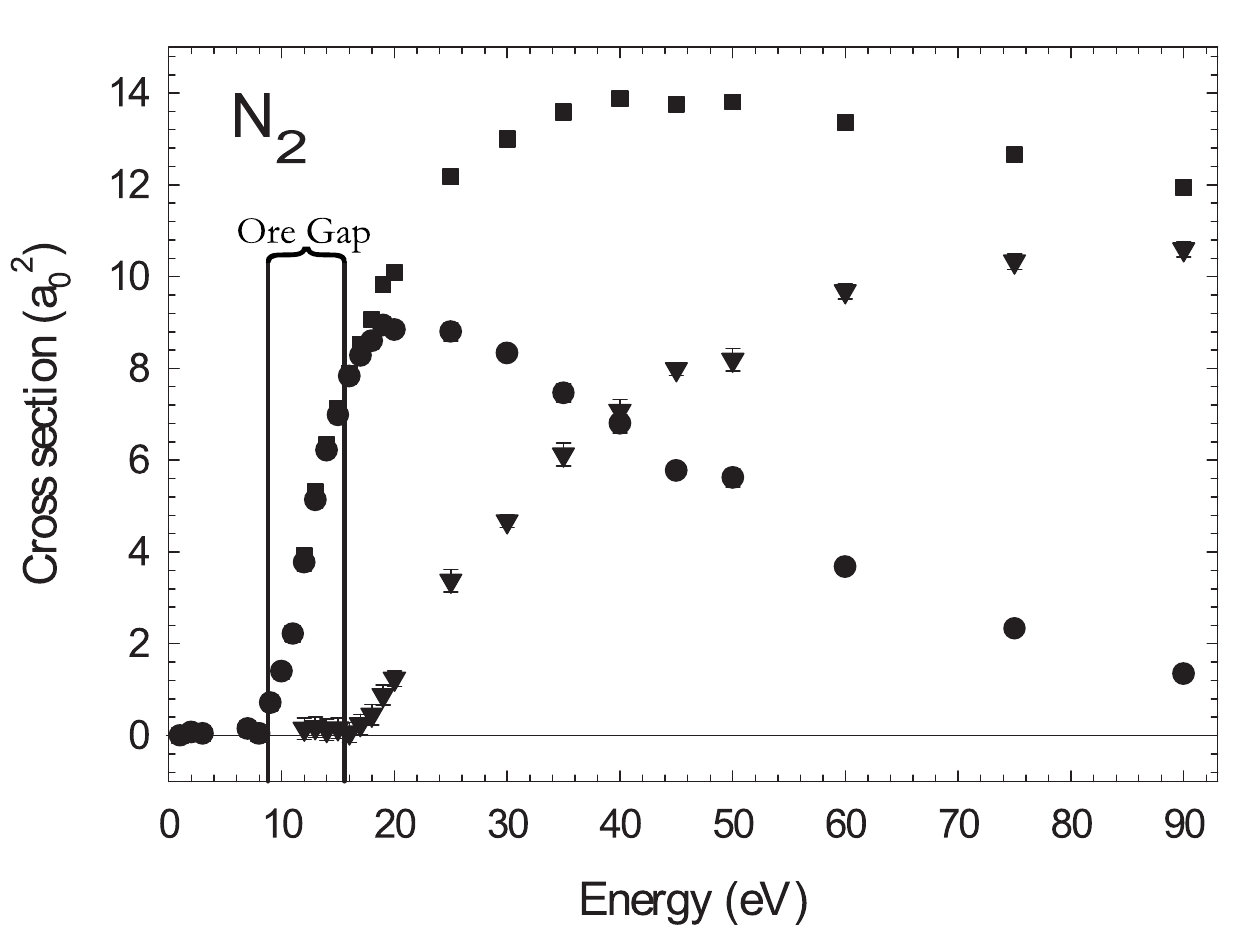}
 \caption{Wirkungsquerschnitte von totaler Ionisation ($\blacksquare$), direkter Ionisation ($\blacktriangledown$) und Positroniumsbildung ($\bullet$) in Stickstoff (aus \cite{Marler2005PHD}).}
 \label{fig:Wqs-IonisationenundPs}
\end{figure}

Für die oben vorgestellten Wechselwirkungen der Positronen im Gas werden nun die Wirkungsquerschnitte in Stickstoff dargelegt und erläutert. Abbildung \ref{fig:TotalerWqs-N2} zeigt den totalen Wirkungsquerschnitt\footnote{Der Wirkungsquerschnitt wird hier in Einheiten von $a_0^2 = 2,8 \cdot 10^{-21}$\,m$^2$ angegeben. $a_0$ ist dabei der Bohrsche Radius.} in Einheiten von $a_0^2$ als Funktion der Energie der Positronen. Im totalen Wirkungsquerschnitt sind alle Wechselwirkungen zwischen Gas und Positronen enthalten. Mit ihm lassen sich daher keine Aussagen über die Moderation von Positronen im Gas treffen. Daher sollen zunächst die Wirkungsquerschnitte der elektronischen Anregungen und der Ionisation (inkl. Positroniumsbildung) betrachtet werden. Abbildung \ref{fig:Wqs-IonisationenundPs} zeigt die Wirkungsquerschnitte von Positroniumsbildung, direkter Ionisation und der Summe aus beiden, der totalen Ionisation. Die vertikalen Linien zeigen die Schwellwerte für Positroniumsbildung (8,78\,eV) und Ionisation (15,58\,eV). Somit können Positronen mit Energien zwischen diesen beiden Schwellenenergien - diesen Bereich nennt man auch \enquote{Ore Gap} \cite{Ore1949} - das Gas nur durch Positroniumsbildung ionisieren, und gehen damit für die Moderation verloren. Erst ab etwa 40\,eV  übertrifft der Wirkungsquerschnitt der direkten Ionisation den der Positroniumsbildung.

\begin{figure}
 \centering
 \includegraphics[width=0.65\textwidth]{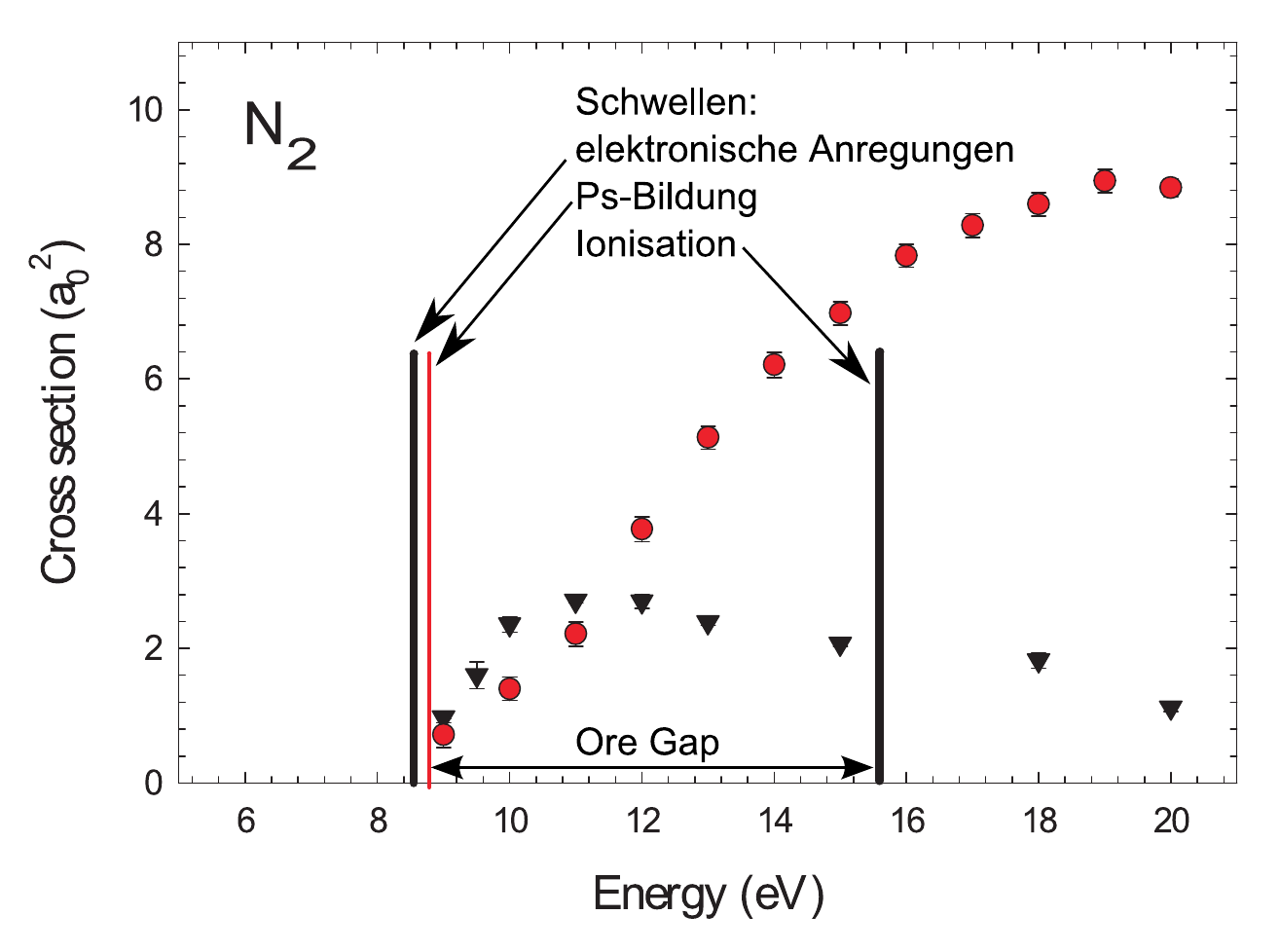}
 \caption{Wirkungsquerschnitte von Positroniumsbildung ({\color{red}$\bullet$}) und elektronischen Anregungen ($\blacktriangledown$) in Stickstoff (aus \cite{Marler2005PHD}). Die vertikalen Linien geben die Energieschwellen der beiden Anregungen an.}
 \label{fig:Wqs-PsundelektronischeAnregungen}
\end{figure}
\begin{figure}
 \centering
 \includegraphics[width=0.65\textwidth]{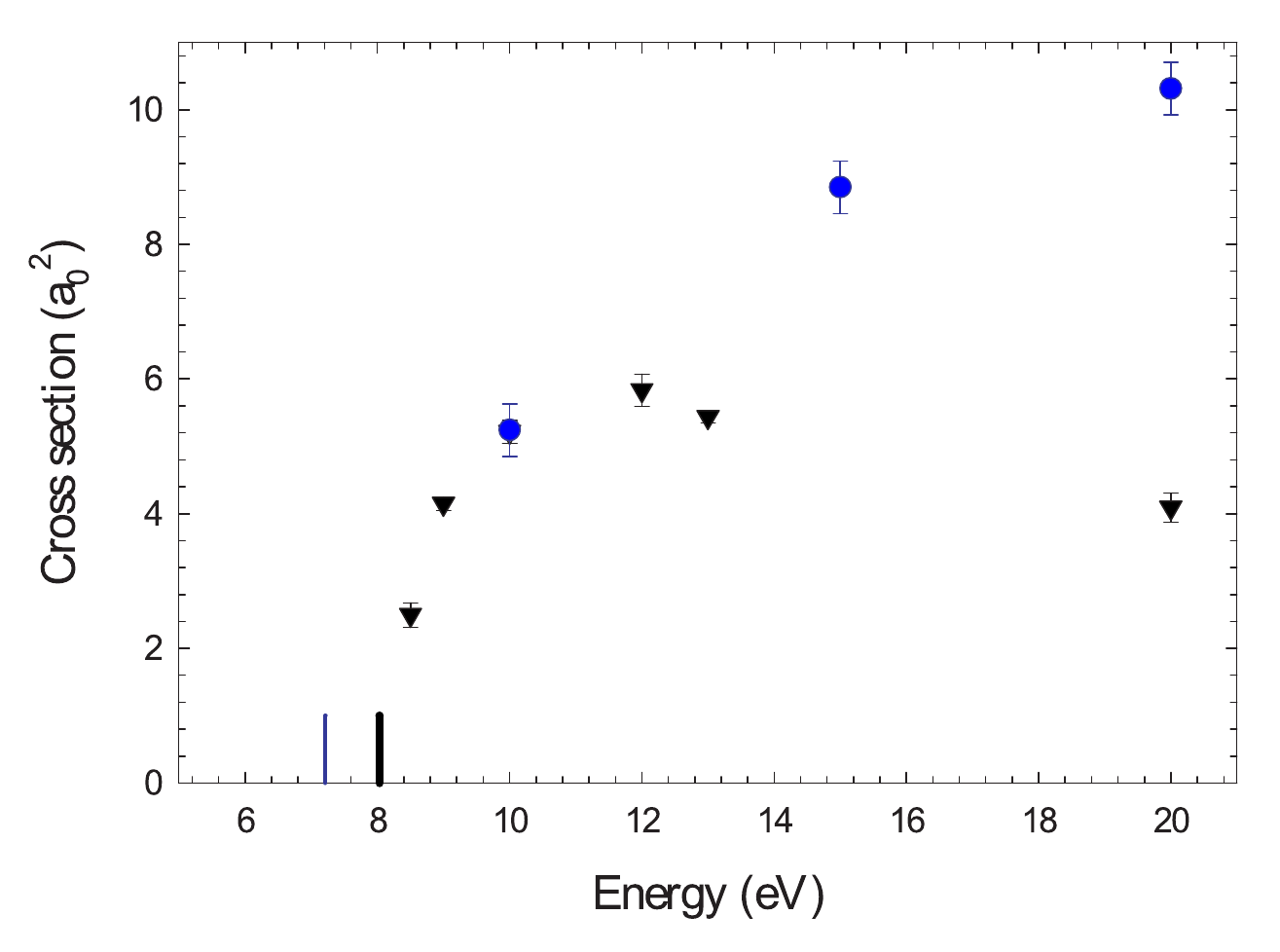}
 \caption{Wirkungsquerschnitte von Positroniumsbildung ({\color{blue}$\bullet$}) und elektronischen Anregungen ($\blacktriangledown$) in CO (aus \cite{Marler2005PHD}). Die vertikalen Linien geben die Energieschwellen der beiden Anregungen an (blau: Positroniumsbildung, schwarz: elektr. Anregungen).}
 \label{fig:Wqs-PsundelektronischeAnregungen-CO}
\end{figure}
Vergleicht man den Wirkungsquerschnitt der elektronischen Anregungen mit dem der Positroniumsbildung (siehe Abbildung \ref{fig:Wqs-PsundelektronischeAnregungen}), stellt man fest, dass Positronen deren Energien im \enquote{Ore Gap} liegen, diesen Bereich durch elektronische Anregung eines Stickstoffmoleküls wieder verlassen können, da sie pro Stoß eine Energie von 8,5 bis 10\,eV verlieren. Der Wirkungsquerschnitt für diesen Prozess liegt jedoch um einen Faktor von bis zu 4,5 unter dem für Ps-Bildung. Somit wird ein Großteil der Positronen in diesem Energiebereich verloren gehen. Eine positive Eigenschaft des Stickstoffs wird jedoch deutlich, wenn man die Grenzwerte beider Prozesse vergleicht. So liegt die Energieschwelle für Positroniumsbildung (8,78\,eV) über der der tiefsten elektronischen Anregung von 8,55\,eV \cite{Marler2005PHD}. Bei anderen Molekülen wie CO oder O$_2$ ist dies nicht der Fall. Bei Sauerstoff liegt die Schwelle für Positroniumsbildung z.\,B. um 1,65\,eV unter der für elektronische Anregungen; bei CO um 0,86\,eV (siehe Abbildung \ref{fig:Wqs-PsundelektronischeAnregungen-CO}) \cite{Marler2005}. Damit geht ein Großteil der in dieses Energiefenster heruntergestreuten Positronen verloren, da ein weiteres Abbremsen der Positronen dort hauptsächlich nur noch durch Vibrationsanregungen möglich ist. Der durch Vibrationsanregungen erreichbare Energieverlust liegt jedoch im Bereich von einigen $10^{-1}$\,eV, weshalb mehrere Stöße notwendig sind, um unter die Positroniumsbildungsschwelle zu gelangen. Da die Wirkungsquerschnitte für beide Prozesse in derselben Größenordnung liegen, wird mit steigender Anzahl an Stößen die Wahrscheinlichkeit für die Annihilation der Positronen durch Positroniumsbildung immer größer. Darüber hinaus kommt beim Stickstoff hinzu, dass im Bereich von 8,55 bis etwa 11,5\,eV der Wirkungsquerschnitt für elektronische Anregungen größer ist als der für Positroniumsbildung, was bei CO beispielsweise nicht der Fall ist.

Bisher wurden Vibrations- und Rotationsanregung, sowie die elastische Streuung vernachlässigt, da deren Beitrag zur Moderation (Energieübertrag $\le 10^{-1}$\,eV) im Vergleich zu den oben genannten vernachlässigbar klein ist. Im Energiebereich unterhalb der Schwelle für elektronische Anregungen sind jedoch hauptsächlich Vibrationsanregungen für die Moderation der Positronen im Stickstoff verantwortlich, bis die Positronen auch diese Energieschwelle unterschritten haben ($E_{vib}=0,290$\,eV, \cite{Coleman1981}). Wirkungsquerschnitte für diesen Prozess sind bisher noch nicht gemessen worden.

\begin{figure}
 \centering
 \includegraphics[width=0.6\textwidth]{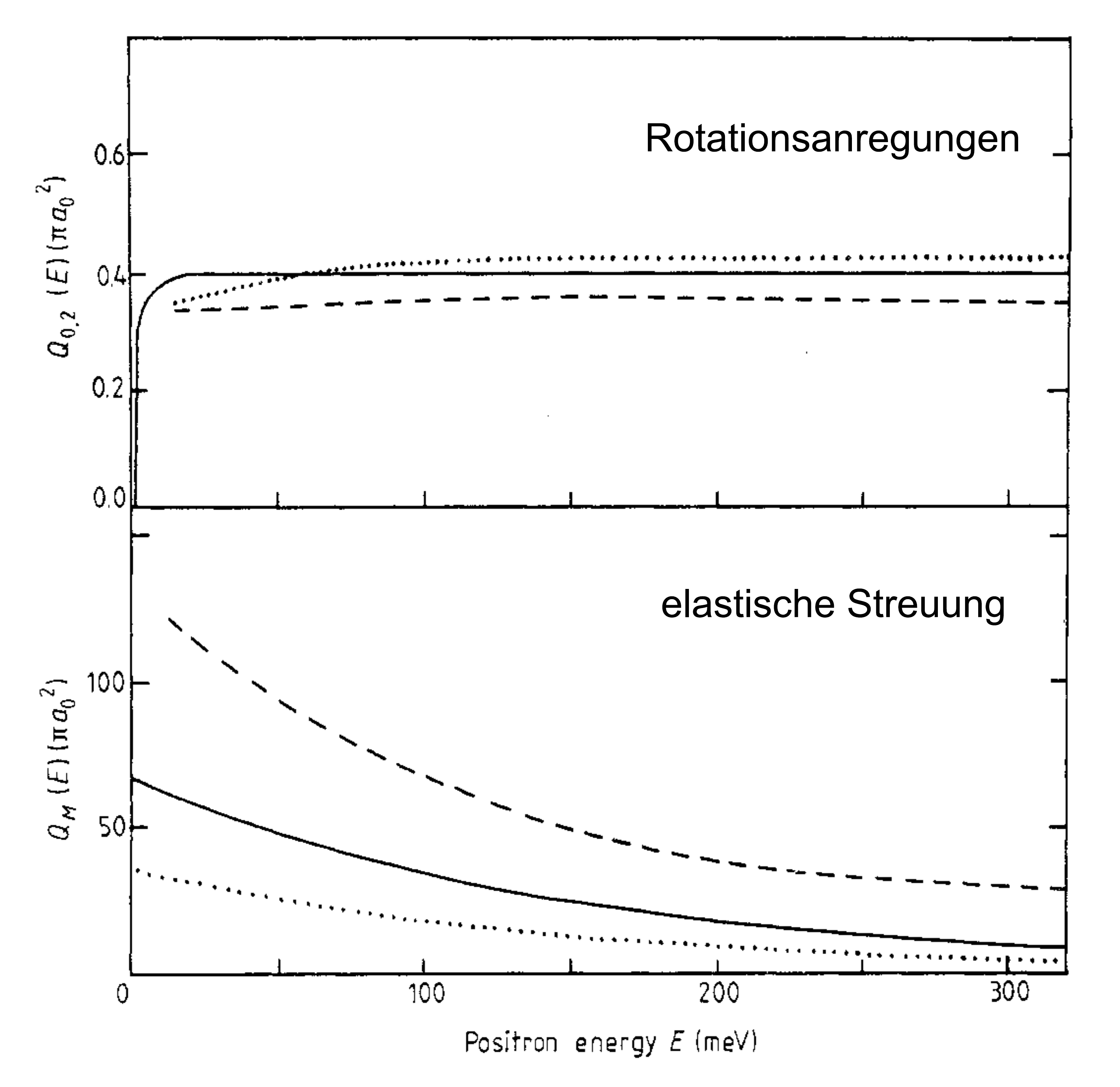}
 \caption{Wirkungsquerschnitte von Rotationsanregungen (oben) und elastischer Streuung (unten) (aus \cite{Coleman1981})}
 \label{fig:Wqs-Rotation-Elastisch}
\end{figure}
Für Energien unterhalb von $E_{vib}$ bis zur Thermalisierung werden schließlich elastische Streuung und Rotationsanregungen relevant. Abbildung \ref{fig:Wqs-Rotation-Elastisch} zeigt beide Wirkungsquerschnitte im Bereich einiger 100\,meV. Trotz des etwa um den Faktor 200 höheren Wirkungsquerschnitts für elastische Streuung hat dieser Prozess jedoch nur einen Anteil von $\approx 25$\,\% an der Thermalisierung unterhalb von $E_{vib}$. Dies liegt an den in Abschnitt \ref{elastischeStreuung} erläuterten geringen Energieverlusten der Positronen bei der elastischen Streuung.

\section{Abschätzung der maximalen Ausbeute an (re)moderierten Positronen}
\begin{figure}
 \centering
 \includegraphics[width=0.7\textwidth]{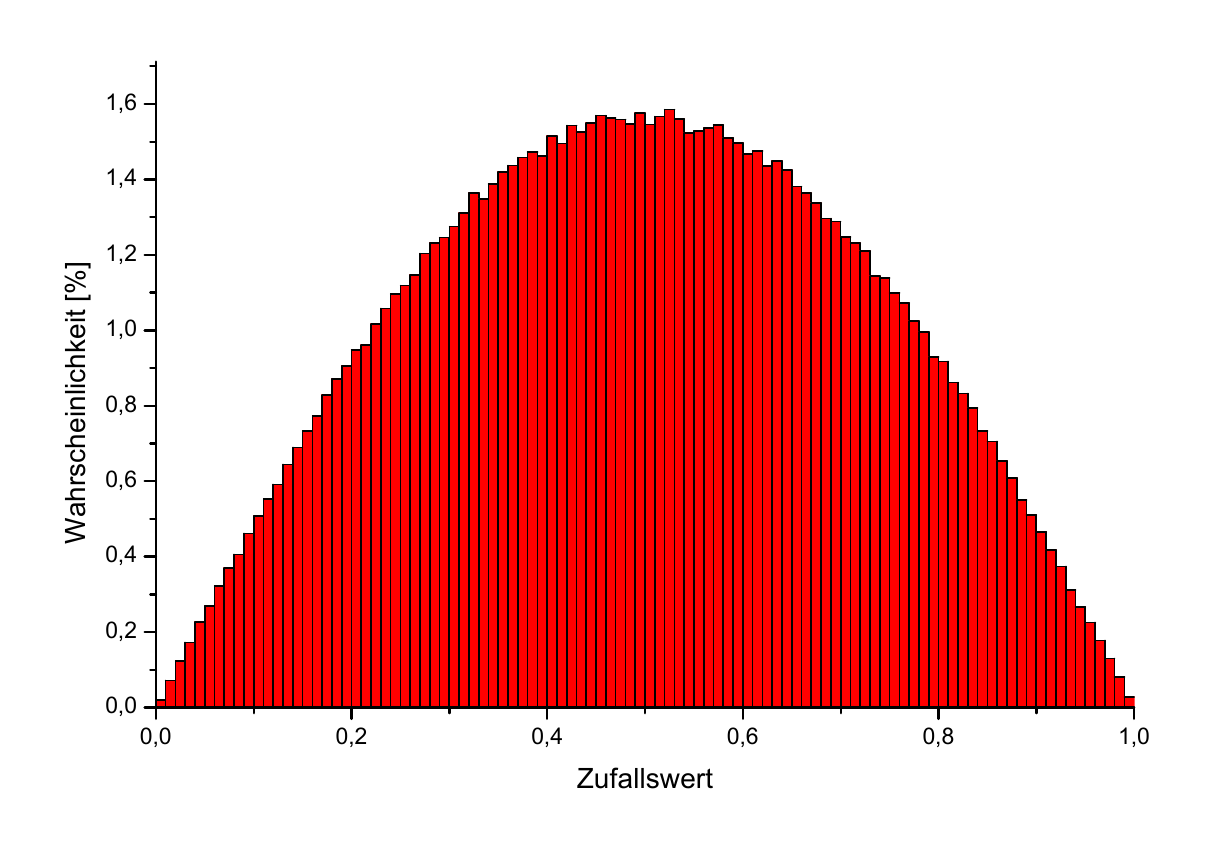}
 \caption{Histogramm einer Sinusverteilung, gebildet aus einer Million Zufallswerten.}
 \label{fig:Sinusverteilung}
\end{figure}

Zur Abschätzung der Effizienz der Moderation im Gas wurden die Wechselwirkungen der Positronen im Stickstoff mit einer, im Rahmen dieser Diplomarbeit erstellten, Monte-Carlo-Simulation am Computer untersucht. Dabei werden die einzelnen Stöße und die dadurch bedingten Energieverluste eines Positrons simuliert. Mit Hilfe der Wirkungsquerschnitte lässt sich, abhängig von der momentanen Energie des Positrons, eine Wahrscheinlichkeit für jeden Wechselwirkungsprozess berechnen. Unter Berücksichtigung dieser Wahrscheinlichkeiten bestimmt ein Zufallsgenerator vor jedem Stoß die Art der Wechselwirkung, mit der das Positron mit dem Gasmolekül wechselwirkt. Davon abhängig verliert das Positron durch den Stoß entweder eine bestimmte Energie oder bildet zusammen mit einem Elektron Positronium. Dies wird für jedes Positron solange wiederholt, bis es entweder eine Energie unterhalb der Schwelle für Positroniumsbildung erreicht oder eben durch Positroniumsbildung verloren geht. Aus dem Anteil der Positronen, die unter die Schwelle für Positroniumsbildung gestreut werden und der primären Positronenintensität lässt sich wiederum eine Moderationseffizienz berechnen. Für jede Anfangsenergie eines Positrons muss diese Simulation nun mehrere Tausend bis Millionen Mal durchgeführt werden, damit die Effizienz aufgrund des Gesetzes der großen Zahlen gegen einen Wert konvergiert. Bei der Simulation wurden folgende Vereinfachungen gemacht:
\begin{itemize}
 \item Die Simulation eines Positrons wurde gestoppt und das e$^+$ als moderiert eingestuft, sobald es eine Energie unterhalb der Ps-Bildungsschwelle von 8,78\,eV hatte.
 \item Es wurden nur Ps-Bildung, Ionisation, elektrische Anregungen und elastische Streuung simuliert. Vibrations- und Rotationsanregungen wurden mit dem Wirkungsquerschnitt für elastische Streuung zusammengefasst, wobei der Energieverlust vernachlässigt wurde.
 \item Der Energieverlust durch elektronische Anregungen (Abbildung \ref{fig:elektronischeAnregungenN2}) wurde durch eine Gleichverteilung genähert. Somit konnte der Einfachheit halber ein gleichverteilter Zufallswert zwischen 8,5 und 10\,eV verwendet werden.
 \item Der Energieübertrag durch Ionisation wurde durch einen sinusförmig verteilten Zufallswert bestimmt. Dies bedeutet, dass die Wahrscheinlichkeit für einen Energieübertrag, der gerade der Ionisationsenergie von 15,58\,eV entspricht, sowie für den Übertrag der gesamten kinetischen Energie des Positrons gegen Null geht. Das Maximum der Wahrscheinlichkeit liegt in der Mitte zwischen diesen beiden Werten. Abbildung \ref{fig:Sinusverteilung} zeigt die Sinusverteilung, wobei ein Wert von 0 gerade einem Energieübertrag von 15,58\,eV und ein Wert von 1 den Übertrag der kompletten kinetischen Energie des Positrons auf das  Elektron entspricht.
\end{itemize}

\begin{figure}
 \centering
 \includegraphics[width=0.98\textwidth]{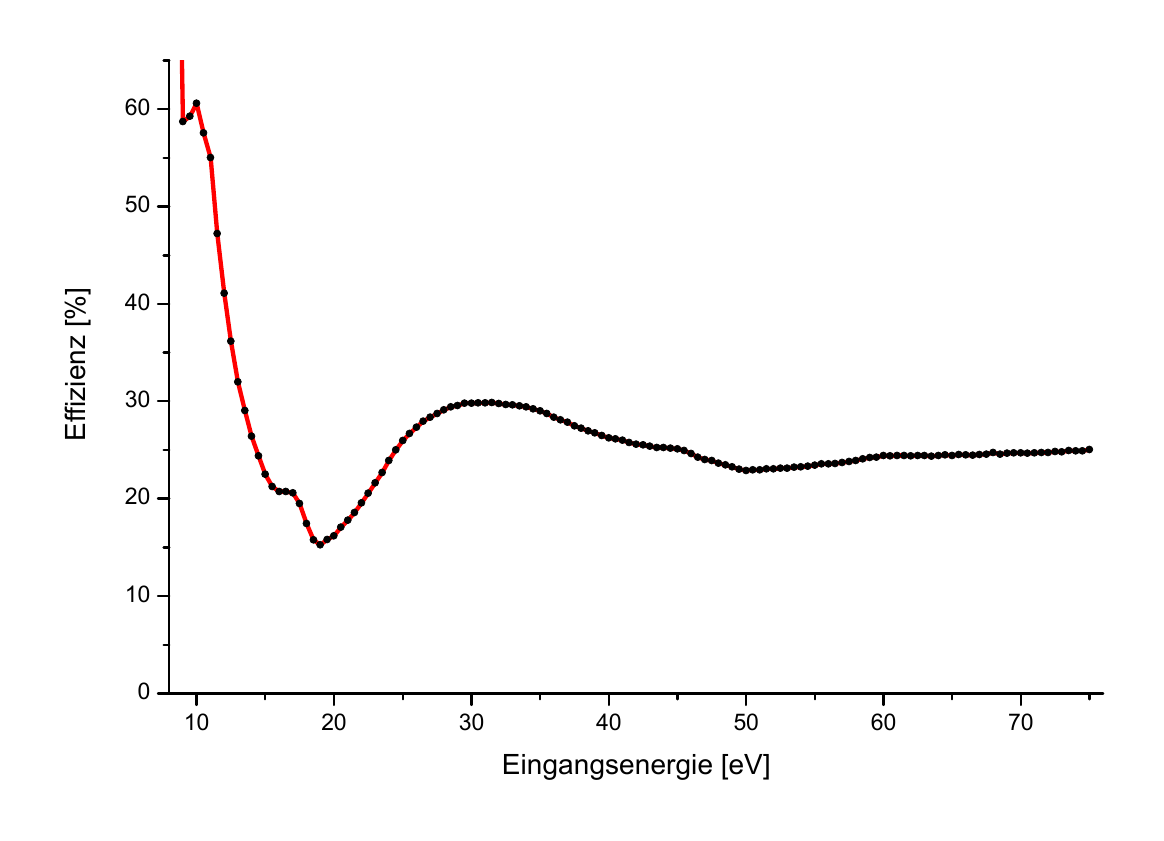}
 \caption{Moderationseffizienz als Funktion der Eingangsenergie. Für jeden Energiewert wurden eine Million Positronen simuliert.}
 \label{fig:MonteCarloSim}
\end{figure}

Abbildung \ref{fig:MonteCarloSim} zeigt die berechnete Effizienz als Funktion der Eingangsenergie der Positronen. Von C. M. Surko et al. wurde eine Transmission eines 40\,eV Strahls in Stickstoff von 25\,\% gemessen \cite{Surko1989}. Der hier für 40\,eV simulierte Wert von ca. 26\,\% stimmt damit sehr gut überein.
\clearpage{}
\clearpage{}\chapter{Entwicklung des Gasmoderators}
In diesem Kapitel wird die Entwicklung und der Aufbau des Moderators erläutert.

\section{Prinzip}
\begin{figure}
 \centering
 \includegraphics[width=0.98\textwidth]{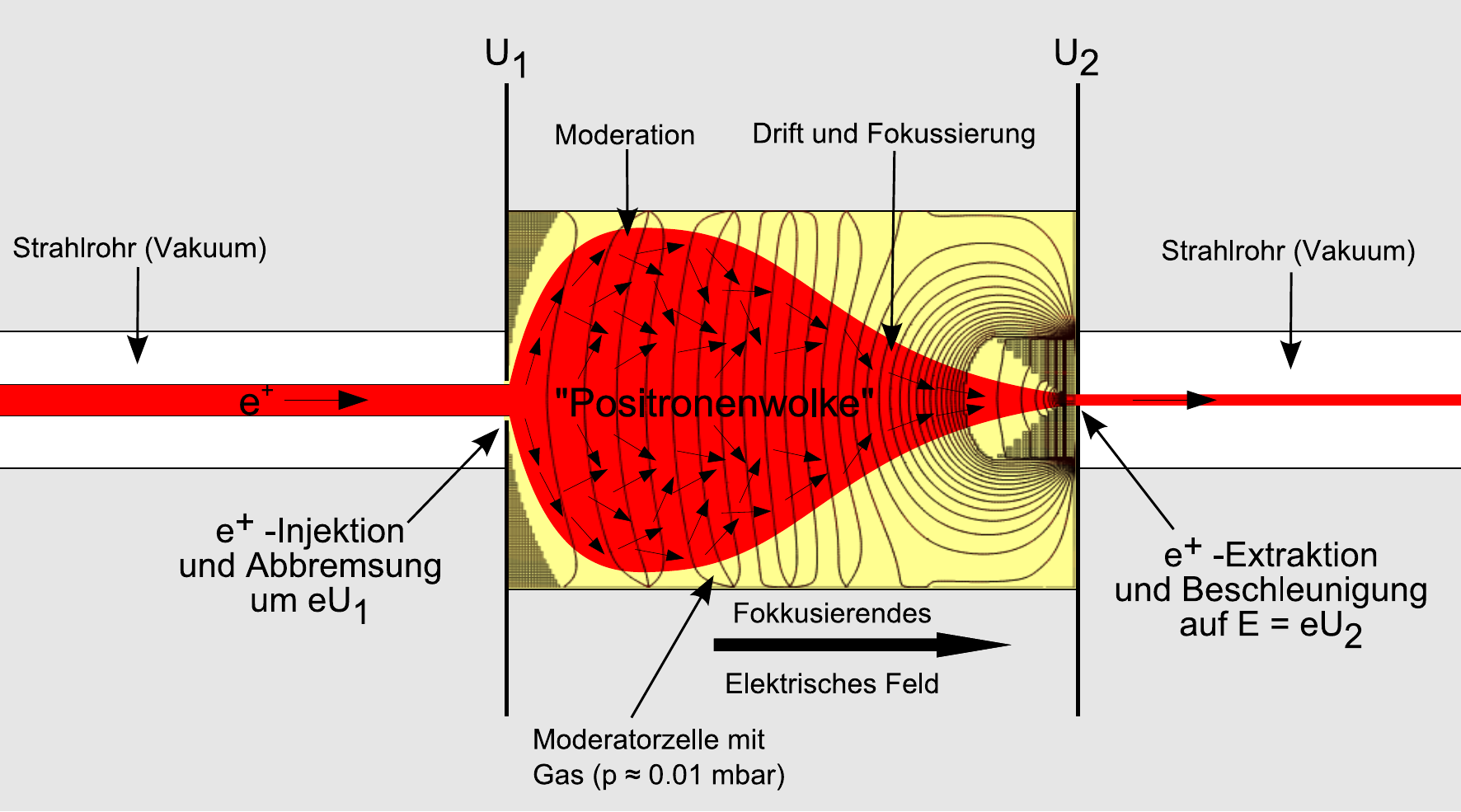}
 \caption{Prinzipskizze des Gasmoderators}
 \label{fig:Prinzipskizze}
\end{figure}
Abbildung \ref{fig:Prinzipskizze} zeigt den prinzipiellen Aufbau des Gasmoderators. Von der linken Seite ankommende Positronen werden am Eingang der Moderatorzelle durch eine Bremsspannung $U_1$ abgebremst und gelangen durch eine kleine Öffnung in das Innere der Zelle. Dort stoßen die Positronen mit den Gasmolekülen, werden abgebremst und driften durch weitere Stöße entlang den Feldlinien eines fokussierenden elektrischen Feldes Richtung Ausgang. Nach dem Passieren des Ausgangs werden die nun thermalisierten Positronen durch ein Potential $U_2$ wiederum beschleunigt. Durch differentielles Pumpen am Eingang und Ausgang muss dabei sicher gestellt werden, dass der Druck im Strahlrohr so gering ist, dass dort keine Positronen streuen und verloren gehen. In dieser Arbeit soll das Einkoppeln der Positronen in die Gaszelle realisiert und deren Energie am Ausgang analysiert werden.

\section{Experimenteller Aufbau}
\subsection{Mechanische Konstruktion}
\begin{figure}
 \centering
 \includegraphics[width=0.98\textwidth]{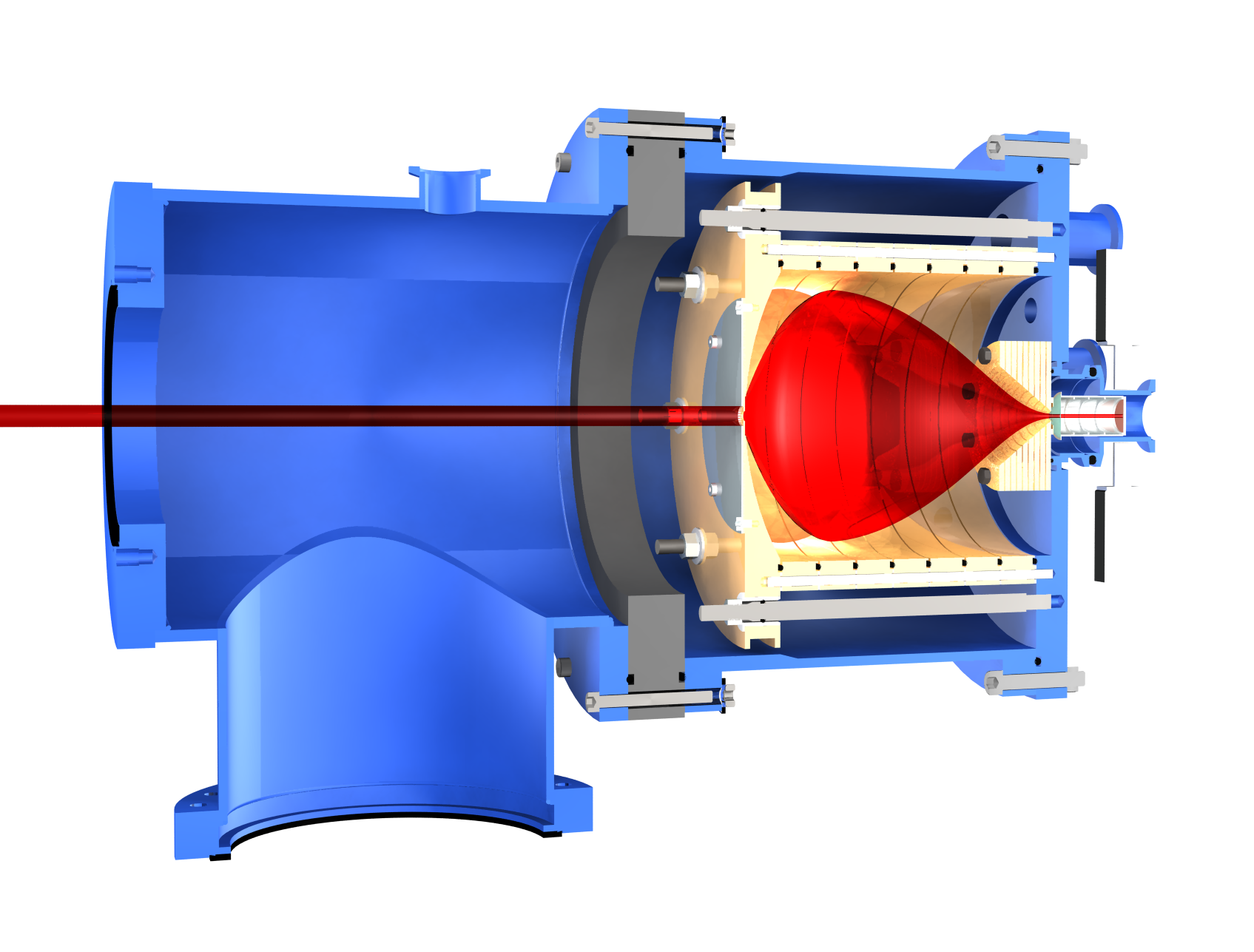}
 \caption{Schnittdarstellung des Versuchsaufbaus}
 \label{fig:AufbaumitPumprohr}
\end{figure}

Abbildung \ref{fig:AufbaumitPumprohr} zeigt eine Schnittdarstellung des gesamten Versuchsaufbaus. Dieser lässt sich in zwei Bereiche teilen: Positronen passieren zunächst ein sogenanntes Pumprohr, an dem an der Unterseite eine Turbomolekularpumpe (nicht eingezeichnet) angeflanscht ist. Durch einen (grau eingezeichneten) elektrisch isolierenden PVC-Flansch ist rechts an diesem Pumprohr die Moderatorzelle befestigt. Das Strahlrohr der \textsc{Nepomuc}-Quelle ist an der linken Seite des Pumprohrs montiert.

\begin{figure}
 \centering
 \includegraphics[width=0.98\textwidth]{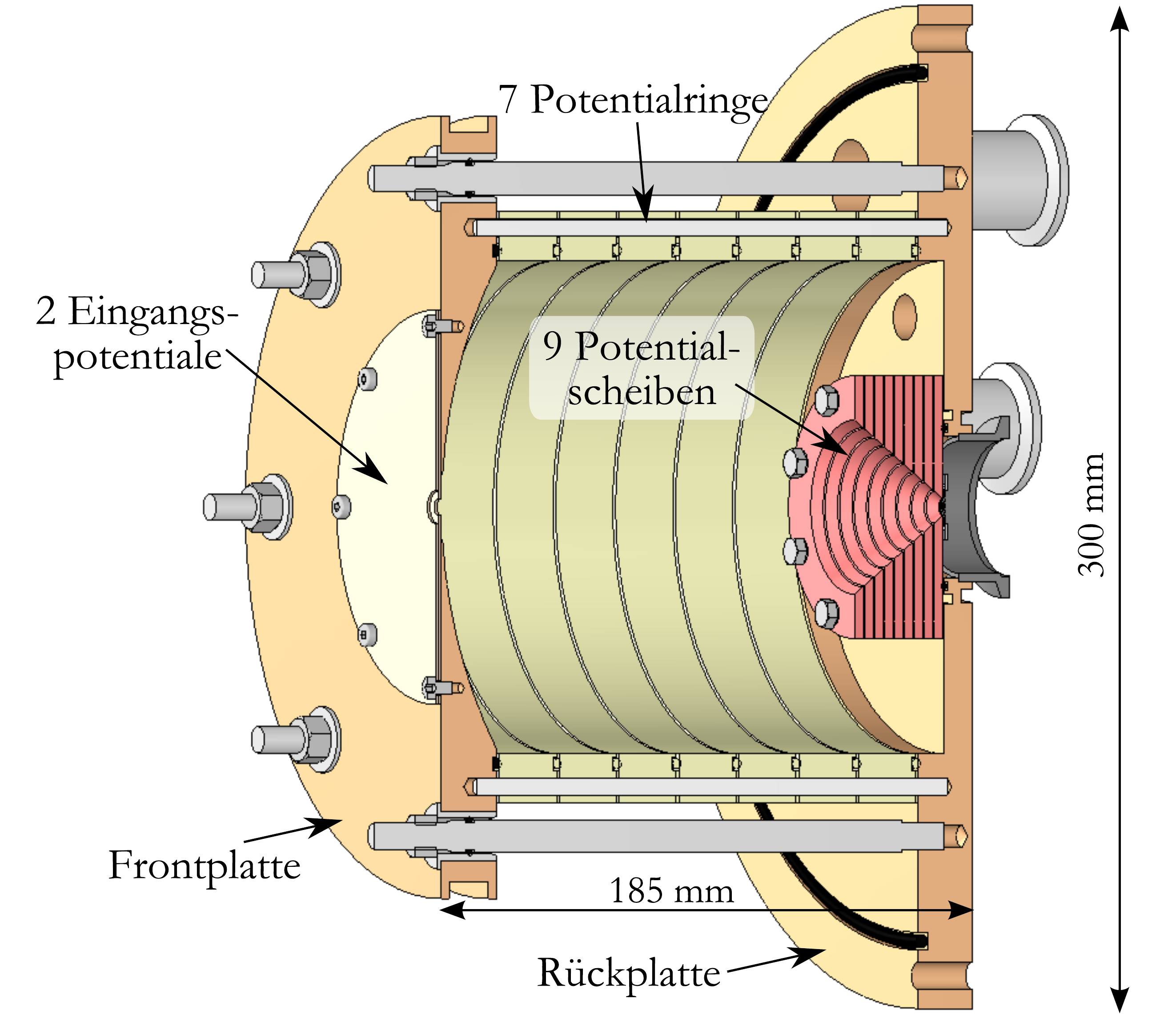}
 \caption{Schnittdarstellung der Gaszelle}
 \label{fig:Gaszelle}
\end{figure}
Eine vergrößerte Darstellung der Gaszelle zeigt Abbildung \ref{fig:Gaszelle}. Die gesamte Zelle ist auf der rechts im Bild sichtbaren Rückplatte montiert. Aufgebaut ist die Zelle aus sieben Metallringen, an denen später die elektrischen Potentiale angelegt werden und daher untereinander durch O-Ringe aus Nitrilkautschuk elektrisch isoliert sind. Gleichzeitig dichten diese O-Ringe die Gaszelle auch zum Hochvakuumbereich nach außen ab. Auf der linken Seite wird die Gaszelle durch die Frontplatte abgeschlossen. Zum Einkoppeln der Positronen befindet sich in der Frontplatte ein 20\,mm großes Loch. Zur Verbesserung des Drucks im Pumprohr wurde jedoch die Größe dieses Loch reduziert, indem ein Metallblech mit einem Lochdurchmesser von nur 3\,mm vor die Öffnung in der Frontplatte montiert wurde. Vor dieses Blech wurde, getrennt durch eine isolierende Teflonfolie, ein weiteres Metallblech mit einem etwas größeren Loch (5\,mm) befestigt. Dieses Blech wurde auf Masse gesetzt. Der Sinn dieses Blechs besteht darin, die Positronen in einem möglichst homogenen elektrischen Feld abzubremsen. Zur Verbesserung der Feldhomogenität wurden in die Öffnungen beider Bleche außerdem noch feine Drahtgitter eingebaut.

\begin{figure}
 \centering
 \includegraphics[width=0.5\textwidth]{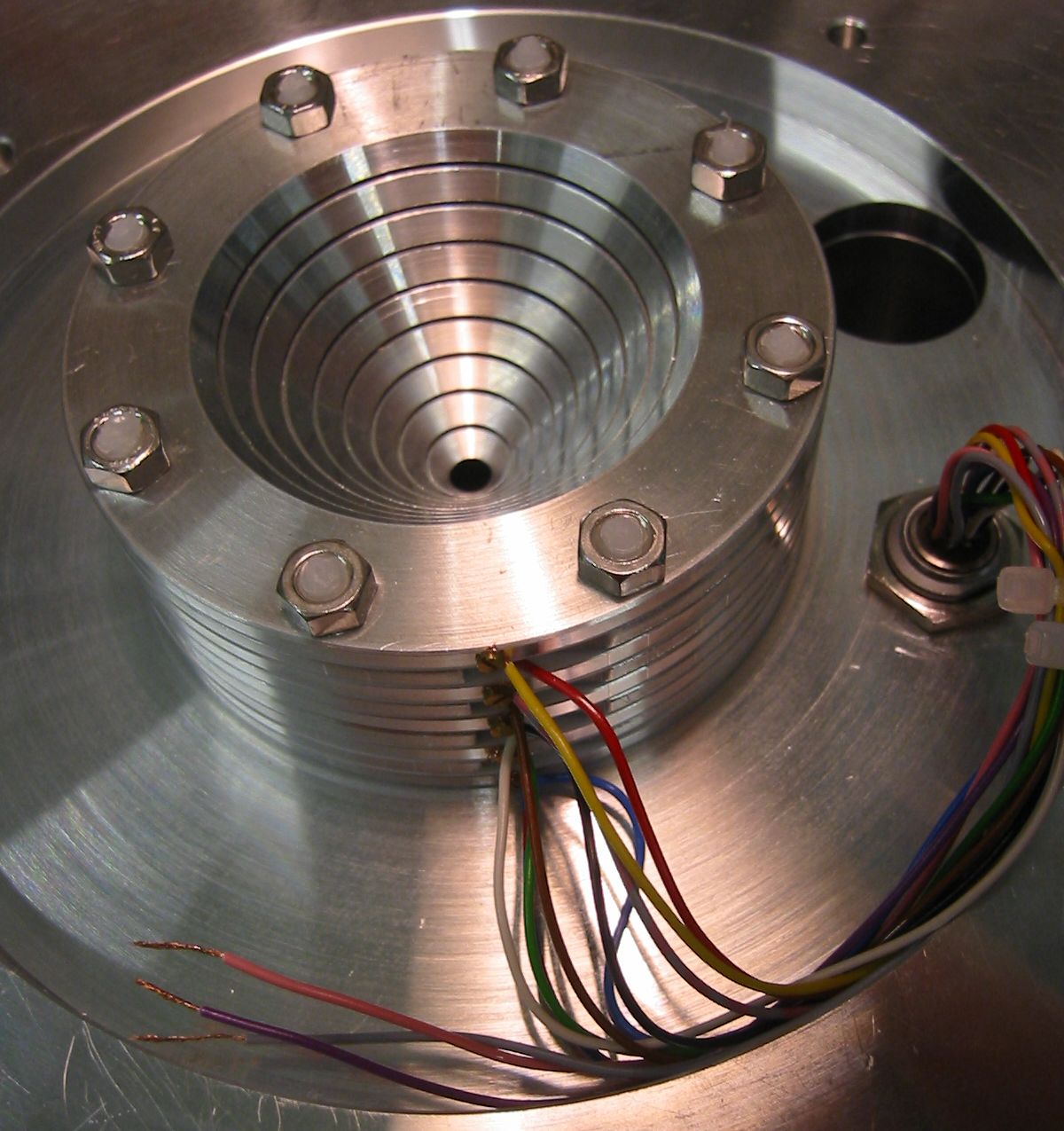}
 \caption{Foto der auf der Rückplatte befestigten, konisch zulaufenden Potentialscheiben zur Fokussierung des Strahls. Die Scheiben sind untereinander durch Teflonfolien isoliert und werden durch Kunststoffschrauben zusammengehalten. Zwischen den Scheiben gelangt das Gas in die Moderatorzelle (siehe Abschnitt \ref{Gassystem}). Das gewölbte Gitter ist hier noch nicht montiert.}
 \label{fig:FotoTrichter}
\end{figure}

Um die Positronen durch ein elektrisches Feld auf den Ausgang fokussierend zu driften wurden neun konisch zulaufende voneinander isolierte Potentialscheiben direkt auf die Rückplatte montiert (siehe Abbildung \ref{fig:FotoTrichter}). Die Ausgangsöffnung der kleinsten Scheibe hat einen Durchmesser von 5\,mm. Hinter dieser Öffnung wurde wiederum ein Gitter befestigt, das von der letzten Scheibe isoliert ist. Zusätzlich wurde ein nach außen gewölbtes Gitter über die 53\,mm große Eingangsöffnung gelegt um die Fokussierung des Feldes zu verstärken.

\subsection{Elektrische Potentiale} \label{simion-potentiale}
\begin{figure}
 \centering
 \includegraphics[width=0.6\textwidth]{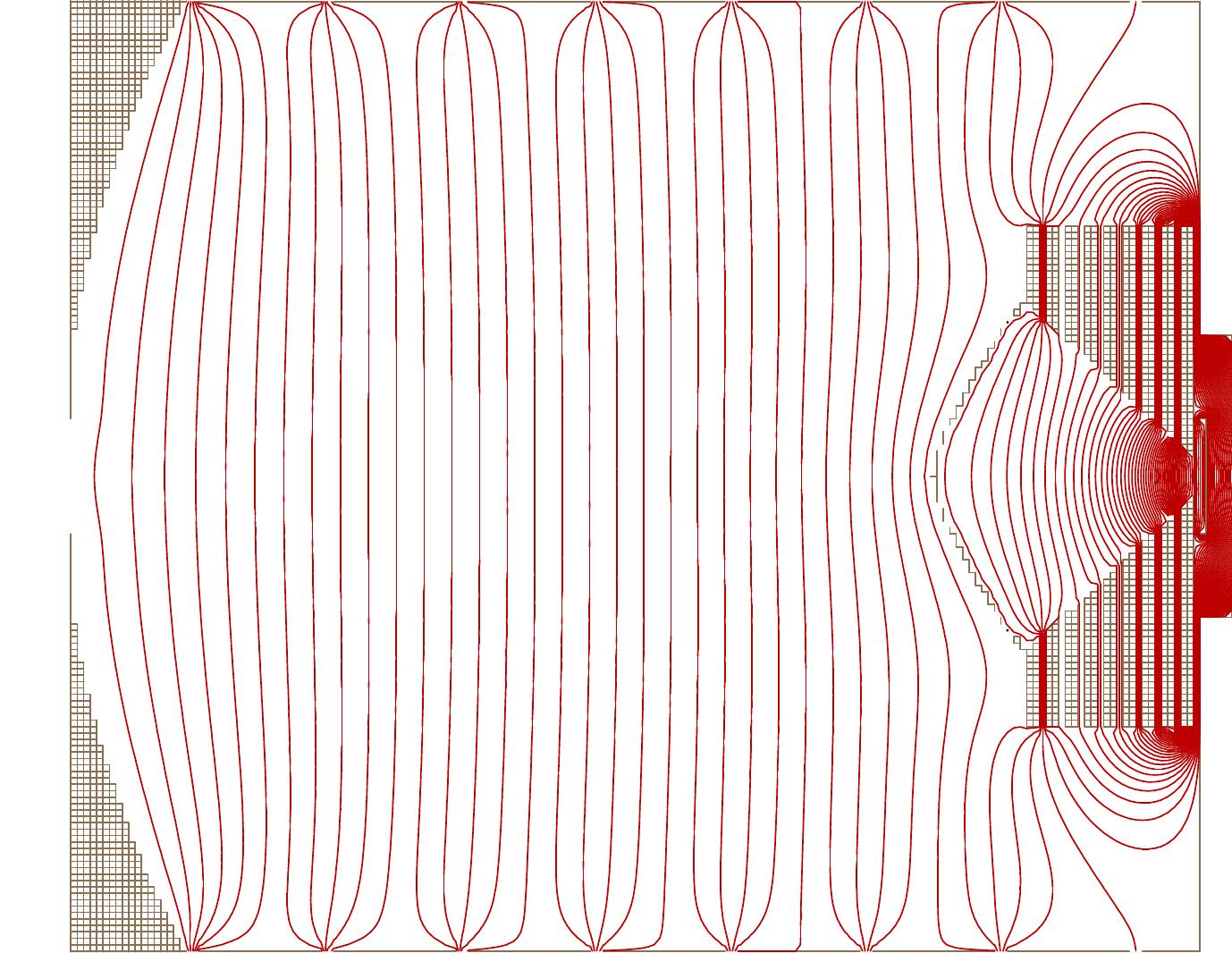}
 \caption{Elektrisches Feld in der Gaszelle}
 \label{fig:E-FeldZelle}
\end{figure}

Da sich alleine im Inneren der Gaszelle 16 Elektroden befinden, wurde das elektrische Feld mithilfe des Simulationsprogramms \textsc{Simion 3D} optimiert. Abbildung \ref{fig:E-FeldZelle} zeigt das elektrische Feld wie es bei den Messungen in Abschnitt \ref{spektrenmitgas} eingestellt wurde. Die roten Linien stellen die Äquipotentiallinien dar. Die Richtung der elektrostatischen Kraft ist senkrecht dazu und insgesamt nach rechts gerichtet. Von der Frontplatte (links) nehmen die Spannungen an den Potentialringen linear von 48\,Volt auf 46,6\,Volt ab. Im Trichter dagegen nehmen die Spannungen zum Fokus hin in etwa mit $r^{-1}$ ab. Das gewölbte Gitter liegt dabei auf etwa 46,7\,Volt und das Gitter am Ausgang auf 44,5\,Volt. Die auf diese Weise erzeugten Äquipotentialflächen entsprechen etwa derjenigen einer für das Positron attraktiven Punktladung am Ausgang.

\subsection{Positronenstrahlführung}
\begin{figure}
 \centering
 \includegraphics[width=0.6\textwidth]{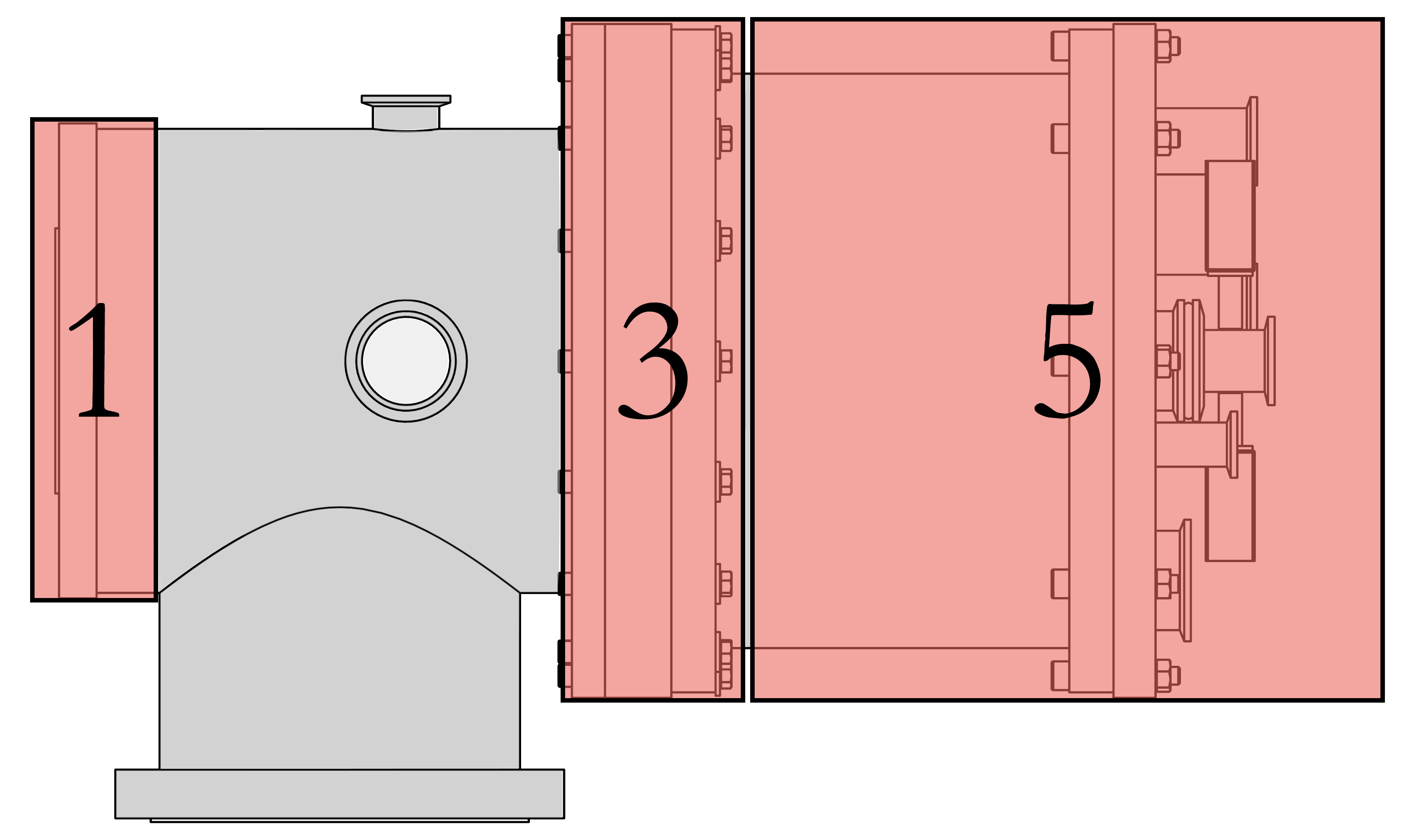}
 \caption{Übersicht der Solenoidspulen am Versuchsaufbau}
 \label{fig:AufbauSpulen}
\end{figure}
Bis zum Eingang des Pumprohrs werden die Positronen durch Solenoidspulen, die auf das Strahlrohr gewickelt sind, geführt. Abbildung \ref{fig:AufbauSpulen} zeigt die am Versuchsaufbau angebrachten Spulen.\footnote{Stromstärken siehe Abschnitt \ref{StroemeModerator}. Spule 2 und 4 sind nicht eingezeichnet da sie bei den Messungen nicht benutzt wurden. Spule 2 befindet sich links von Spule 3 und Spule 4 unter Spule 5 zwischen den Flanschen.} Die Richtung des erzeugten longitudinalen Magnetfeldes zeigt in Strahlrichtung. Um zu verhindern, dass die Positronen durch Streuung in der Gaszelle auseinander laufen wurde in der Gaszelle ein stärkeres B-Feld von ca. 21\,mT eingestellt.

\subsection{Gas- und Vakuumsysteme} \label{Gassystem}
Eine besondere Herausforderung des Aufbaus liegt im differentiellen Pumpen des Moderatorgases Stickstoff. Damit ist gemeint, dass das aus der Eintrittsöffnung der Gaszelle strömende Gas möglichst effektiv aus dem Bereich vor der Zelle abgepumpt werden muss. Dies wurde durch einen dreistufigen Pumpstand realisiert: Während es normalerweise ausreicht eine Turbopumpe nur mit einer Vorpumpe zu betreiben, wurde hier zusätzlich noch eine leistungsstarke Wälzkolbenpumpe zwischen Vor- und Turbopumpe geschaltet. Darüber hinaus wurde eine besonders starke Turbomolekularpumpe mit einem großem Nenndurchmesser von 160\,mm beschafft. Bei einem Druck von $2,3 \cdot 10^{-2}$\,mbar in der Moderatorzelle, wie bei den Messungen in Abschnitt \ref{spektrenmitgas}, lag der Druck im Pumprohr bei guten $6,5 \cdot 10^{-5}$\,mbar, was einer mittleren freien Weglänge für Positronen von etwa 8\,m entspricht. Der Druck wurde somit durch differentielles Pumpen, relativ zur Moderatorzelle, um einen Faktor von ca. 300 reduziert.

\begin{figure}
 \centering
 \includegraphics[width=0.7\textwidth]{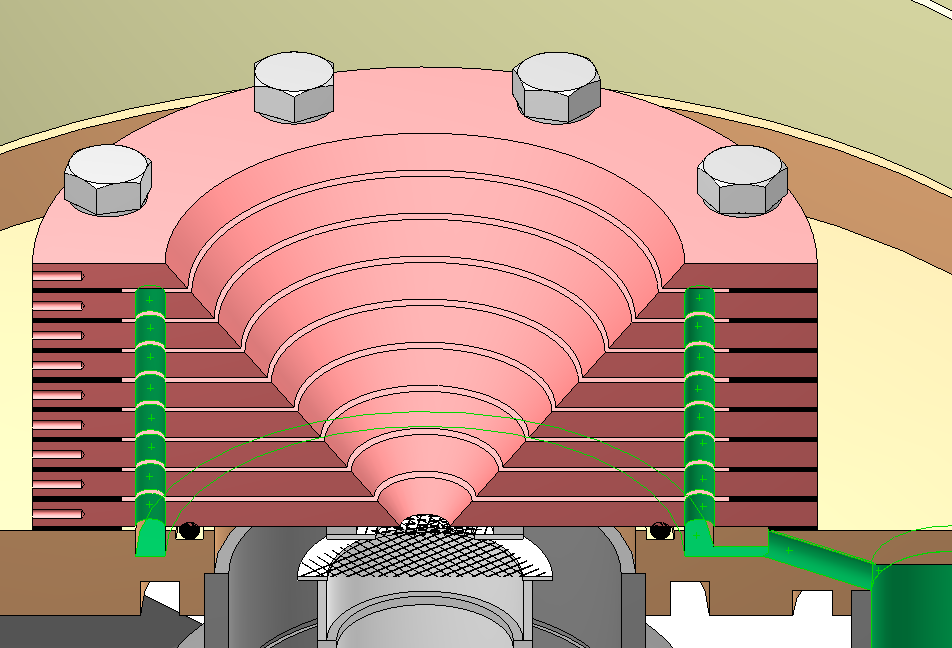}
 \caption{Schnittdarstellung des Gaseinlasses}
 \label{fig:Gaseinlass}
\end{figure}

Eingelassen wird das Moderatorgas durch ein fein einstellbares Nadelventil an der Rückplatte des Moderators. Durch die in Abbildung \ref{fig:Gaseinlass} grün eingezeichneten Kanäle gelangt es schließlich zwischen den konisch zulaufenden Potentialscheiben in die Zelle. Zusätzlich besteht die Möglichkeit über einen anderen Anschluss an der Rückplatte Gas direkt hinter der Frontplatte einzulassen, was aber in dieser Arbeit nicht genutzt wurde.

\section{Nachweis der Positronen im System}
\subsection{Analyse der moderierten Positronen}
\begin{figure}
 \centering
 \includegraphics[width=0.6\textwidth]{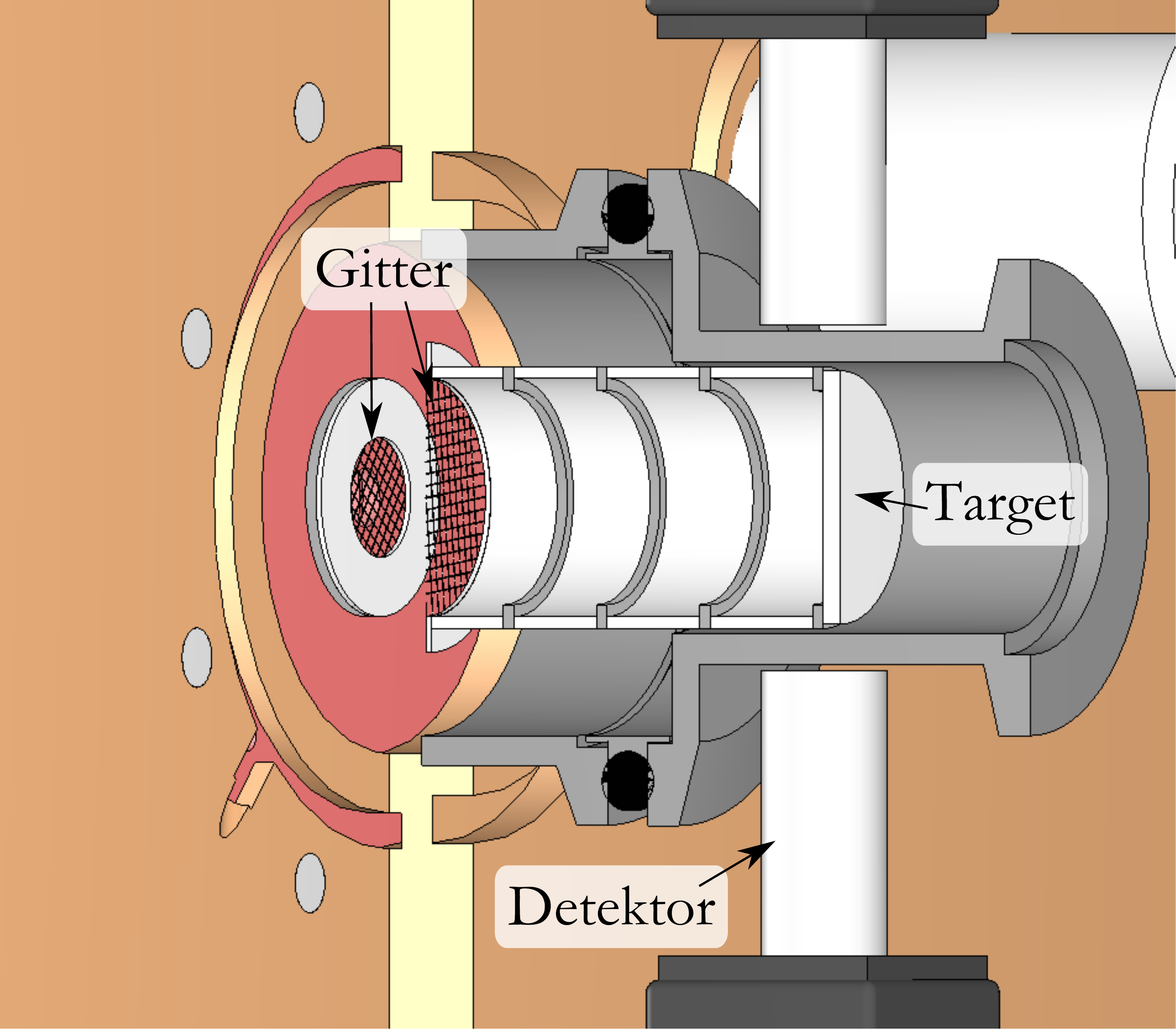}
 \caption{Schnittdarstellung des Analysators}
 \label{fig:SchnittAnalysator}
\end{figure}

\begin{figure}
 \centering
 \includegraphics[width=0.6\textwidth]{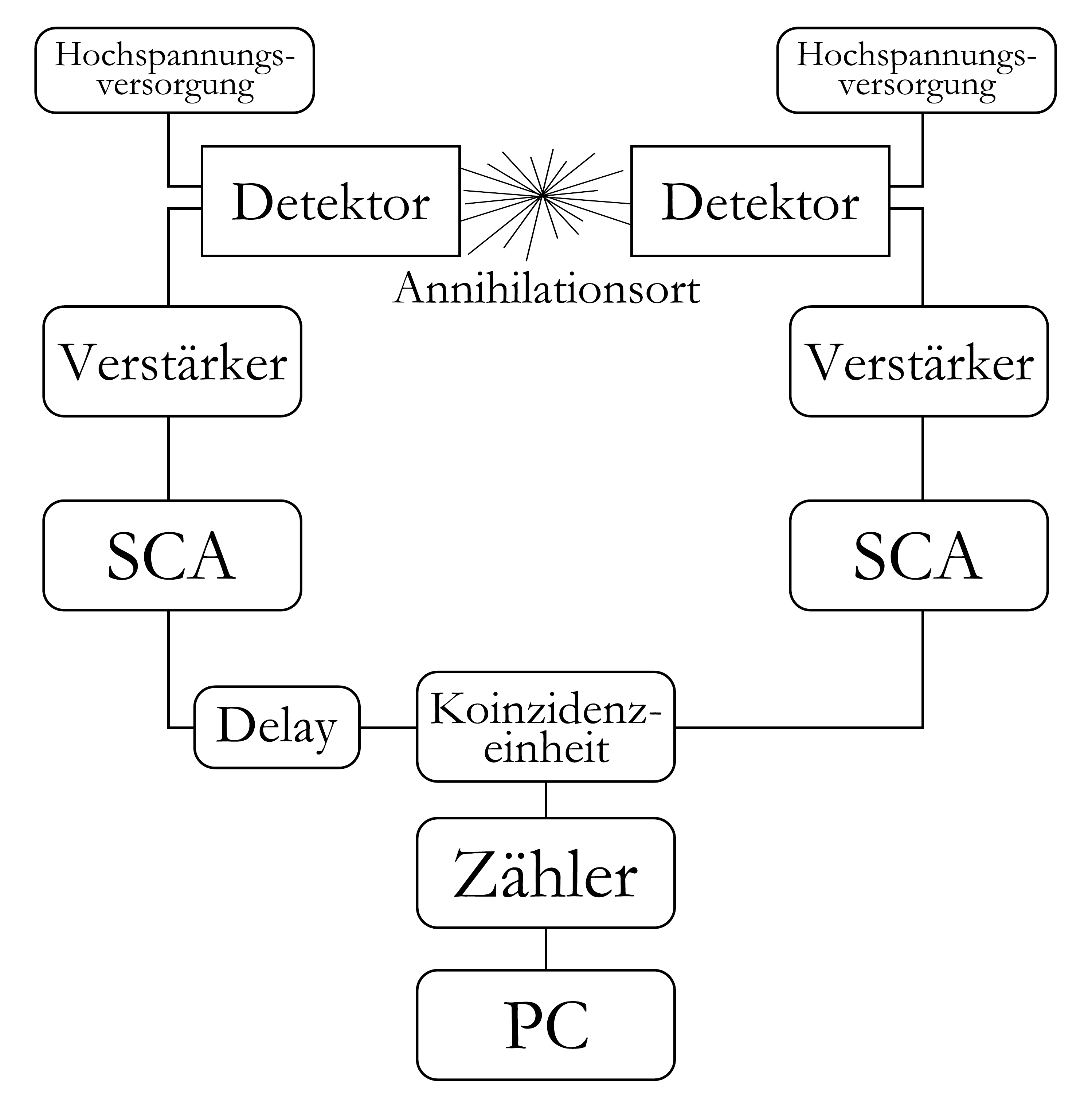}
 \caption{Blockschaltbild der Elektronik zum Nachweis der Annihilationsstrahlung}
 \label{fig:Diagramm-Elektronik}
\end{figure}

Um die Energie der remoderierten Positronen zu analysieren, wurde hinter dem Ausgangsgitter des Gasmoderators ein Bremsgitter installiert dessen Spannung computergesteuert variiert werden kann. Positronen, deren Energie hoch genug ist, um das Bremsgitter zu passieren, werden durch vier Potentialringe hindurch auf ein ca. 40\,mm dahinter liegendes Annihilationstarget beschleunigt (siehe Abbildung \ref{fig:SchnittAnalysator}). Die zwei bei der Annihilation emittierten charakteristischen 511\,keV $\gamma$-Quanten werden durch zwei BGO-Szintillatoren in Koinzidenz nachgewiesen (siehe Abbildung \ref{fig:Diagramm-Elektronik}). Daran angeschlossene Photomultiplier wandeln das im Szintillator entstandene Licht in einen Spannungsimpuls um. Dieser wird verstärkt und an einen Single Channel Analyzer (SCA) weitergeleitet. Der SCA liefert an seinem Ausgang ein logisches Signal, wenn am Eingang ein Spannungssignal vorliegt, das innerhalb eines einstellbaren (Spannungs-)Fensters liegt. Auf diese Weise kann eine bestimmte Energie selektiert werden (hier: 511\,keV). Das logische Signal beider SCAs liefert in einer Koinzidenzeinheit ein logisches Signal, wenn beide eingehenden Signale zur gleichen Zeit anliegen. Durch einen Zähler können die nachgewiesenen Ereignisse wiederum von einem PC registriert werden. Nimmt man nun die Zählrate als Funktion der Bremsspannung am Gitter auf, so misst man integral das Energiespektrum. Leitet man dieses nach der Spannung ab, so erhält man das sog. Bremsspektrum.

\subsection{Ortsaufgelöste Detektion der annihilierenden Positronen} \label{Linearscanner}
\begin{figure}
 \centering
 \includegraphics[width=0.8\textwidth]{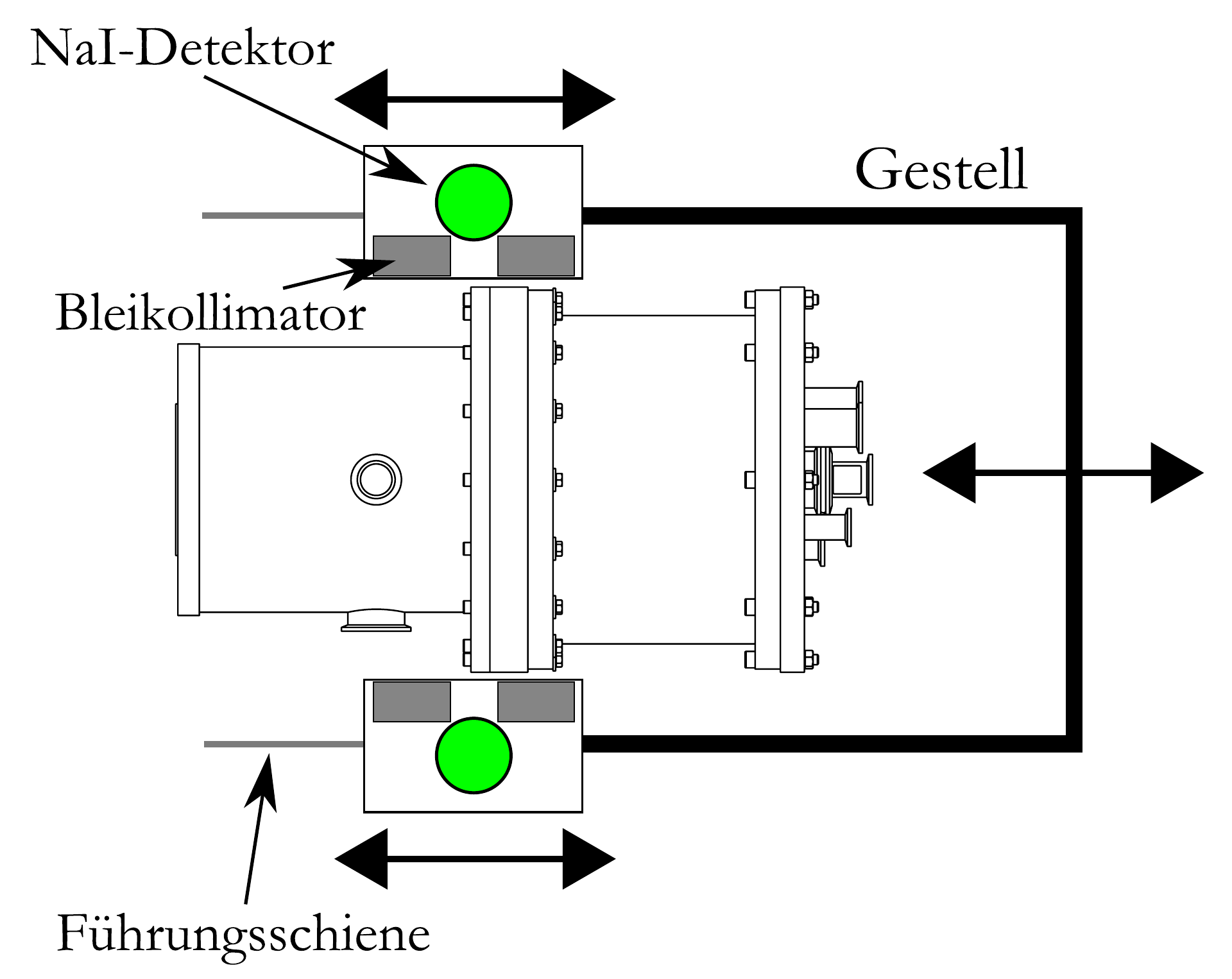}
 \caption{Schema des Linearscanners, Ansicht von oben auf den Moderatoraufbau}
 \label{fig:Schema-Linearscanner}
\end{figure}

Zur Optimierung der Parameter wie Druck, Spulenströme und der Potentiale ist es essentiell zu wissen, an welcher Stelle im Aufbau die Positronen annihilieren. Dazu wurden an den Seiten des Moderators zwei in Strahlrichtung verfahrbare NaI-Detektoren installiert (siehe Abbildung \ref{fig:Schema-Linearscanner}). Durch Bleiziegel vor den Detektoren lässt sich ein Spalt einstellen, der die Ortsauflösung dieses Linear-Scanners bestimmt. Zu beachten ist allerdings, dass ein zu kleiner Spalt die Zählrate verringert und eine dementsprechend längere Messzeit erfordert. Auch hier werden die Annihilationsquanten in Koinzidenz nachgewiesen, um sicherzustellen, dass nur Ereignisse auf der Verbindungslinie zwischen beiden Detektoren nachgewiesen werden. Über ein LabVIEW-Programm (siehe Abschnitt \ref{LabVIEW-Scanner}) lässt sich ein in die Scanvorrichtung eingebauter Schrittmotor steuern, der beide Detektoren parallel verschieben kann. Somit lässt sich die Zählrate ortsabhängig darstellen.
\clearpage{}
\clearpage{}\chapter{Messungen und Ergebnisse}
Zur Durchführung der Messungen stand der von einem Wolframkristall bereits remoderierte Strahl der Positronenquelle \textsc{Nepomuc} am FRM II zur Verfügung. Die Spannung am Wolframkristall betrug 90,9\,V. Zusammen mit der Austrittsarbeit von -3,0\,eV \cite{Hugenschmidt2002} ergibt sich somit eine Strahlenergie von 93,9\,eV. Als Moderatorgas wurde Stickstoff verwendet.

\section{Bestimmung der ortsabhängigen \texorpdfstring{$\upgamma$}{Gamma}-Absorption im Moderatoraufbau}
Um qualitative Aussagen über die ortsaufgelöste Annihilation der Positronen im Moderator treffen zu können, ist es nötig, das Absorptionsverhalten des Aufbaus für 551\,keV-Gammaquanten zu kennen. Über
\begin{equation}
	I(d)=I_0 \cdot e^{-\mu d}
\end{equation}
bzw.
\begin{equation}
	\frac{I(d)}{I_0} = e^{-\mu d} \label{equ:Int2}
\end{equation}
lässt sich mit $\mu=\mu_{Al}\approx0,227\text{\,cm}^{-1}$ die Intensität $I$ der transmittierten 511\,keV-$\upgamma$-Quanten in Abhängigkeit der durchquerten Wandstärke von Aluminium $d$ bestimmen.

Aufgrund der durch verschiedene Kupferspulen und anderen Aufbauten komplexen Geometrie, wurde das Absorptionsverhalten mit Hilfe des Linearscanners (siehe Abschnitt \ref{Linearscanner}) gemessen.

\begin{figure}
 \centering
 \includegraphics[width=0.98\textwidth]{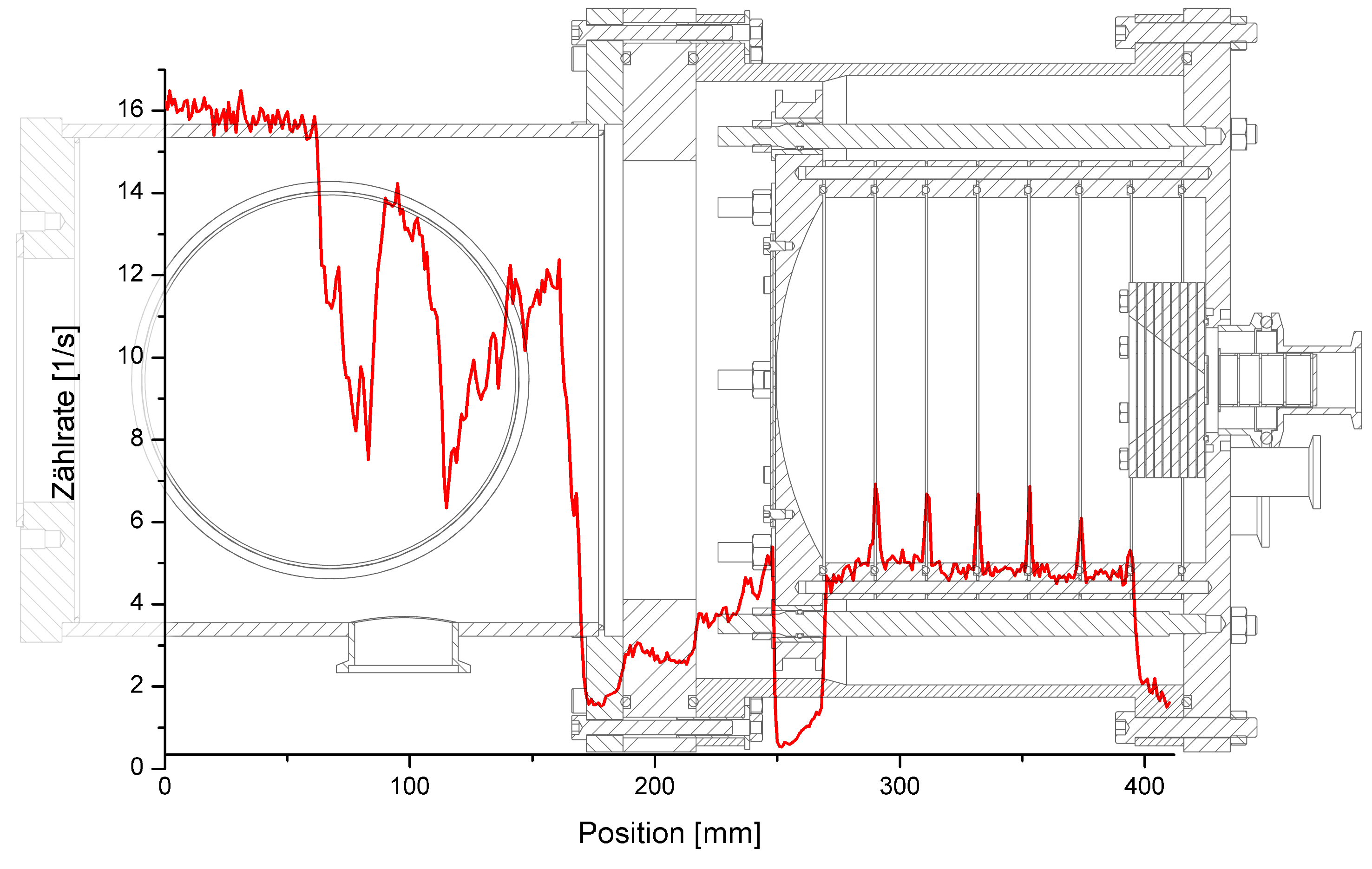}
 \caption{Referenzscan mit einer punktförmigen Na$^{22}$-Quelle durch den Moderator. Aufgetragen ist die Zählrate als Funktion des Ortes, bei ca. 240\,s pro Messpunkt. Die Spaltbreite betrug 5\,mm und der Abstand zwischen zwei Messpunkten 1\,mm. Mit Hilfe der sechs Peaks im hinteren Bereich konnte die maßstäbliche CAD-Zeichnung exakt über den Plot gelegt werden.}
 \label{fig:refscan}
\end{figure}

Für den Referenzscan wurde eine punktförmige Na$^{22}$-Quelle auf der Mittelachse des Moderators zwischen beiden Detektoren platziert und gemeinsam mit diesen verfahren. Abbildung \ref{fig:refscan} zeigt die ortsabhängige Koinzidenzzählrate beim Scannen durch den Moderator. Folgende qualitativen Beobachtungen können gemacht werden:

\begin{itemize}
  \item Die beiden Einbrüche zwischen 60 und 155\,mm lassen sich durch einen dort angebrachten Flansch erklären.
  \item Der Abfall der Zählrate ab 162\,mm liegt nicht nur am dort beginnenden Flansch, sondern auch an der über diesen geschobenen Führungsspule (nicht eingezeichnet).
  \item Der Anstieg der Zählrate, bei gleichzeitiger Zunahme der Wandstärke, bei etwa 190\,mm  liegt daran, dass der Flansch, der Pumprohr und Moderatorzelle miteinander verbindet, aus PVC besteht.
	\item Deutlich zu sehen ist der starke Einbruch der Zählrate bei 248\,mm, der von der bis zu 109\,mm dicken Frontplatte am Eingang der Gaszelle verursacht wird.
	\item Erkennbar sind auch sechs Peaks in regelmäßigen Abständen zwischen 290 und 394\,mm. Dies sind die Spalte zwischen den Potentialringen.
\end{itemize}

Gleichung \ref{equ:Int2} kann als Wahrscheinlichkeit betrachtet werden, dass ein $\upgamma$-Quant durch das Material der Dicke $d$ transmittiert wird. Misst man in Koinzidenz quadriert sich diese Wahrscheinlichkeit, da beide Gammas das Material passieren müssen\footnote{Das gilt natürlich nur bei symmetrischem Aufbau. Das heißt, dass beide $\upgamma$-Quanten das selbe Material mit gleicher Dicke passieren müssen.}. Somit gilt für die Koinzidenzwahrscheinlichkeit $P_{Koi}(d)$ mit den Koinzidenzzählraten $J(d)$ und $J_0$:\footnote{Normalerweise müsste man zum Berechnen einer Zählrate aus einer Intensität noch Raumwinkel und Nachweiswahrscheinlichkeit im Detektor mit einberechnen. Diese kürzen sich jedoch hier heraus.}
\begin{equation}
	P_{Koi}(d) = \frac{J(d)}{J_0} = \left( e^{-\mu\cdot d}\right)^2 = e^{-2\mu\cdot d} \label{Koinzcount}
\end{equation}
Daraus folgt, mit $d$ als Funktion des Ortes $d(x)$:
\begin{equation}
 \frac{J(d(x))}{J_0} =  e^{-2\mu\cdot d(x)} = \alpha(x)
\end{equation}

\section{Bestimmung der Einkopplungseffizienz}
\label{einkopplung}
Aufgrund des differentiellen Pumpens wurde eine Eingangsöffnung von 3\,mm gewählt. Da der Durchmesser des Strahl jedoch etwa 5\,mm beträgt, verliert man bereits Positronen die an der Frontplatte annihilieren. Um diesen Verlust zu bestimmen wurde der Positronenstrahl zunächst ohne Moderatorgas\footnote{d.\,h. bei einem Druck von unter $5 \cdot 10^{-4}$\,mbar} in den Moderator eingefädelt, d.\,h. die Korrekturfelder an der Beamline wurden auf eine möglichst hohe Zählrate am Target am Ausgang des Moderator optimiert. Danach wurde mit den beiden NaI-Detektoren der Scanvorrichtung die Zählrate der am Eingang annihilierenden Positronen bestimmt. Schließlich wurde der Strahl mit Hilfe der Korrekturfelder gegen die Frontplatte gelenkt, so dass sämtliche Positronen dort annihilierten und wiederum die Zählrate gemessen. Aus dem Verhältnis der beiden ermittelten Zählraten konnte der Anteil der in die Gaszelle eingekoppelten Positronen zu 50$\pm$4,5\,\% bestimmt werden.

\section{Spektrum des einfallenden Strahls}
\begin{figure}
 \centering
 \includegraphics[width=0.8\textwidth]{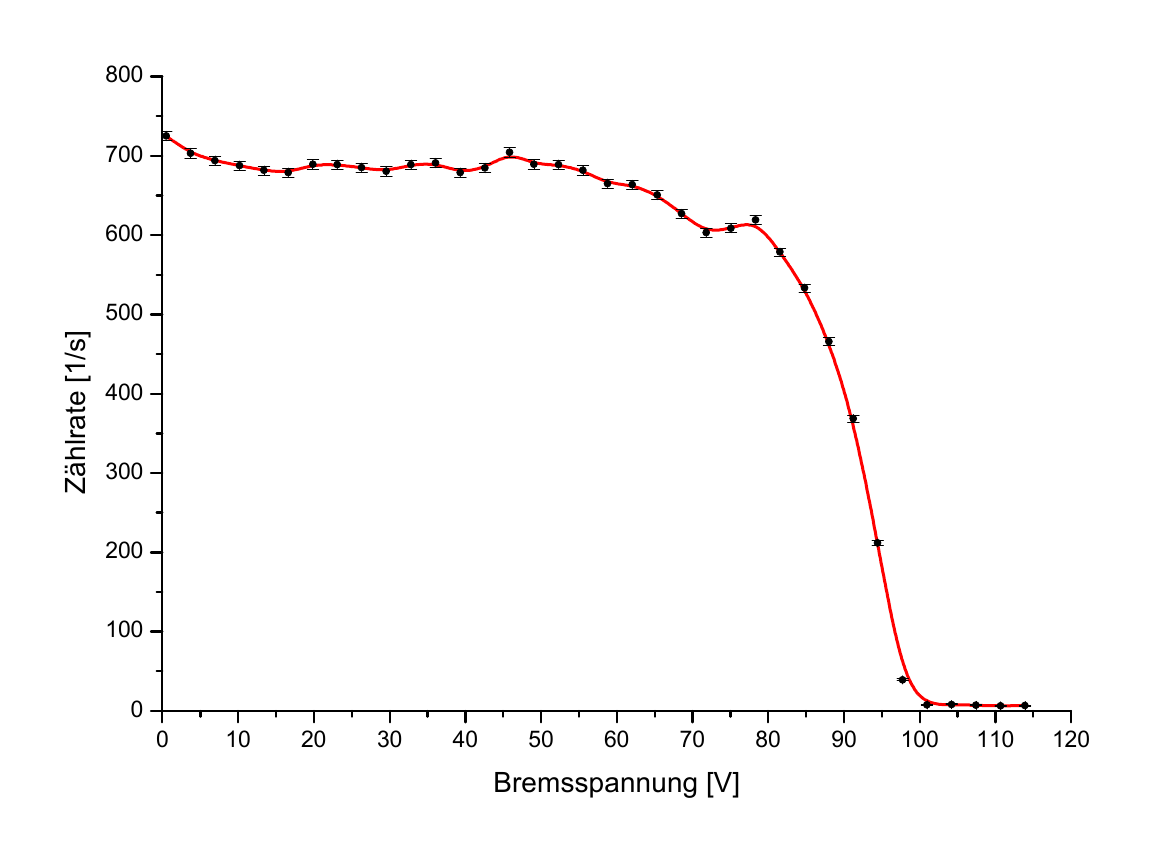}
 \caption{Zählrate am Target als Funktion der Spannung}
 \label{fig:intspekohnegas}
\end{figure}

\begin{figure}
 \centering
 \includegraphics[width=0.8\textwidth]{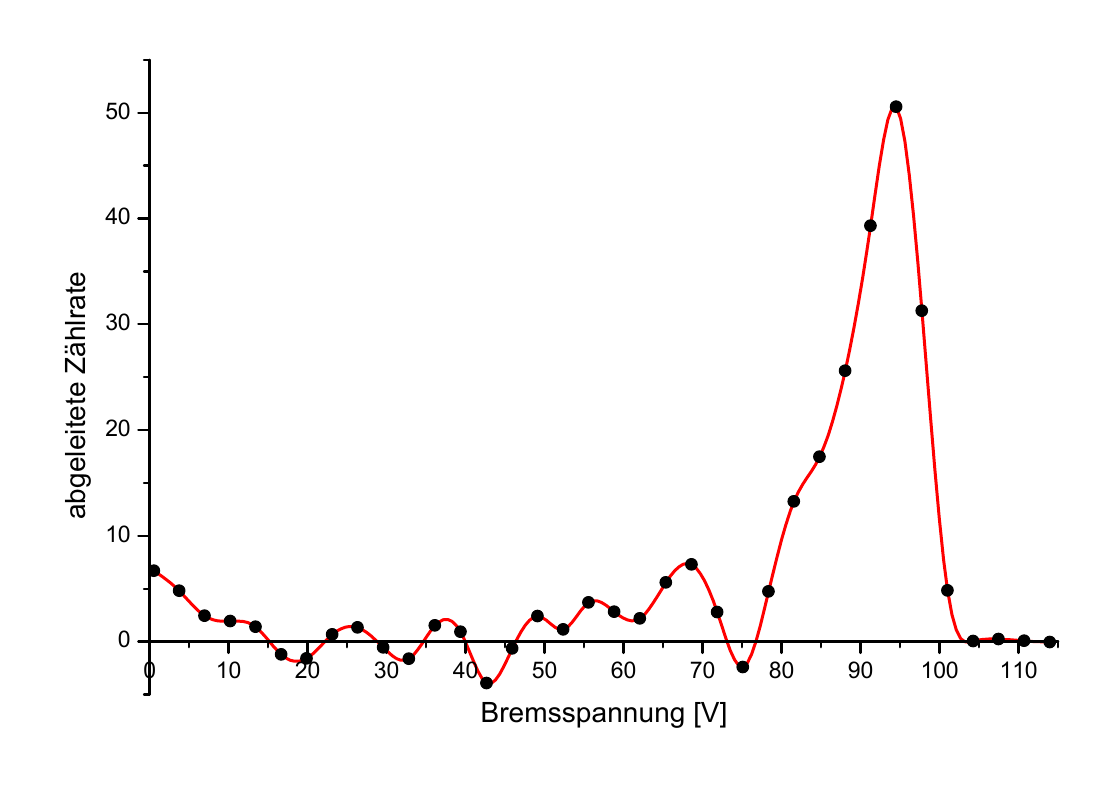}
 \caption{Bremsspektrum des 93,9\,eV \textsc{Nepomuc}-Strahls (Ableitung der Zählrate aus Abbildung \ref{fig:intspekohnegas})}
 \label{fig:spekohnegas}
\end{figure}

Zunächst wurde das Spektrum des \textsc{Nepomuc}-Strahls ohne Gas in der Moderatorzelle vermessen, um die Funktionsweise des Energieanalysators zu testen. Dafür wurde die Spannung am Bremsgitter in ca. 3\,V Schritten von 1-114\,V durchgefahren und jeweils 20 Zählraten aufgenommen. Abbildung \ref{fig:intspekohnegas} zeigt die Zählrate als Funktion der Bremsspannung und Abbildung \ref{fig:spekohnegas} das davon abgeleitete Energiespektrum. Die Tatsache, dass die Zählrate bereits bei etwa 80\,eV beginnt einzubrechen, lässt sich damit erklären, dass ein Teil der Positronen nicht senkrecht auf das Bremsgitter auftrifft. Das bedeutet, dass diese Positronen einen Transversalimpuls besitzen und somit gyrieren.

\section{Aufnahme der Zählrate in Abhängigkeit des Drucks}
\begin{figure}
 \centering
 \includegraphics[width=0.75\textwidth]{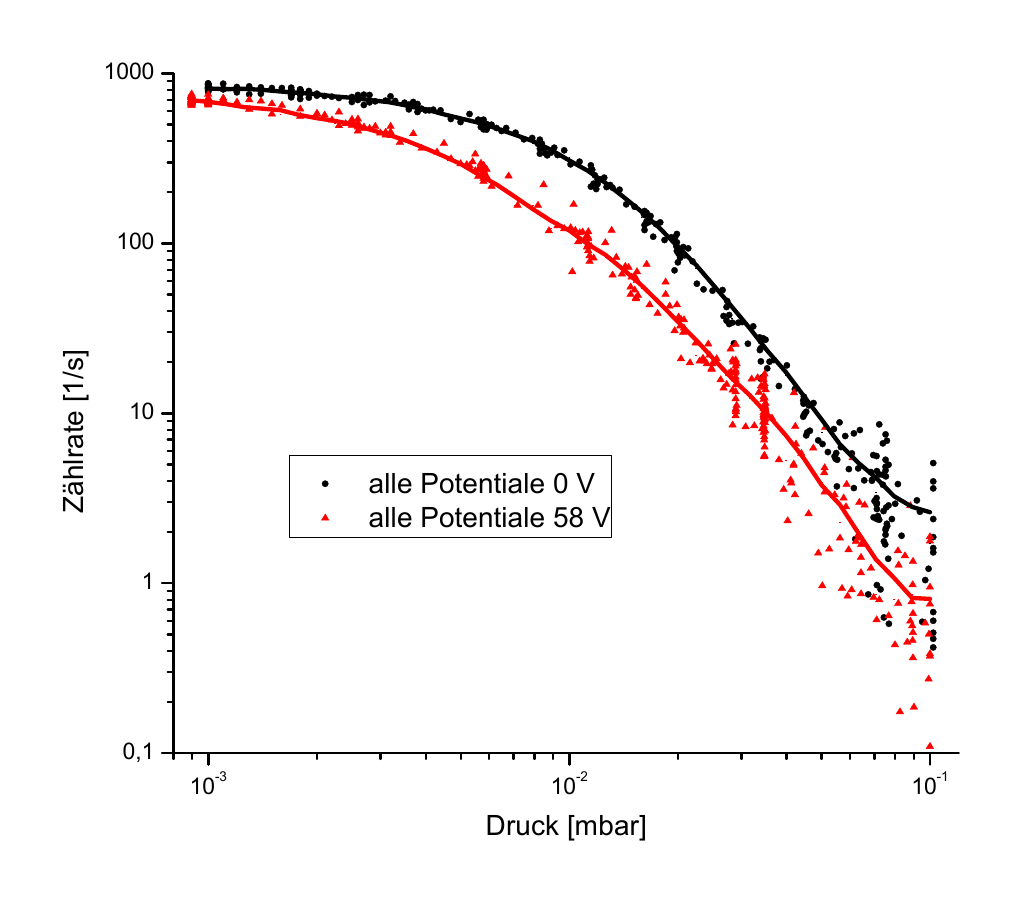}
 \caption{Zählrate am Target in Abhängigkeit des Drucks in der Gaszelle}
 \label{fig:druckzaehlrate}
\end{figure}
Mithilfe des Nadelventils wurde nun Stickstoff in die Moderatorzelle eingelassen. Dabei wurde der Druck und gleichzeitig die Koinzidenzzählrate am Target mit dem Computer mitgeschrieben (siehe Abbildung \ref{fig:druckzaehlrate}). Dies wurde zunächst bei auf Masse gesetzten Potentialen durchgeführt. Danach wurden alle Potentiale auf 58\,V eingestellt, so dass die Positronen am Eingang um 58\,eV abgebremst wurden, und die Messung wiederholt. Man erkennt, dass im Fall der auf 58\,V gesetzten Potentiale die Zählrate, gegenüber dem Fall ohne Spannung an den Potentialringen, schneller einbricht. Das ist qualitativ verständlich, da die Positronen bei angelegtem Potential mit nur noch etwa 36\,eV in die Gaszelle gelangen und dadurch mit durchschnittlich weniger Stößen in den Energiebereich der verstärkten Positroniumsbildung (\enquote{Ore Gap}) herunter streuen können.

\section{Spektrum des moderierten Strahls bei verschiedenen Drücken} \label{spektrenmitgas}
\begin{figure}
 \centering
 \includegraphics[width=0.7\textwidth]{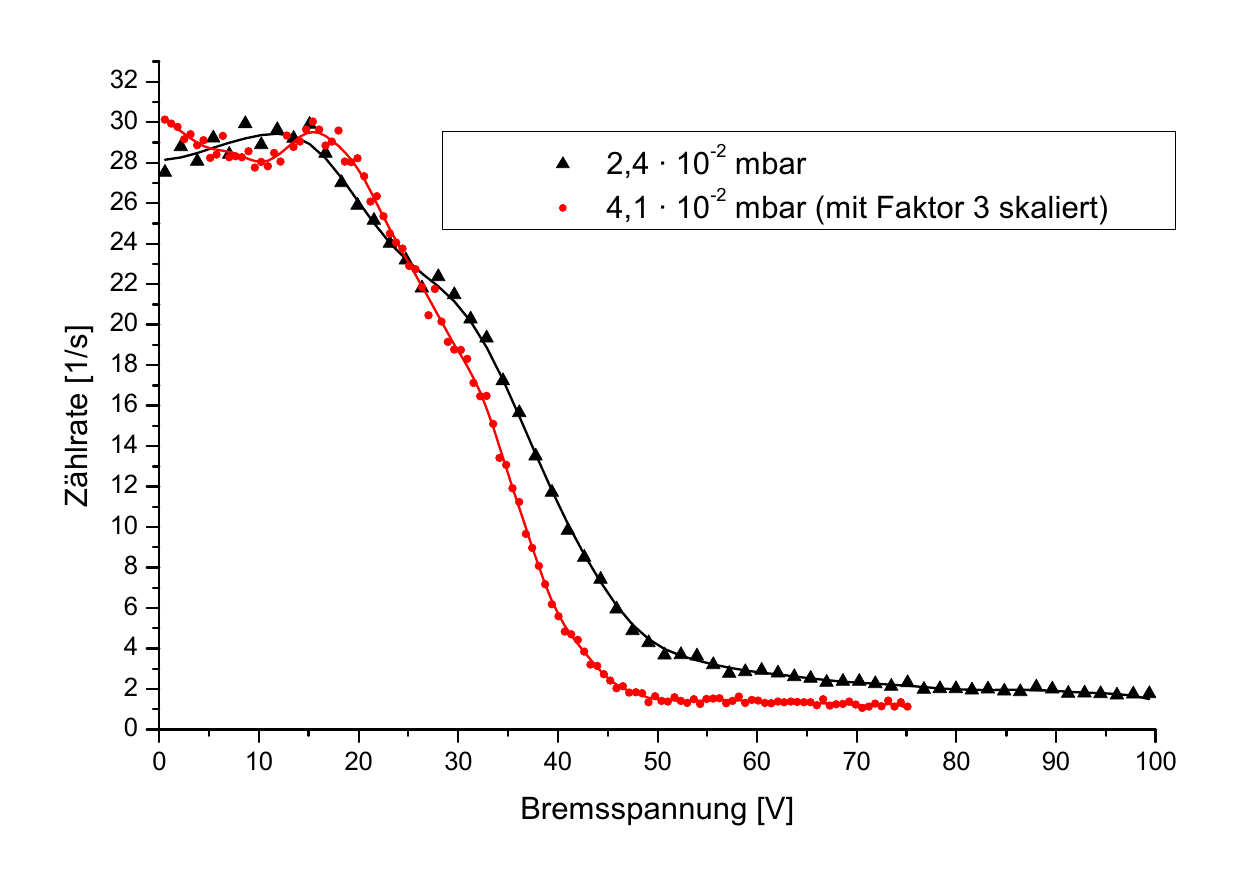}
 \caption{Integrale Spektren des Strahls bei zwei verschiedenen Drücken und einem Potentialverlauf von 48\,Volt am Eingang bis 20\,Volt am Ausgang und somit einem Driftpotential von 28\,Volt.}
 \label{fig:intspekmitgas1-20-48V}
\end{figure}
\begin{figure}
 \centering
 \includegraphics[width=0.7\textwidth]{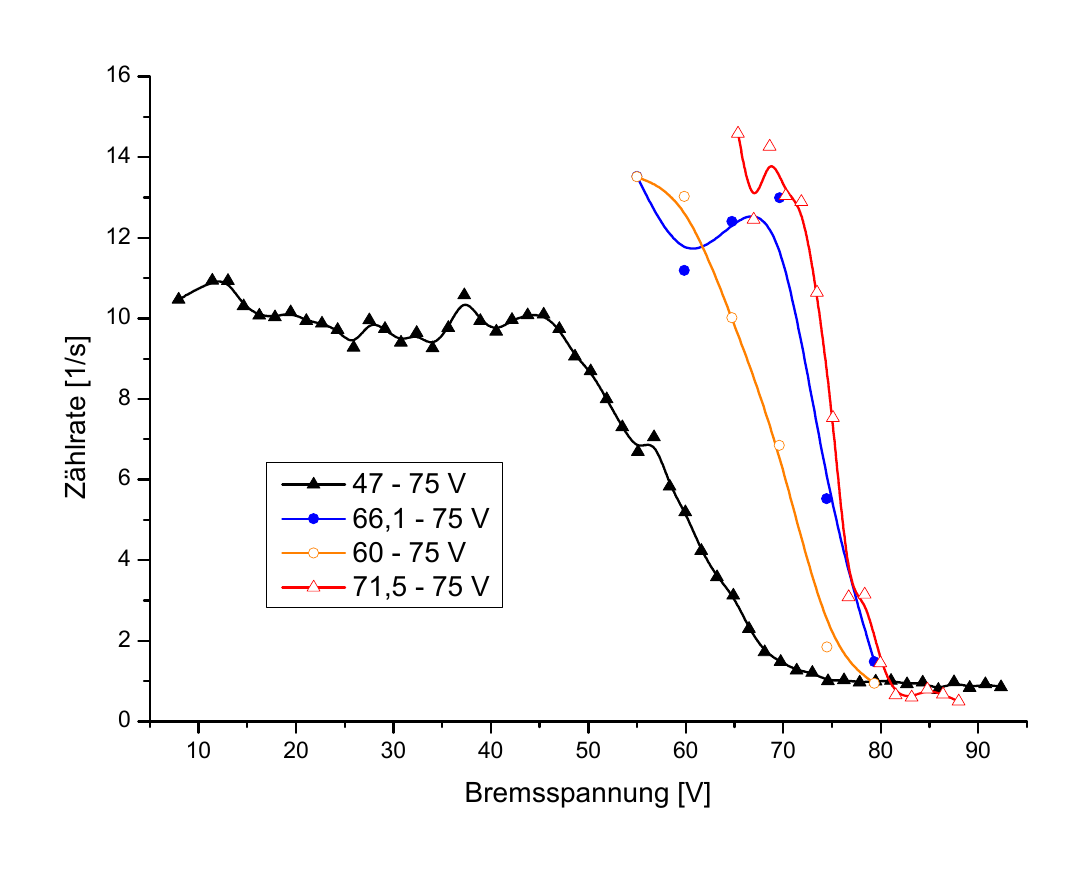}
 \caption{Integrale Spektren bei einem Druck von $4,2\cdot 10^{-2}$\,mbar und vier verschiedenen Potentialeinstellungen}
 \label{fig:intspekmitgas2-bis75V}
\end{figure}
Nun wurden verschiedene elektrische Felder, entsprechend der unter \ref{simion-potentiale} durchgeführten Simulationen, in der Gaszelle eingestellt und das Spektrum der moderierten Positronen mit Hilfe des Bremsgitters am Ausgang aufgenommen. Trotz unterschiedlicher Spannungen an den Potentialringen und -scheiben wurde stets ein auf den Ausgang fokussierendes elektrisches Feld erreicht. Somit kann man über die Spannung am Eingang des Moderators die Eingangsenergie der Positronen festlegen. Hiervon und von der Spannungsdifferenz über den Moderator hängt schließlich die Spannung am Ausgangsgitter und damit die Austrittsenergie der vollständig moderierten Positronen ab. Man erwartet also, dass die Zählrate beim Durchfahren der Spannung am Bremsgitter oberhalb der Ausgangsspannung abnimmt. Je stärker die Zählrate dabei einbricht, desto größer ist der Anteil thermalisierter Positronen.

Abbildung \ref{fig:intspekmitgas1-20-48V} zeigt die Zählrate als Abhängigkeit der Bremsspannung bei einem Potentialverlauf von 48\,Volt (am Eingang des Moderators) bis 20\,Volt (am Ausgang) bei zwei verschiedenen Drücken. Man erkennt, dass die Zählrate in beiden Fällen über einen weiten Bereich von etwa 30\,V abfällt.

Um zu untersuchen ob diese Energieverschmierung von der Spannungsdifferenz über die Moderatorzelle verursacht wird, wurde in den folgenden Messungen diese Differenz sukzessive halbiert. In den Messungen in Abbildung \ref{fig:intspekmitgas2-bis75V} wurde die Spannung am Eingang der Zelle konstant auf 75\,Volt gehalten, während die Spannung am Ausgang (und somit auch die Spannungen im inneren der Zelle) von Messung zu Messung erhöht wurde. Von 47\,Volt am Ausgang bei der ersten Messung wurde die Spannung schließlich auf 71,5\,Volt erhöht und damit eine Spannungsdifferenz von nur noch 3,5\,Volt erreicht. Wie man sieht, konnte diese Optimierung des elektrischen Driftfelds die Energieverschmierung der Positronen stark reduzieren.
\begin{figure}
 \centering
 \includegraphics[width=0.75\textwidth]{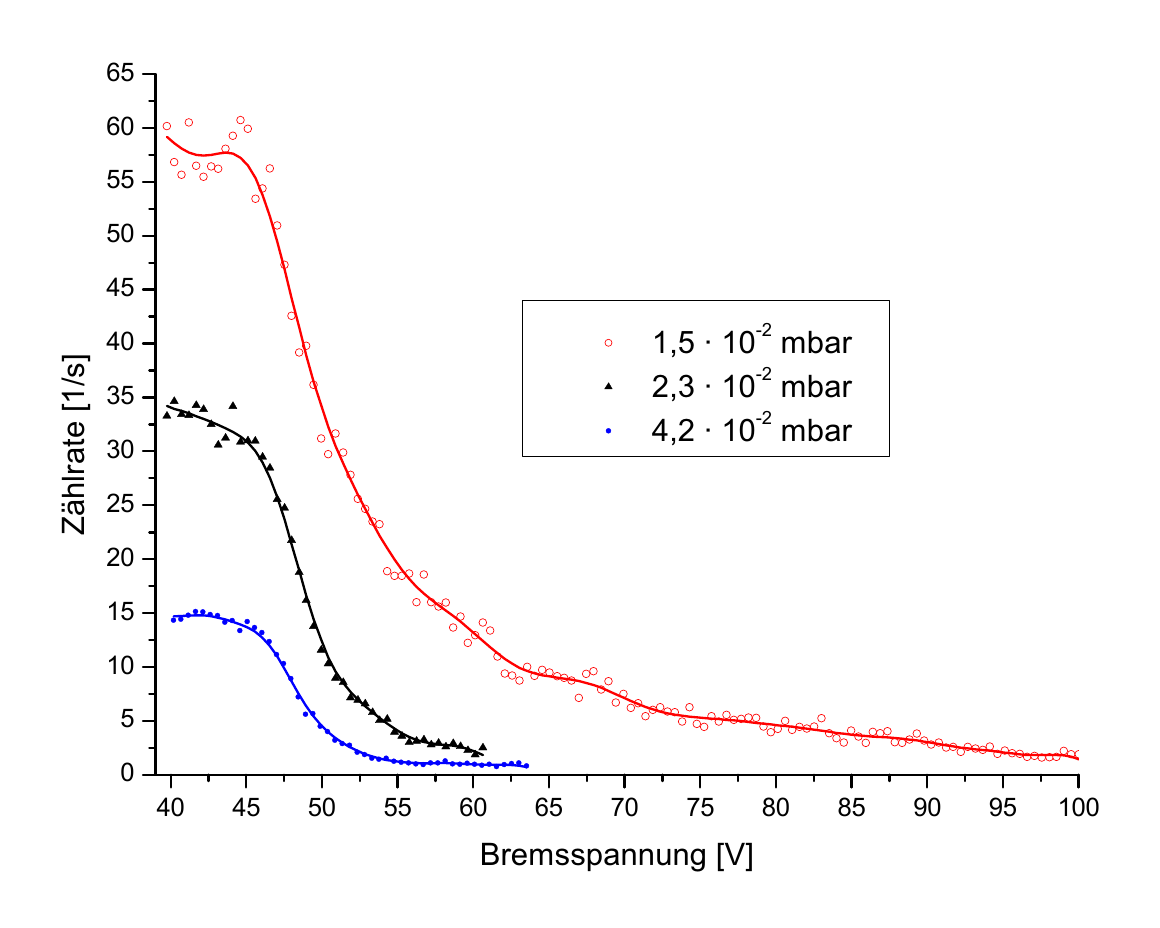}
 \caption{Integrale Spektren bei drei verschiedenen Drücken in der Moderatorzelle.  Die Spannung betrug 48\,Volt am Eingang und 44,5\,Volt am Ausgang.}
 \label{fig:intspekmitgas3-bis48V}
\end{figure}
\begin{figure}
 \centering
 \includegraphics[width=0.75\textwidth]{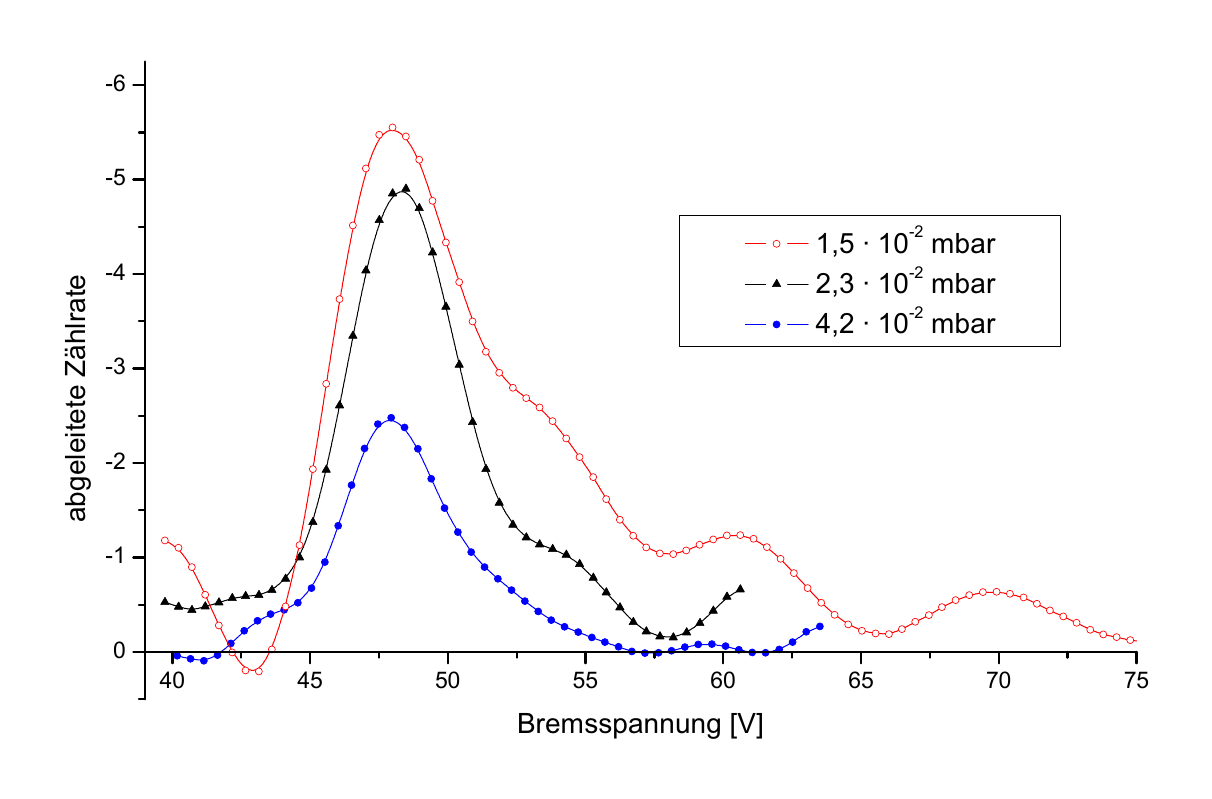}
 \caption{Spektren des moderierten Strahls (Ableitung von Abbildung \ref{fig:intspekmitgas3-bis48V}). Die Spannung am Ausgang der Gaszelle betrug 44,5\,Volt}
 \label{fig:spekmitgas1-bis48V}
\end{figure}
\begin{figure}
 \centering
 \includegraphics[width=0.6\textwidth]{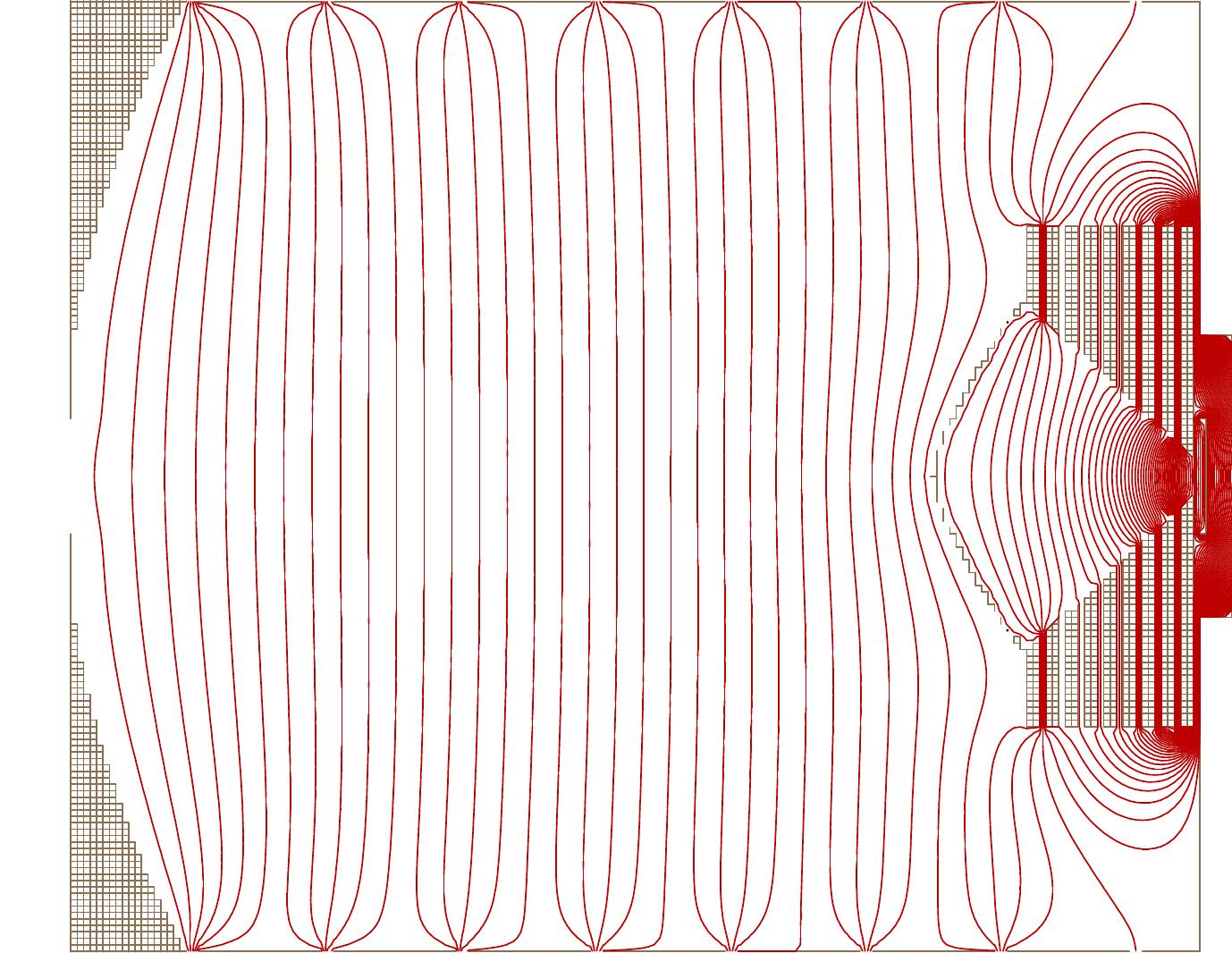}
 \caption{Elektrisches Feld in der Gaszelle. Die Spannung am Eingang (links) betrug 48\,V; am Ausgang 44,5\,V.}
 \label{fig:PotentialeMessung}
\end{figure}

Schließlich wurden die Spannungen am Moderator um 27\,Volt abgesenkt, so dass das Potential am Eingang 48\,Volt und am Ausgang 44,5\,Volt betrug und somit die Driftspannung von 3,5\,Volt konstant blieb (siehe Abbildung \ref{fig:PotentialeMessung}). Die vorherigen Messungen wurden nun mit besserer Statistik und bei drei verschiedenen Drücken wiederholt (siehe Abbildung \ref{fig:intspekmitgas3-bis48V}). Aus den daraus abgeleiteten Spektren in Abbildung \ref{fig:spekmitgas1-bis48V} erkennt man bei allen drei Drücken einen Peak bei ca. 48-48,5\,Volt. Da die Spannung am Austrittsgitter auf 44,5\,Volt lag, betrug die kinetische Energie der Positronen in diesem Peak somit 3,5-4\,eV bei einer Halbwertsbreite von etwa 4,5\,eV. Unterschiede zeigten sich jedoch in den Bereichen höherer Energie. Während bei $2,3\cdot 10^{-2}$ und $4,2\cdot 10^{-2}$\,mbar das Spektrum im Bereich zwischen 52 und 60\,Volt schnell gegen Null ging, zeigte das Spektrum bei $1,5\cdot 10^{-2}$\,mbar deutlich, dass auch nicht vollständig moderierte Positronen das Target erreichten. Dies erkennt man an den beiden Peaks bei 60,5 und 69,5\,Volt, deren Abstand von 9\,Volt genau der Energie einer Anregung des Stickstoffs entspricht. Und auch bei etwa 52\,Volt kann man einen weiteren Peak ausmachen, der als Verbreiterung des Hauptpeaks (der vollständig moderierten Positronen) in Erscheinung tritt.

\section{Berechnung der Moderationseffizienzen} \label{Moderationseffizienz}
Im Folgenden soll nun die Effizienz des Moderators bei diesen drei Drücken berechnet werden. Dabei unterscheidet man einerseits zwischen der reinen Moderationseffizienz, die nur das Verhältnis der Anzahl an moderierten Positronen zur Anzahl der in den Moderator eingekoppelten Positronen darstellt. Die Anzahl der in den Moderator eingekoppelten Positronen erhält man direkt aus der Zählrate am Target, ohne dass Gas in die Zelle eingelassen wurde. Andererseits muss beachtet werden, dass durch die nur 3\,mm große Öffnung am Eingang des Moderators nicht alle Positronen des ankommenden Strahls in die Moderatorzelle eingekoppelt werden können. Somit muss in die Gesamteffizienz des Systems noch die unter Abschnitt \ref{einkopplung} bestimmte Einkopplungseffizienz mit einberechnet werden. Die Anzahl der moderierten Positronen erhält man durch die Bestimmung des Abfalls der Zählrate im integralen Spektrum (Abbildung \ref{fig:intspekmitgas3-bis48V}) zwischen 44,5 und 55\,Volt -- was in etwa der Breite des Peaks im Spektrum entspricht. Tabelle \ref{tbl:Moderationseffizienz} zeigt die berechneten Effizienzen bei den drei verschiedenen Drücken sowie die dafür benötigten Zählraten.
\begin{table}
 \centering
 \caption{Moderationseffizienzen bei verschiedenen Drücken}
 \label{tbl:Moderationseffizienz}
 \begin{tabular}{|l|c|c|c|}
  \hline
  \textbf{Druck [mbar]} & $1,5 \cdot 10^{-2}$ & $2,3\cdot 10^{-2}$ & $4,2\cdot 10^{-2}$ \\
  \hline
  \hline
  Eingekoppelte e$^+$/s & \multicolumn{3}{|c|}{702} \\
  \hline
  Ausgekoppelte e$^+$/s & 60,7 & 34,0 & 15,1 \\
  \hline
  Moderierte e$^+$/s & 41,2 & 30,0 & 13,9 \\
  Anteil am ausgekoppelten Strahl & 68\,\% & 88\,\% & 92\,\% \\
  \hline
  \textbf{Effizienz des Moderators} & \textbf{5,9\,\%} & \textbf{4,3\,\%} & \textbf{2,0\,\%} \\
  \hline
  \hline
  Einkopplungseffizienz & \multicolumn{3}{|c|}{50\,\%} \\
  \hline
  \textbf{Gesamteffizienz des Aufbaus} & \textbf{2,9\,\%} & \textbf{2,1\,\%} & \textbf{1,0\,\%} \\
  \hline
 \end{tabular}
\end{table}

\section{Untersuchung der Positronenverluste im Moderator}
\begin{figure}
 \centering
 \includegraphics[width=0.98\textwidth]{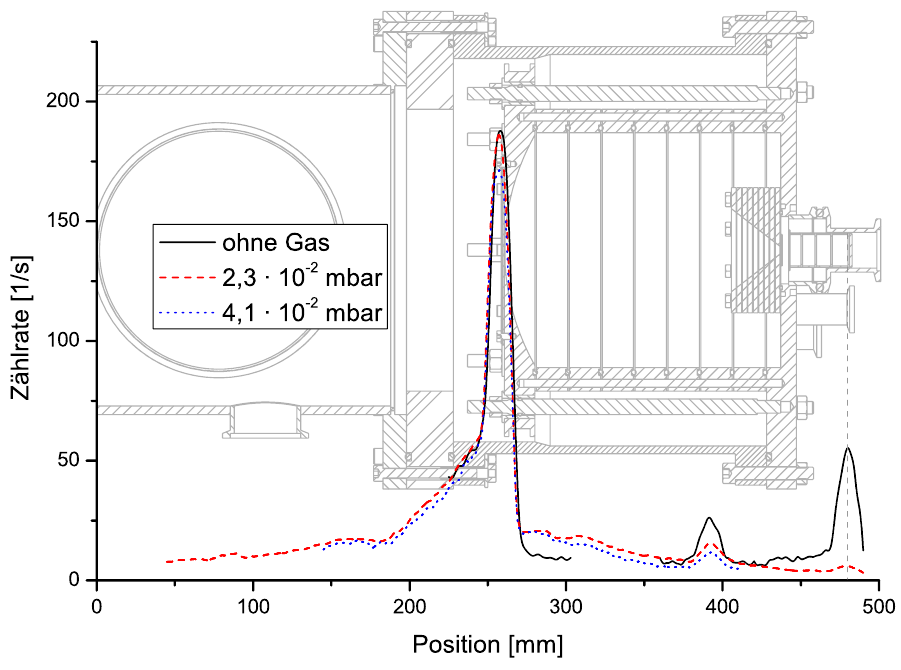}
 \caption{Ortsaufgelöster Scan der Annihilationsstrahlung}
 \label{fig:scan1mitaufbau}
\end{figure}

Um die Verluste an Positronen im Moderator zu untersuchen, wurde mit Hilfe der 1D-Scanapparatur die Annihilationsstrahlung ortsabhängig detektiert (siehe Abbildung \ref{fig:scan1mitaufbau}). Diese Messung wurde zunächst ohne Stickstoff und darauf bei den Drücken $2,3 \cdot 10^{-2}$\,mbar und $4,1 \cdot 10^{-2}$\,mbar mit Stickstoff durchgeführt. Zunächst fällt auf, dass der Peak am Eingang der Moderatorzelle bei $2,3 \cdot 10^{-2}$\,mbar und ohne Gas die selbe Höhe hat\footnote{Die Abweichung beträgt ca. 1\,\% und liegt somit innerhalb des Messfehlers.}. Und selbst der Peak bei $4,1 \cdot 10^{-2}$\,mbar liegt gerade einmal um 9\,\% niedriger als der Peak ohne Gas. Dies bedeutet, dass die Einkopplungseffizienz praktisch druckunabhängig ist.

Bemerkenswerter Weise liegen alle Messwerte (links des Peaks am Eingang) des Scans bei $4,1 \cdot 10^{-2}$\,mbar um etwa 9\,\% niedriger als die des vorherigen Scans bei $2,3 \cdot 10^{-2}$\,mbar. Da man einen systematischen Fehler bei dieser Messung praktisch ausschließen kann, ergibt sich folgende mögliche Erklärung: Durch den bei diesem Scan auch im Pumpflansch höheren Druck ($1,2 \cdot 10^{-4}$\,mbar) streut und annihiliert ein Teil der Positronen schon weit vor dem Moderator. Der auf der linken Seite zwischen etwa 180 und 250\,mm verbreiterte Peak lässt sich durch Rückstreuung der Positronen von der Eingangsblende und anschließender Annihilation an der Gehäusewand erklären.

Rechts des Peaks (ab 272\,mm) stellt man jedoch eine bei Stickstoff bis zu doppelt so hohe Zählrate fest wie ohne Gas. Dies wird verursacht durch die in Abschnitt \ref{Positroniumsbildung} erläuterte Positroniumsbildung.

\begin{figure}
 \centering
 \includegraphics[width=0.7\textwidth]{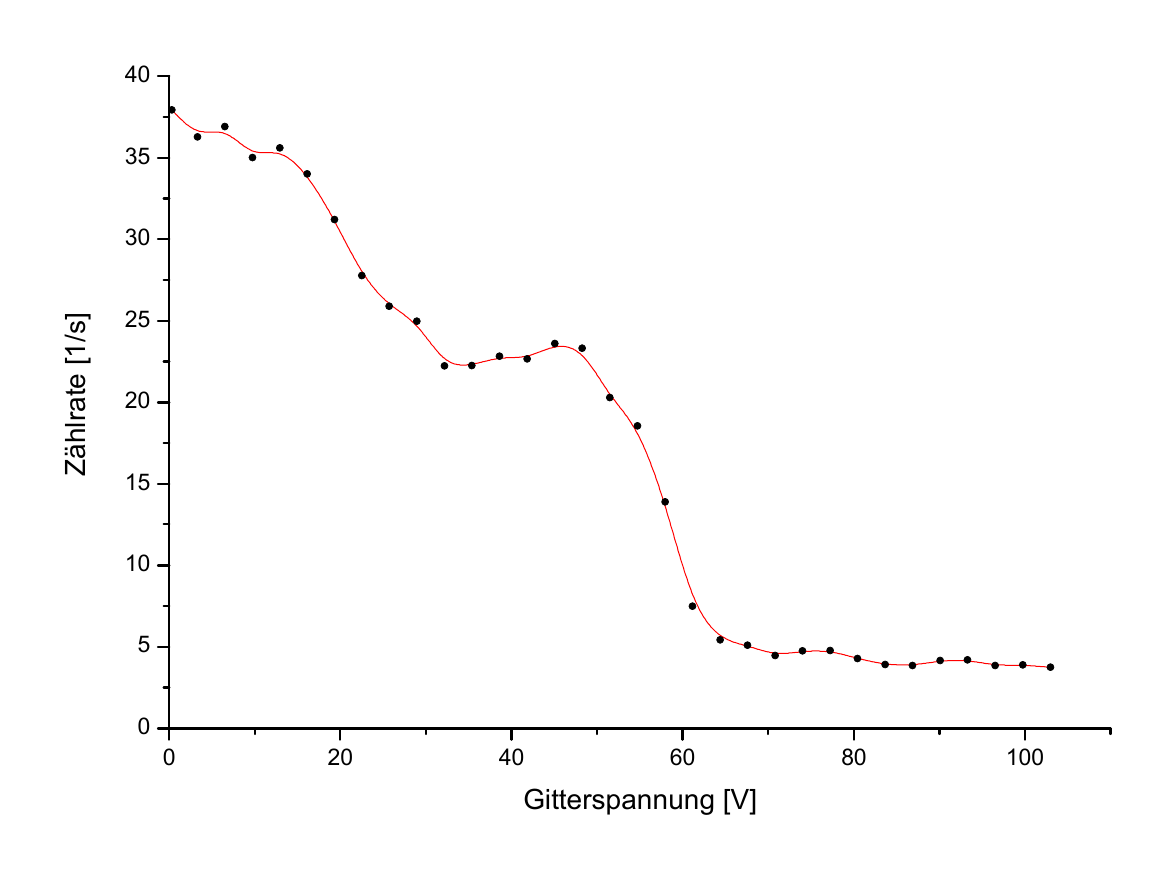}
 \caption{Integrales Spektrum, gemessen auf Höhe des gewölbten Gitters in der Moderatorzelle am Eingang des Trichters}
 \label{fig:intspek-gitter-10-58V}
\end{figure}

Ein sehr großer Teil der Positronen zerstrahlt am gewölbten Gitter, das am Eingang des Trichters angebracht ist (etwa bei 390\,mm). Während der Scan ohne Gas (nach Einberechnung der Absorption) zeigt, dass in etwa die selbe Anzahl an Positronen am Target annihiliert wie am gewölbten Gitter, sieht man, dass bei einem Druck von $2,3 \cdot 10^{-2}$\,mbar beinahe zehn mal mehr Positronen am Gitter annihilieren als am Target. Leider ist es aufgrund der großen Wandstärken im Inneren des Trichters und am Ausgang des Moderators nicht möglich zu untersuchen, wie groß der Anteil der Positronen ist, die in diesem Bereich verloren gehen. Es ist jedoch möglich abzuschätzen, wie hoch der Anteil an zerstrahlenden Positronen am Gitter ist. Dazu wird die Scanvorrichtung an die Position des Gitters gefahren und durch Variation der Spannung am Gitter ein Spektrum aufgenommen. Abbildung \ref{fig:intspek-gitter-10-58V} zeigt ein solches integrales Spektrum bei einem Spannungsverlauf von 58\,Volt (am Eingang des Moderators) bis 10\,Volt (am Ausgang) und einem Druck von $4,5 \cdot 10^{-2}$\,mbar. Deutlich erkennt man drei Stufen: Die hohen Zählraten am Gitter bei niedrigen Spannungen ($\le 10$\,Volt) kommen Zustande, da das Gitter von beiden Seiten ein attraktives Potential für die Positronen bildet. So annihilieren selbst solche Positronen am Gitter die es bereits passiert haben. Im mittleren Spannungsbereich (30--50\,Volt) wird ein Teil der Positronen durch das Gitter transmittiert, während ab 58\,Volt beinahe alle Positronen reflektiert werden.

\begin{figure}
 \centering
 \includegraphics[width=0.7\textwidth]{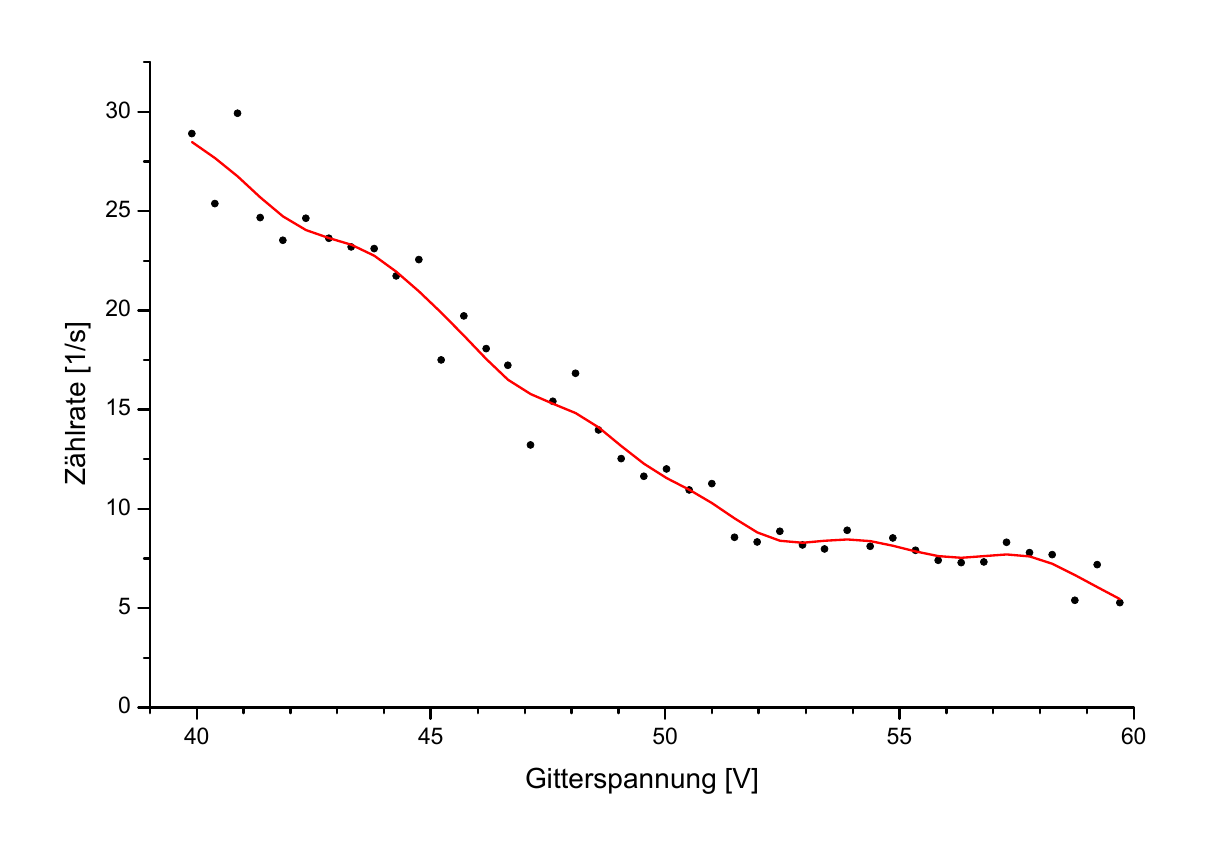}
 \caption{Integrales Spektrum, gemessen am gewölbten Gitter am Eingang des Trichters}
 \label{fig:intspek-gitter-44,5-48V}
\end{figure}

Die Messung wurde nun bei dem in Abschnitt \ref{spektrenmitgas} festgestellten optimalen Spannungsverlauf (44,5--48\,Volt) wiederholt. Abbildung \ref{fig:intspek-gitter-44,5-48V} zeigt wieder das integrale Spektrum, aufgenommen am Gitter. Bei der am Gitter normalerweise eingestellten Spannung von 46,7\,Volt (wie bei den Messungen im vorherigen Abschnitt) erhält man eine Zählrate von 15,6\,s$^{-1}$.\footnote{Dieser Wert wurde aufgrund besserer Statistik aus Abbildung \ref{fig:scan1mitaufbau} entnommen.} Für niedrigere Spannungen am Gitter steigt die Zählrate bis auf 29\,s$^{-1}$ an, möglicherweise sogar noch weiter da die niedrigste gemessene Spannung bereits 39,9\,Volt betrug. Somit kann man feststellen, dass mindestens 46\,\% der am Gitter ankommenden Positronen selbiges passieren können und somit zu einem großen Teil im Trichter oder am Ausgang zerstrahlen müssen, ohne das Target zu erreichen.

\section{Diskussion der Messergebnisse und Ausblick}
Mit den hier vorgestellten Messungen konnte gezeigt werden, dass eine kontinuierliche Moderation eines Positronenstrahls in Stickstoff möglich ist. Die Messung bei einem Druck von $2,3 \cdot 10^{-2}$\,mbar stellte dabei einen guten Kompromiss aus Energieschärfe (inklusive apparativer Auflösung etwa 5\,eV Halbwertsbreite) und Moderationseffizienz (ca. 4\,\% ohne Einkopplungsverluste) dar. Bei niedrigeren Drücken war ein deutlich nachweisbarer Anteil an nicht vollständig moderierten Positronen vorhanden, während die Moderationseffizienz bei höheren Drücken deutlich niedriger war.

Es zeigte sich, dass die Spannungsdifferenz über die Moderatorzelle den entscheidenden Einfluss auf die Energieschärfe der moderierten Positronen hatte. Offensichtlich bewirkte eine größere Spannungsdifferenz ein nachträgliches Beschleunigen der bereits moderierten Positronen. Eine genaue Erklärung dieses Effekts bedarf jedoch weiterer Untersuchungen. Eine weitere Verringerung der Spannungsdifferenz sollte eine Verbesserung der Energieschärfe ermöglichen.

Durch den ortsaufgelösten Scan des Moderators konnten die Verluste innerhalb des Aufbaus lokalisiert werden. Der Bereich zwischen den konusförmig zulaufenden Potentialscheiben und dem Ausgang des Moderators konnte jedoch auf Grund der großen Wandstärken nur unzureichend analysiert werden. Es ist jedoch sehr wahrscheinlich, dass dort ein Großteil der Positronen annihiliert. Für weitere Untersuchungen sollten große Wandstärken in Bereichen vermieden werden, welche für die Verlustidentifikation von Interesse sind.

Für spätere Arbeiten ist geplant, durch differentielles Pumpen am Ausgang, den Strahl aus dem Moderator zu extrahieren und in einem Strahlrohr weiter zu führen.
\clearpage{}
\clearpage{}\chapter{Zusammenfassung}
Im Rahmen dieser Diplomarbeit wurde das Konzept eines Moderators für Positronen untersucht, welches auf der Abbremsung und dem Drift von Positronen in gasförmigem Stickstoff beruht. Dazu wurde ein Versuchsaufbau konstruiert, aufgebaut und in Betrieb genommen. Ein Positronenstrahl wird durch eine kleine Blende in eine Gaszelle eingekoppelt, dort durch Stöße mit den Gasmolekülen abgebremst, durch ein elektrisches Feld auf den Ausgang fokussiert und dahinter analysiert. Durch ausführliche Simulationen mit dem Programm \textsc{Simion 3D} konnte das elektrische Feld im Inneren der Gaszelle optimiert werden. Zur Untersuchung der Energie der moderierten Positronen wurde ein Bremsspannungsanalysator angefertigt und am Ausgang der Zelle installiert. Die Lokalisierung der im Versuchsaufbau zerstrahlenden Positronen wurde durch zwei in Koinzidenz geschaltete Gammadetektoren ermöglicht, welche gegenüber liegend längs des Moderators auf Schienen verfahrbar waren.

Die Software zur Steuerung des Positronenanalysators, der Potentiale im Inneren der Zelle und des Linearscanners sowie diverse Programme zum Aufnehmen der Messdaten (beispielsweise Druck und Zählrate) wurde eigens in der Entwicklungsumgebung LabVIEW erstellt. Dabei wurde Wert auf einen modularen Aufbau der Programme gelegt, damit auch eine spätere Verwendung der Unterprogramme, wie beispielsweise der Druckauslese, für andere Zwecke leicht möglich ist.

Messungen am Gasmoderator konnten an der Positronenquelle \textsc{Nepomuc} am FRM~II während zweier Messzyklen zu jeweils 10 Tagen durchgeführt werden. Dabei wurde der Positronenstrahl in den Moderator eingekoppelt und die Energie der am Target annihilierten Positronen analysiert. Dies wurde bei unterschiedlichen Drücken in der Moderatorzelle und verschiedenen elektrischen Potentialen über den Moderator wiederholt. Mit Hilfe der Scannapparatur wurden die Positronenverluste im Aufbau lokalisiert.

In dem Experiment konnte die Moderation eines Positronenstrahls in Stickstoff nachgewiesen werden. Es wurde eine Ausbeute an moderierten Positronen am Ausgang der Zelle -- abhängig vom Stickstoffdruck -- von bis zu 6\,\% beobachtet. Dieser Wert liegt noch unter dem theoretischen Wert von 26\,\%, wie er im Rahmen der vorliegenden Arbeit durch eine Monte-Carlo-Simulation ermittelt worden ist. Ortsaufgelöste Scans der Annihilationsstrahlung mittels der koinzidenten Gammadetektoren deuten darauf hin, dass ein Großteil der Positronen im Bereich des Fokus verloren geht. Eine genauere Untersuchung bedarf einer umfangreichen Änderung des Versuchsaufbaus und soll zukünftig in Verbindung mit der Extraktion des moderierten Strahls aus dem Gas in ein Strahlrohr durchgeführt werden. Auch die Abhängigkeit der Energieschärfe der moderierten Positronen von der am Moderator angelegten Driftspannung sollte im Zuge der Weiterentwicklung dieser Methode untersucht werden.

\nocite{DiplJakob}
\nocite{DiplThomas}

\nocite{Hugenschmidt2006} \nocite{Murphy1992} \nocite{Kaupilla1982} \nocite{Hoffman1982} \nocite{Surko2005} \nocite{Coleman2000} \nocite{Cassidy2005} \nocite{Cassidy2006-2} \nocite{GrossPhysikIV}
\clearpage{}

\appendix
\clearpage{}\chapter{Anhang}

\section{Aufbau}
\begin{figure}[h]
 \centering
 \includegraphics[width=0.98\textwidth]{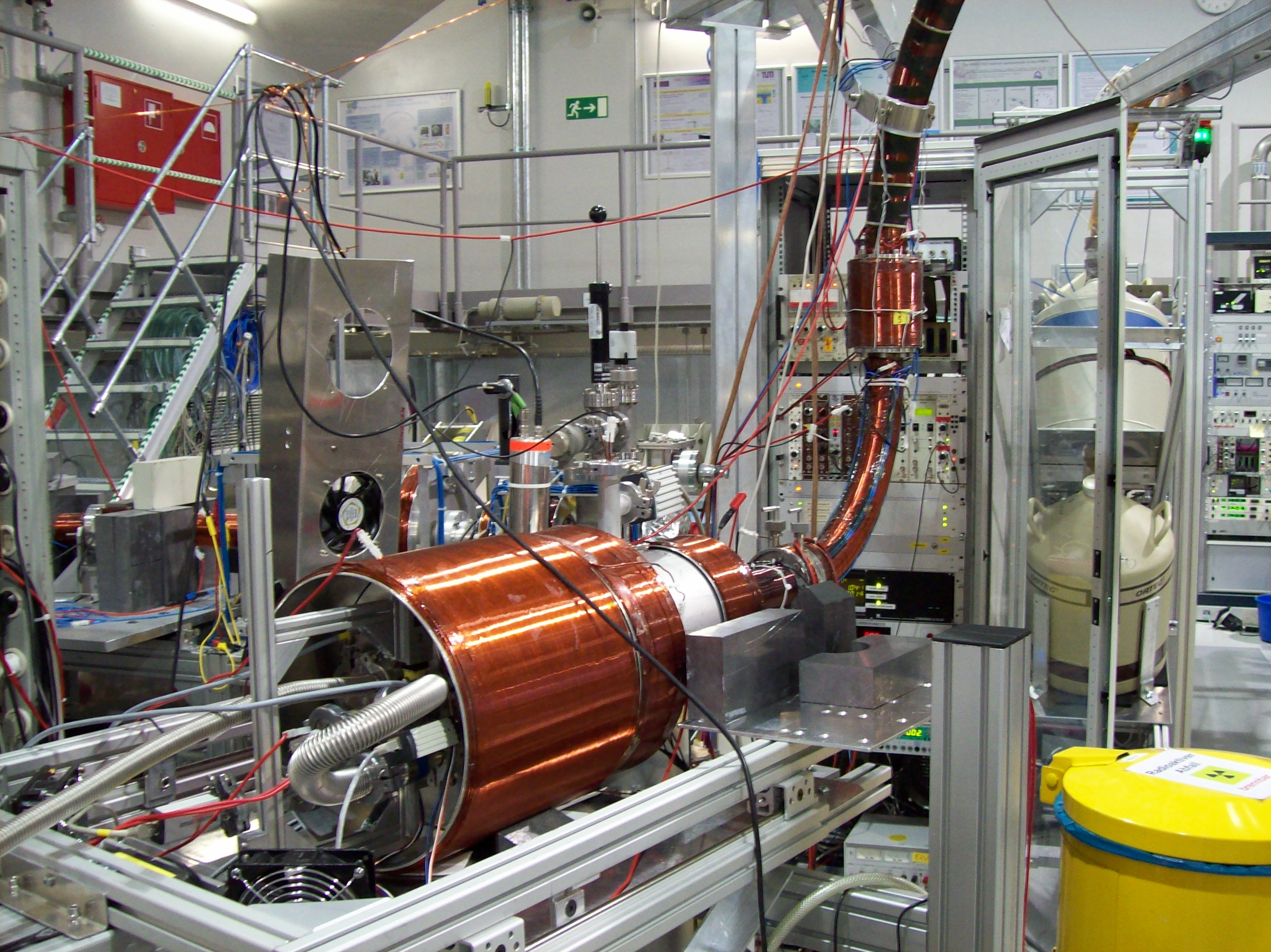}
 \caption{Übersichtsfoto des Moderatoraufbaus: Der Positronenstrahl wird von oben kommend in die waagrecht liegende Moderatorkammer, im Solenoid, eingekoppelt. Parallel zur Kammer läuft auf Schienen der Linearscanner zum ortsaufgelösten Nachweis der annihilierten Positronen.}
 \label{fig:Übersichtsfoto}
\end{figure}

\section{Spulenströme}

\subsection*{Spulenströme am Gasmoderator} \label{StroemeModerator}

\subsubsection*{Führungsfelder}
\begin{tabular}{|c|l|l|}
\hline
\textbf{Spule (Nr.)} & \textbf{Beschreibung} & \textbf{Stromstärke [A]}\\
\hline
\hline
1 & Eingang  Pumprohr & 10,0\\
\hline
2 & Schwarzer  Draht & 0,0\\
\hline
3 & Eingang  Gasmoderator & 8,1\\
\hline
4 & Gasmoderator  Innenraum & 0,0\\
\hline
5 & Große  Spule & 7,0\\
\hline
\end{tabular}

\subsection*{Spulenströme an der Beamline}

\subsubsection*{Führungsfelder}
\begin{tabular}{|c|l|l|}
\hline
\textbf{Spule (Nr.)} & \textbf{Beschreibung} & \textbf{Stromstärke [A]}\\
\hline
\hline
F21 & Balg 300 & 8,4\\
\hline
F22 & & 7,6\\
\hline
F23-25 & & 8,6\\
\hline
F26-28 & & 10,1\\
\hline
F29 & & 10,1\\
\hline
F30 & Weiche & 5,5\\
\hline
F31 & & 10,2\\
\hline
F32 & Bogen über Weiche & 10,11\\
\hline
F34 & Gerades Stück oben & 9,98\\
\hline
F35 & Bogen oben am Open Beam Port & 10,0\\
\hline
F36 & Balg & 11,5\\
\hline
F37 & Bogen unten am Open Beam Port & 9,0\\
\hline
& Beide Spulen am Nedflansch & 10,2\\
\hline
& Zwischenstück vor Moderator & 9,0\\
\hline
\end{tabular}

\subsubsection*{Korrekturfelder}
\begin{tabular}{|c|l|l|}
\hline
\textbf{Spule (Nr.)} & \textbf{Beschreibung} & \textbf{Stromstärke [A]}\\
\hline
\hline
KP16V & \multirow{2}{*}{Bogen unter Strahlweiche} & -3,67\\
\cline{1-1}
\cline{3-3}
KP16H & & 3,57\\
\hline
KP18V & \multirow{2}{*}{Bogen über Strahlweiche} & 0,8\\
\cline{1-1}
\cline{3-3}
KP18H & & -2,7\\
\hline
KP19V & \multirow{2}{7cm}{Bogen oben am Open Beam Port (Polarität unbekannt)}& ca. 3,5 (schwankt)\\
\cline{1-1}
\cline{3-3}
KP19H & & 3,23\\
\hline
KP20V & \multirow{2}{7cm}{Bogen unten am Open Beam Port (Polarität unbekannt)} & 0,52\\
\cline{1-1}
\cline{3-3}
KP20H & & 6,79\\
\hline
\end{tabular}

\section{Messprotokoll}
Auf den drei nächsten Seiten folgt eine Übersicht über die bei den Messungen gewonnenen Daten. Die Daten wurden in übliche ASCII-Dateien mit der Endung \texttt{.txt} gespeichert. Hier der Aufbau der unterschiedlichen Dateien:

Folgende Dateitypen sind mehrspaltig, Beschreibung der Spalten von links nach rechts:
\begin{description}
	\item[Spektren:] Spannung [V], Zählrate [1/s], Standardabweichung d. Zählrate
	\item[Scans:] Position [mm], Zählrate [1/s], Standardabweichung d. Zählrate
	\item[Druckmessungen:] Zeit (Kanal S1) [s], Druck (Kanal S1) [mbar], Zeit (Kanal S2) [s], Druck (Kanal S2) [mbar], Zeit (Kanal ITR) [s], Druck (Kanal ITR) [mbar]
	\item[Druck-gegen-Zählrate:] Druck (Kanal S1) [mbar], Druck (Kanal S2) [mbar], Druck (Kanal ITR) [mbar], Zählrate [1/s]
\end{description}

Folgende Dateitypen sind einspaltig bzw. einzeilig:
\begin{description}
  \item [Spannungswerte DA-AD:] Zeile 1-20: 16bit-Wert des Digital-Analog-Wandlers\\(Ausgangsspannung 10\,Volt)\\Zeile 22-41: Sollspannungen am Ausgang des Spannungsteilers in Volt
  \item [Netzteileinstellungen:] Stromstärken des 20-Kanal-Netzteils, strichpunktgetrennte Werte, Einheit: mA
\end{description}

\begin{landscape}
\begin{figure}
 \centering
 \includegraphics[width=21cm, page=1]{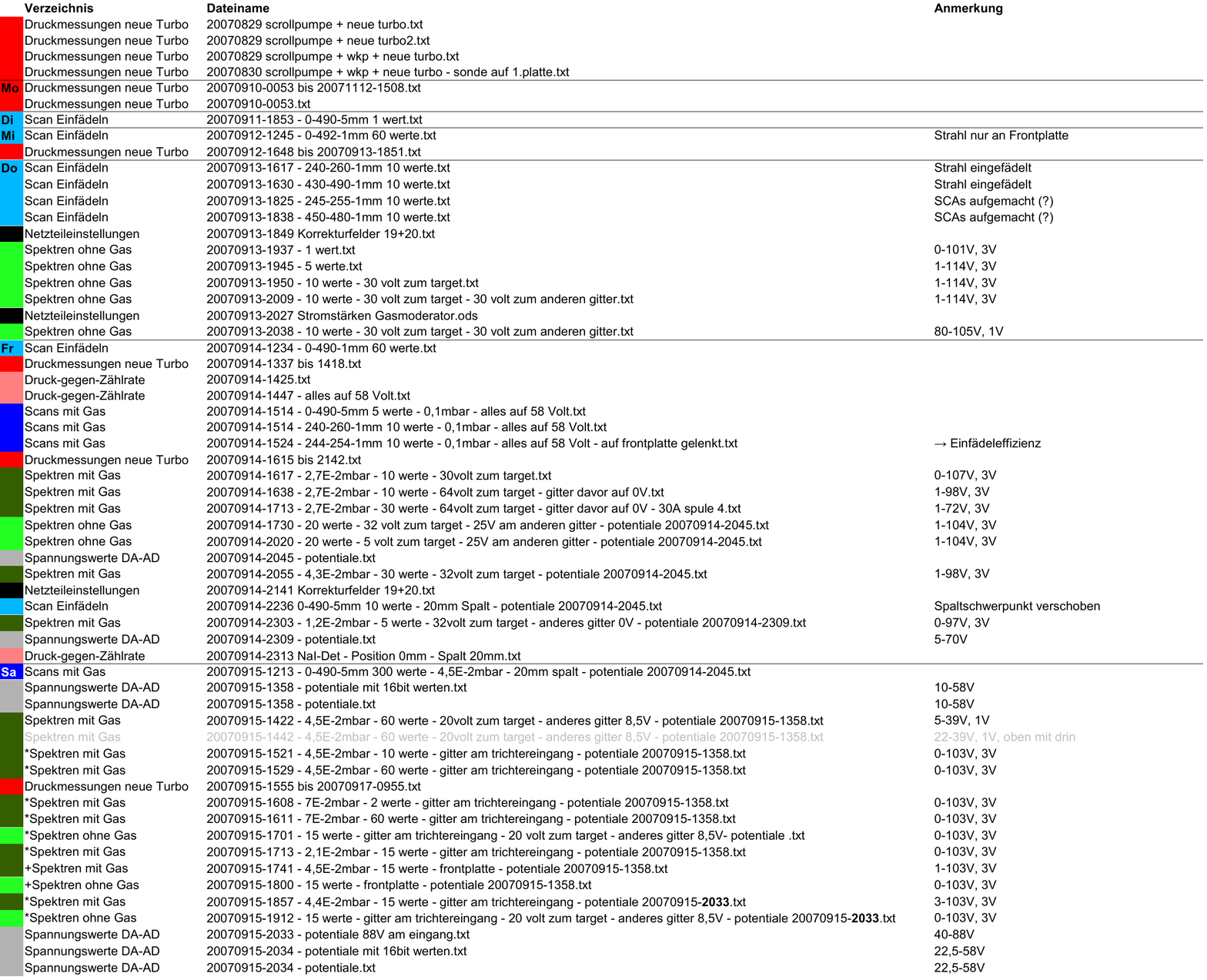}
\end{figure}

\begin{figure}
 \centering
 \includegraphics[width=21cm, page=2]{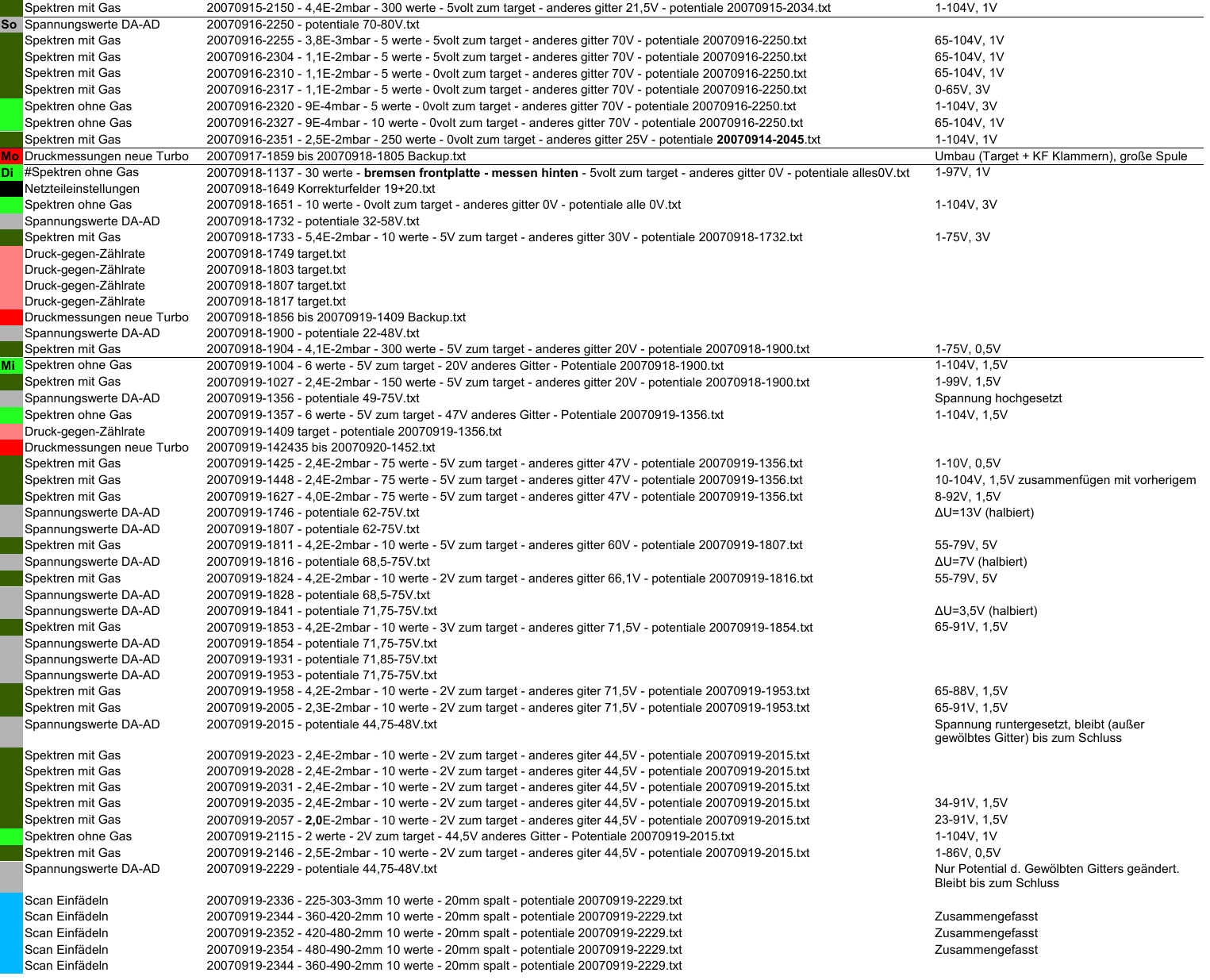}
\end{figure}

\begin{figure}
 \centering
 \includegraphics[width=21cm, page=3]{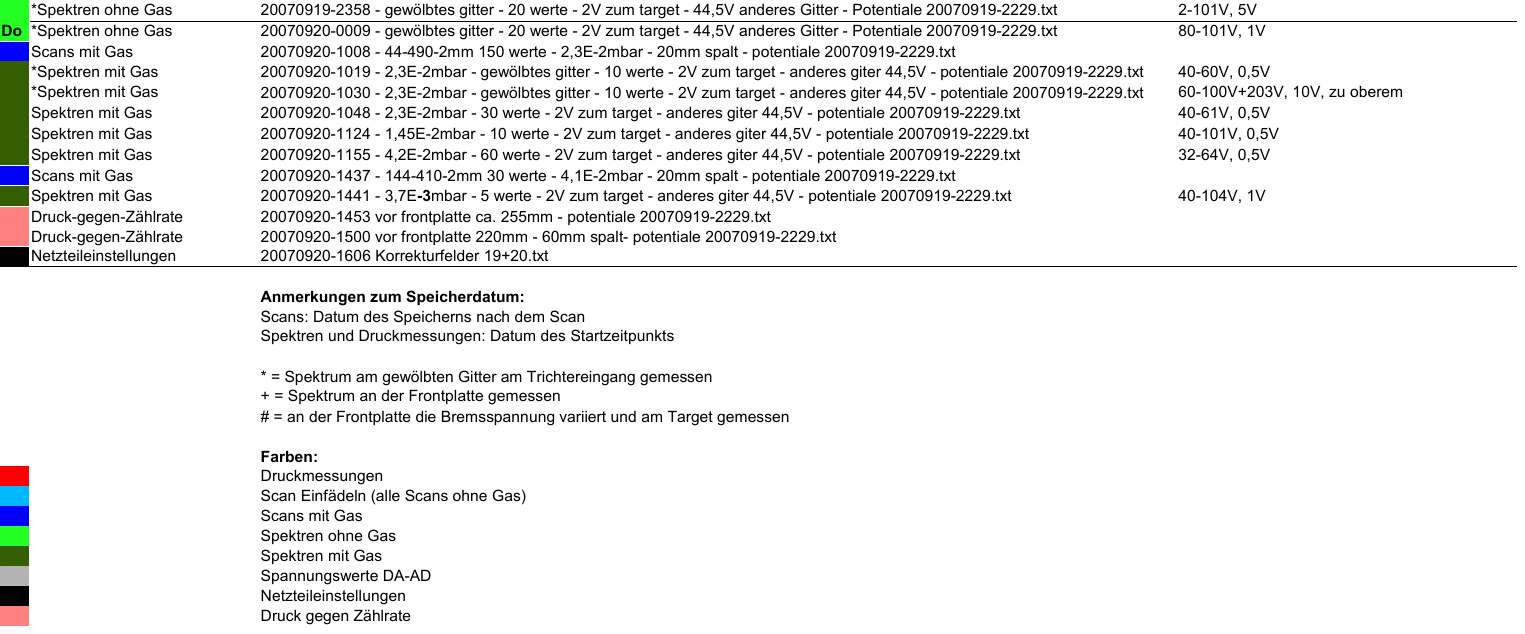}
\end{figure}
\end{landscape}

\section{Übersicht LabVIEW-Programme}
Zur Durchführung der Messungen wurden im Rahmen der Diplomarbeit verschiedene Programme in der graphischen Programmiersprache LabVIEW\footnote{\textbf{Lab}oratory \textbf{V}irtual \textbf{I}nstrumentation \textbf{E}ngineering \textbf{W}orkbench} erstellt. Aus Wikipedia\footnote{\url{http://de.wikipedia.org/wiki/LabVIEW}, Stand: 26.11.2007}:

\begin{quotation}
Haupt-Anwendungsgebiete von LabVIEW sind die Mess- und Automatisierungstechnik. Die Programmierung erfolgt mit einer graphischen Programmiersprache genannt \enquote{G}, nach dem Datenfluss-Modell. Durch diese Besonderheit eignet sich LabVIEW besonders gut zur Datenerfassung und -verarbeitung.

LabVIEW-Programme werden als Virtuelle Instrumente oder einfach VIs bezeichnet. Sie bestehen aus zwei Komponenten: das Frontpanel enthält die Benutzerschnittstelle, das Blockdiagramm den graphischen Programmcode. Dieser wird nicht von einem Interpreter abgearbeitet, sondern kompiliert. Dadurch ist die Performance vergleichbar mit der anderer Hochsprachen.
\end{quotation}

\subsection*{Steuerung des Linear-Scanners} \label{LabVIEW-Scanner}
\begin{figure}
 \centering
 \includegraphics[width=0.6\textwidth]{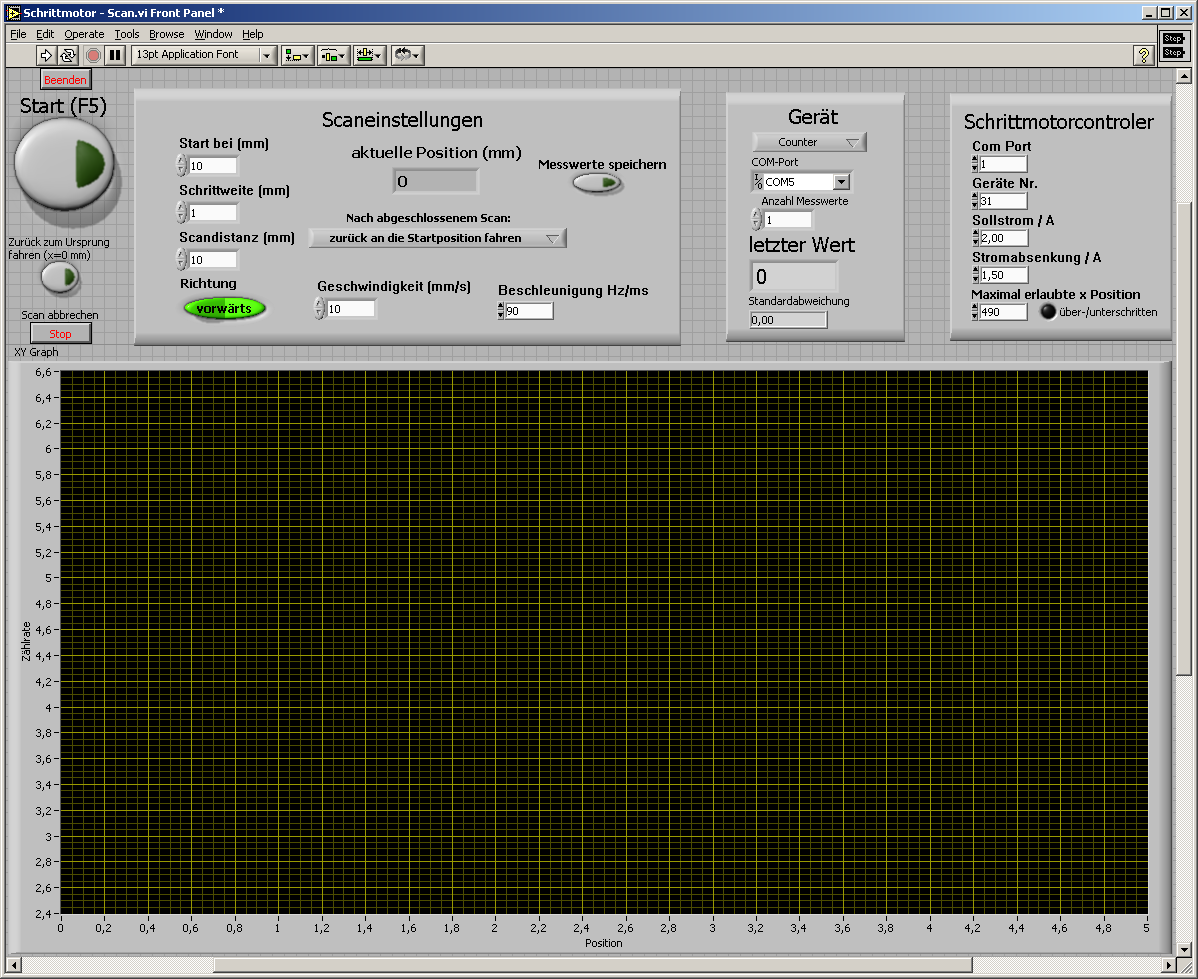}
 \caption{\enquote{Screenshot} des Programms zur Steuerung des Linear-Scanners}
 \label{fig:Schrittmotor}
\end{figure}
Für die Steuerung des Linear-Scanners zur Aufnahme der ortsabhängigen Annihilationsstrahlung wurde ein einfach zu bedienendes Programm entwickelt, dass nach Eingabe eines Startpunktes, einer Schrittweite und der Scandistanz beide NaI-Detektoren selbständig jeden Messpunkt anfährt und dabei eine einstellbare Zeit die Zählrate misst (siehe Abbildung \ref{fig:Schrittmotor}). Das Unterprogramm zum Auslesen des Zählers wurde so konzipiert, dass es sich auch leicht in andere Programme integrieren lässt.

\subsection*{DA/AD-Controller}
\begin{figure}
 \centering
 \includegraphics[width=0.6\textwidth]{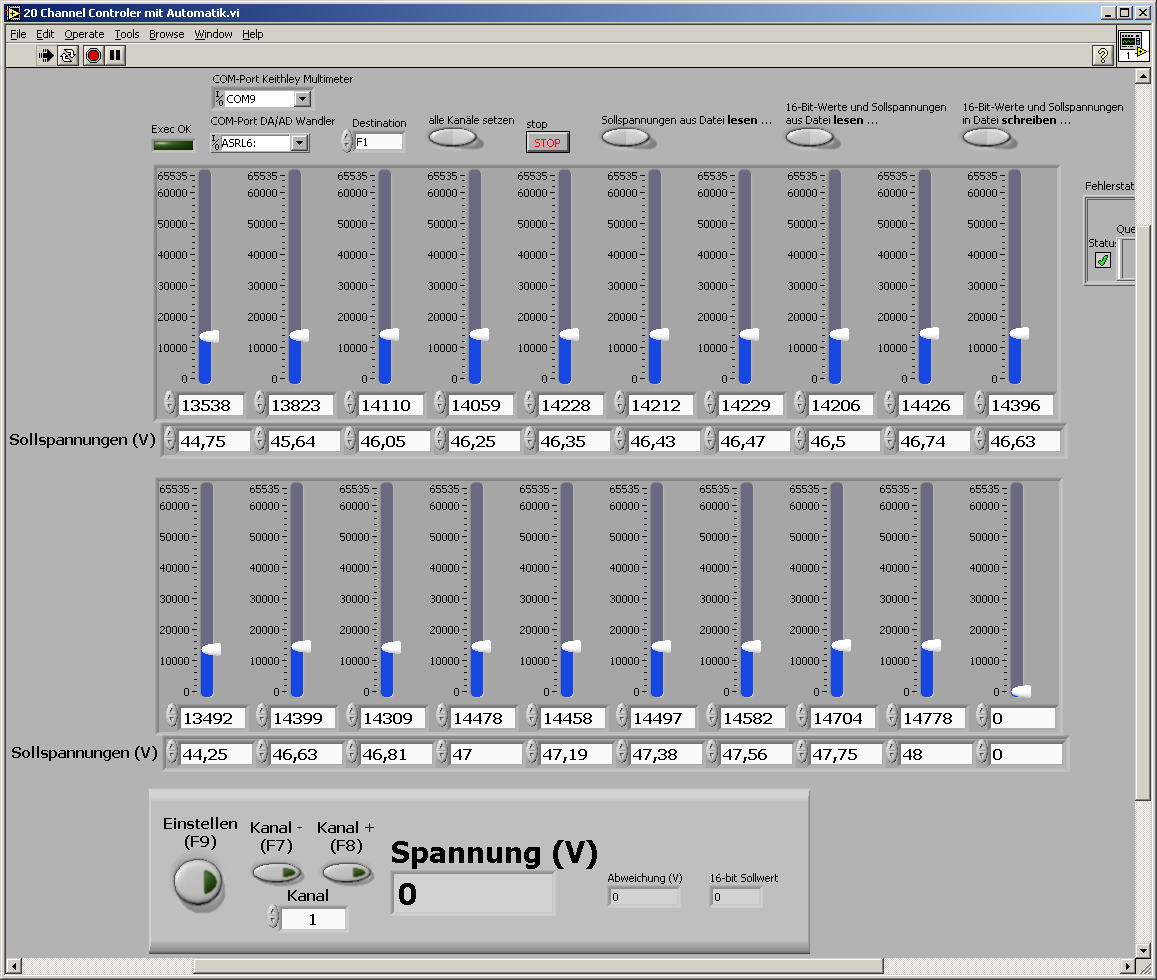}
 \caption{Programm zum Einstellen der Spannungen in der Gaszelle}
 \label{fig:Channelcontroler}
\end{figure}

Mit Hilfe eines Digital-Analog-Wandlers wurde ein Mehrkanal-Spannungsteiler geregelt über den wiederum die Potentiale in der Gaszelle sowie die Bremsspannung zur Analyse der Positronenenergien gesteuert wurden. Es wurde ein Programm erstellt (siehe Abbildung \ref{fig:Channelcontroler}), dass über ein Keithley 2000-Multimeter in der Lage ist die Spannung eines Kanals, entsprechend einer gegebenen Sollspannung, selbständig einzustellen. Für die Aufnahme der Bremsspektren wurde ein separates Programm erstellt, dass das Keithley-Multimeter benutzt um die eingestellte Bremsspannung präzise zu messen.

\subsection*{Druckauslese}
\begin{figure}
 \centering
 \includegraphics[width=0.98\textwidth]{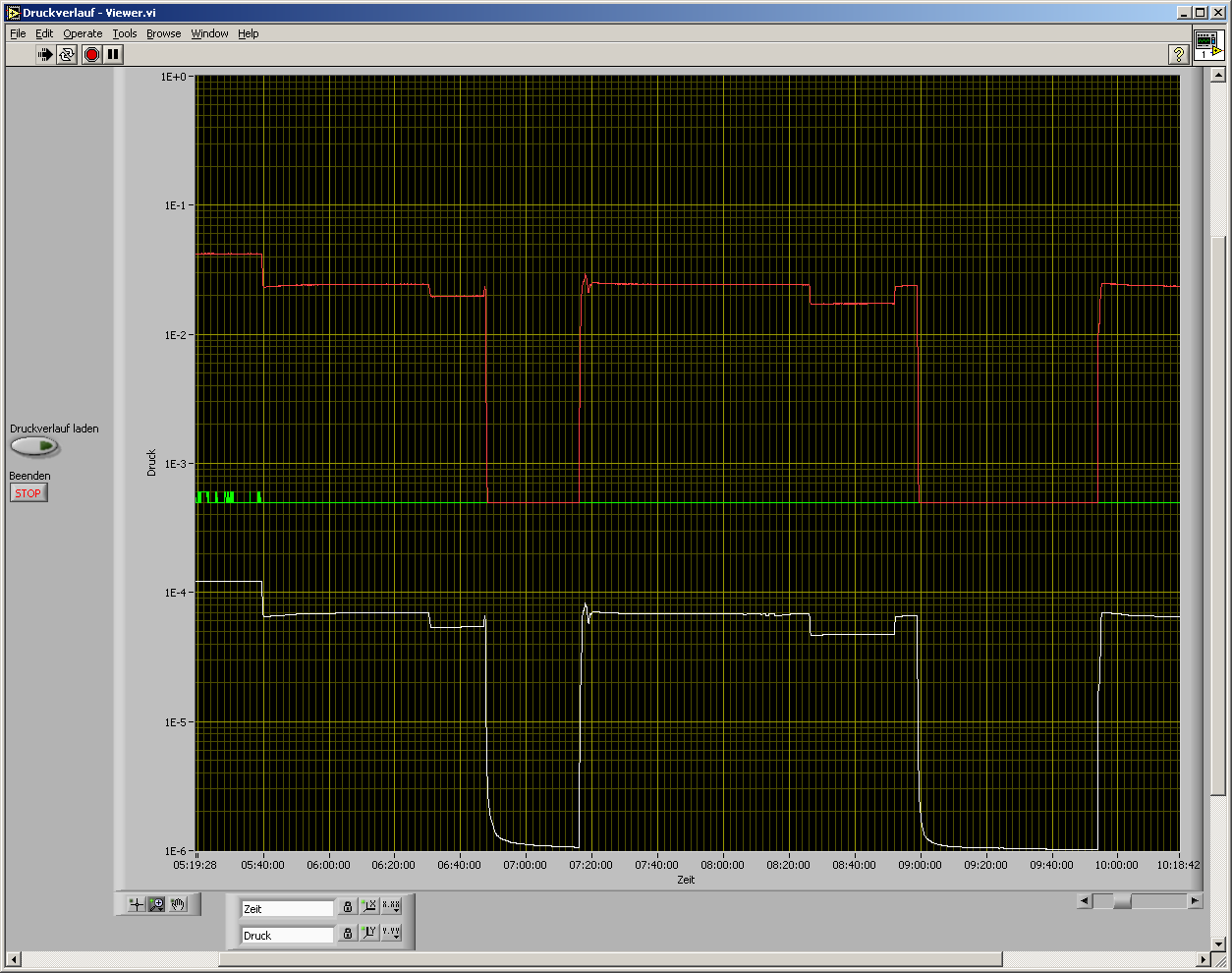}
 \caption{Druckverlauf}
 \label{fig:Druckverlauf}
\end{figure}

Um den Druckverlauf im Inneren der Gaszelle und im Pumprohr zu überwachen wurde ein Programm erstellt, dass in der Lage ist den Druckverlauf von bis zu vier Drucksensoren aufzuzeichnen. Abbildung \ref{fig:Druckverlauf} zeigt einen mit drei Sensoren aufgenommenen Druckverlauf über eine Dauer von etwa 5 Stunden.
\clearpage{}

\bibliographystyle{geralpha} \bibliography{references}

\end{document}